\documentclass[nofootinbib,
 reprint,
 amsmath,amssymb,
 aps]{revtex4-2}

\usepackage{graphicx}
\usepackage[most]{tcolorbox}
\graphicspath{ {./images/} }
\usepackage{tikz}    
\usepackage{float}
\tikzset{mynode/.style={draw, thick, circle, inner sep=1.5pt}}
\usepackage{dcolumn}
\usepackage{bm}
\usepackage[dvipsnames]{xcolor}
\usepackage{cleveref}
\usepackage{amsmath}
\usepackage{relsize}
\usepackage{suffix}
\usepackage{mathtools}

\numberwithin{equation}{section}

\ExplSyntaxOn

\NewDocumentCommand{\pFq}{O{}mmmmm}
 {
  \group_begin:
  \keys_set:nn { hypergeometric } { #1 }
  \hypergeometric_print:nnnnn { #2 } { #3 } { #4 } { #5 } { #6 }
  \group_end:
 }
\NewDocumentCommand{\hypergeometricsetup}{m}
 {
  \keys_set:nn { hypergeometric } { #1 }
 }

\tl_new:N \l_hypergeometric_divider_tl
\tl_new:N \l_hypergeometric_left_tl
\tl_new:N \l_hypergeometric_right_tl

\keys_define:nn { hypergeometric }
 {
  symbol .tl_set:N = \l_hypergeometric_symbol_tl,
  symbol .initial:n = F,
  separator .tl_set:N = \l_hypergeometric_separator_tl,
  separator .initial:n = {},
  skip .tl_set:N = \l_hypergeometric_skip_tl,
  skip .initial:n = 8,
  divider .choice:,
  divider/semicolon .code:n = \tl_set:Nn \l_hypergeometric_divider_tl { \;; },
  divider/bar .code:n = \tl_set:Nn \l_hypergeometric_divider_tl { \;\middle|\; },
  divider .initial:n = semicolon,
  fences .choice:,
  fences/brack .code:n = 
   \tl_set:Nn \l_hypergeometric_left_tl {[}
   \tl_set:Nn \l_hypergeometric_right_tl {]},
  fences/parens .code:n = 
   \tl_set:Nn \l_hypergeometric_left_tl {(}
   \tl_set:Nn \l_hypergeometric_right_tl {)},
  fences .initial:n = brack,
 }

\cs_new_protected:Nn \hypergeometric_print:nnnnn
 {
  {} \sb {#1} \l_hypergeometric_symbol_tl \sb { #2 }
  \left\l_hypergeometric_left_tl
  \genfrac .. 
           {0pt} 
           {} 
           { \__hypergeometric_process:n { #3 } } 
           { \__hypergeometric_process:n { #4 } } 
  \l_hypergeometric_divider_tl
  #5
  \right\l_hypergeometric_right_tl
 }

\cs_new_protected:Nn \__hypergeometric_process:n
 {
  \clist_use:nn { #1 }
   {
    {\l_hypergeometric_separator_tl}
    \mspace { \l_hypergeometric_skip_tl mu }
   }
 }

\ExplSyntaxOff

\DeclarePairedDelimiterX\MeijerM[3]{\lparen}{\rparen}%
{\begin{smallmatrix}#1 \\ #2\end{smallmatrix}\delimsize\vert\,#3}

\newcommand\MeijerG[8][]{%
  G^{\,#2,#3}_{#4,#5}\MeijerM[#1]{#6}{#7}{#8}}

\WithSuffix\newcommand\MeijerG*[7]{%
  G^{\,#1,#2}_{#3,#4}\MeijerM*{#5}{#6}{#7}}

\def\Xint#1{\mathchoice
   {\XXint\displaystyle\textstyle{#1}}%
   {\XXint\textstyle\scriptstyle{#1}}%
   {\XXint\scriptstyle\scriptscriptstyle{#1}}%
   {\XXint\scriptscriptstyle\scriptscriptstyle{#1}}%
   \!\int}
\def\XXint#1#2#3{{\setbox0=\hbox{$#1{#2#3}{\int}$}
     \vcenter{\hbox{$#2#3$}}\kern-.5\wd0}}

\def\dashint{\Xint-}

\begin{document}

\preprint{APS/123-QED}

\title{ Approximating the Fourier Transform of Ring-Like Images: the Focal Expansion }

\author{Filip Niewinski}
    \affiliation{Department of Physics, Harvard University, Cambridge, MA
02138, USA}
\email{fniewinski@g.harvard.edu}

\author{Michael D.\ Johnson }
\affiliation{Center for Astrophysics $\vert$ Harvard \& Smithsonian, 60 Garden Street, Cambridge, MA 02138, USA}
\affiliation{Black Hole Initiative at Harvard University, 20 Garden Street, Cambridge, MA 02138, USA}
\email{mjohnson@cfa.harvard.edu}

\date{\today}

\begin{abstract}

We derive and showcase a novel approach to approximating Fourier transforms in higher dimensions, focusing specifically on the case of 2D radially concentrated (`ring-like') functions. We first reduce the problem to that of evaluating the Hankel transforms of each angular mode of the image and then use our \textit{focal expansion} to approximate these remaining Hankel transforms. Our method provides a single approximation that remains accurate from small to large spatial frequencies, bridging regimes where moment-based or large-frequency asymptotic expansions are individually reliable. 
We explore a series of examples, showing that the leading focal term provides an accurate global approximation for a broad range of functions. We demonstrate the utility of this approximation by examining the interferometric response for toy models of a black hole's ``photon ring,'' highlighting the application to experiments designed to measure this feature such as the Black Hole Explorer.
\end{abstract}

\maketitle

\section{\label{sec:intro} Introduction}

\subsection{Motivation}

Fourier analysis is an indispensable part of the mathematical and physical sciences \cite{Bracewell,Fourier_Engineer,Fourier_Math}. However, analytic Fourier transforms are often unavailable, especially in higher dimensions -- a brief perusal of integral tables will show those containing one dimensional (1D) Fourier transforms to be the most expansive, with the length decreasing as we increase the dimensionality. Here, we focus on analytic approximations for the two-dimensional (2D) Fourier transform, with higher-dimensional generalization relegated to the appendices.

The motivation for our focus on the 2D Fourier transform comes from studies of black hole images, a new field prompted by the successful imaging of the nuclear supermassive black holes in the Milky Way and in the M87 galaxy using astronomical interferometry by the Event Horizon Telescope collaboration \cite[EHT;][]{EHTM87,EHTSgrA}. These images appear as approximately circular rings, but with non-trivial radial and azimuthal structure. Moreover, these images are expected to include a series of exponentially sharper rings from strong gravitational lensing near the black hole \cite{Bardeen_1973,Luminet_1979,Logrings,PhotonRingReview}. This rich ``photon ring'' structure both cleanly encodes the black hole properties and dominates the interferometric response on long baselines, which sample corresponding large spatial wavenumbers of the image, providing a crisp motivation for improved interferometric studies of black holes \citep{PhotonRingReview}. Hence, deriving model-agnostic predictions for the interferometric response and understanding their connection to the image features is of intense interest in shaping and interpreting the results from future experiments with longer baselines such as the Black Hole Explorer \cite{BHEX_Concept}.

Specifically, the data one obtains from an astronomical interferometer is \textit{not} the image as it appears on the sky but is instead a discrete (and often sparse) sampling of the image's two-dimensional Fourier transform \cite{TMS}. Given that we expect images of black holes to be ring-like this raises a natural question: what aspect of the measured Fourier transform carries information about this approximate circular symmetry of the image? And how do small perturbations away from a perfect circle (thickness, shape, etc.) manifest in the interferometric response? While the asymptotic Fourier response is straightforward to compute in certain toy models of black hole images \cite[e.g.,][]{PhotonRingReview,Gralla}, the generalization to more realistic astronomical images with non-trivial radial and azimuthal structure is a surprisingly difficult problem to deal with analytically. In particular, the envelope of the periodic visibility response is especially challenging to compute in general cases of interest.  For an example of some of the difficulties previous investigations have encountered when applying standard approximation techniques to this problem, see \cite{AlexPaper}.

Here, we present a new analytic approximation -- \textit{the focal expansion} -- that resolves some of these difficulties. 
We first split the function into angular modes, in 2D this amounts to expanding $f(r,\theta)$ into a Fourier \textit{series} at each radius 
\begin{align}\label{eq: ang mode decomposition introduction}
    f(r,\theta)=\sum_{n\in\mathbb{Z}}g_n(r) e^{i n \theta}
\end{align}
with the coefficients now dependent on the distance away from the origin. One can then show (see section~\ref{sec: preliminaries} below) that the 2D Fourier transform can be written as
\begin{equation} \label{eq: Hankel tr intro}
    \mathcal{F}[f](\rho, \phi)=2\pi\sum_{n\in\mathbb{Z}}e^{i n (\phi-\pi/2)}\int_0^{\infty}r g_n(r) J_n(2\pi r \rho)\, dr
\end{equation}
where $(\rho, \phi)$ are spherical coordinates in Fourier space and the $J_n$ are Bessel functions. The remaining integrals over the angular modes are the so called Hankel transforms \cite{NIST} and the focal expansion is an analytic method for approximating them. Our approach provides a perturbative series with each term given by a superposition of Bessel functions, each with frequency determined by the radius of the ring and modulated in amplitude by functions encoding the radial profile of the ring. In addition, the focal expansion is \textit{uniformly valid}, meaning it performs well over the \textit{full} range of arguments, without uncontrollably diverging from the true function at either small or large wavenumber, in contrast to Taylor or asymptotic expansions respectively. In fact, although the focal expansion is a series of infinitely many terms, due to this universal validity we found that just the leading order term sufficed in most practical situations.

There are two important caveats to the statements above. The first is that the angular decomposition \cref{eq: ang mode decomposition introduction} might be difficult to perform analytically as it does involve evaluating Fourier series integrals. The second caveat is that we have not escaped taking a Fourier transform altogether -- in the focal expansion the role of modulating the amplitude of the Bessel functions is played by, essentially, \textit{one} dimensional Fourier transforms of the radial profiles $g_n(r)$ of the image. What this means, however, is that our scheme offers a dimensional reduction: evaluating higher dimensional Fourier transforms (at least for functions which manifest approximate spherical symmetry) can be reduced to evaluating 1D Fourier transforms. Alternatively, one can think of the focal expansion as a way to exchange the relatively obscure Hankel transforms of \cref{eq: Hankel tr intro} for 1D Fourier transforms. As we remarked earlier, the 1D Fourier transform is more analytically tractable than either its higher dimensional version or the Hankel transform, so this is already an improvement. 

The outline of the rest of the paper is as follows. We first finish section~\ref{sec:intro} by providing a representative example highlighting the application and properties of our approach. We then derive the full focal expansion in section~\ref{sec: The Method} and provide a series of examples probing the performance of our scheme in Section~\ref{sec: examples}. Finally, we offer some brief concluding remarks in Section~\ref{sec:conclusion}.

\subsection{Demonstration \label{sec: demonstration}}

We now provide an example of our method at work. Consider a distant observer looking at a black hole. Let $(r,\theta)$ denote the polar coordinates in the 2D image plane of the observer, centered on the black hole, and let $(X_1, X_2)$ denote the corresponding cartesian coordinates with $\Vec{X}$ denoting an arbitrary point on the plane. One can show \cite{Luminet_1979,Logrings,PhotonRingReview} that for a black hole with an emissive, optically thin, and geometrically thick accretion disc, the observed light intensity will diverge logarithmically approaching a (very nearly circular) critical curve on the image -- the photon ring. This motivates an interest in the 2D Fourier transform of a ring of radius $a$ for which
\begin{align}\label{eq: intro log approx}
    f(r,\theta)\propto \log \left(\frac{c}{|r-a|}\right)\quad \text{when } r\approx a,
\end{align}
where $c$ is a constant determining how wide the logarithmic divergence is. For concreteness, in this section we will focus on the case $c/a=1$.

\begin{figure*}
    \centering
    \includegraphics[width=\linewidth]{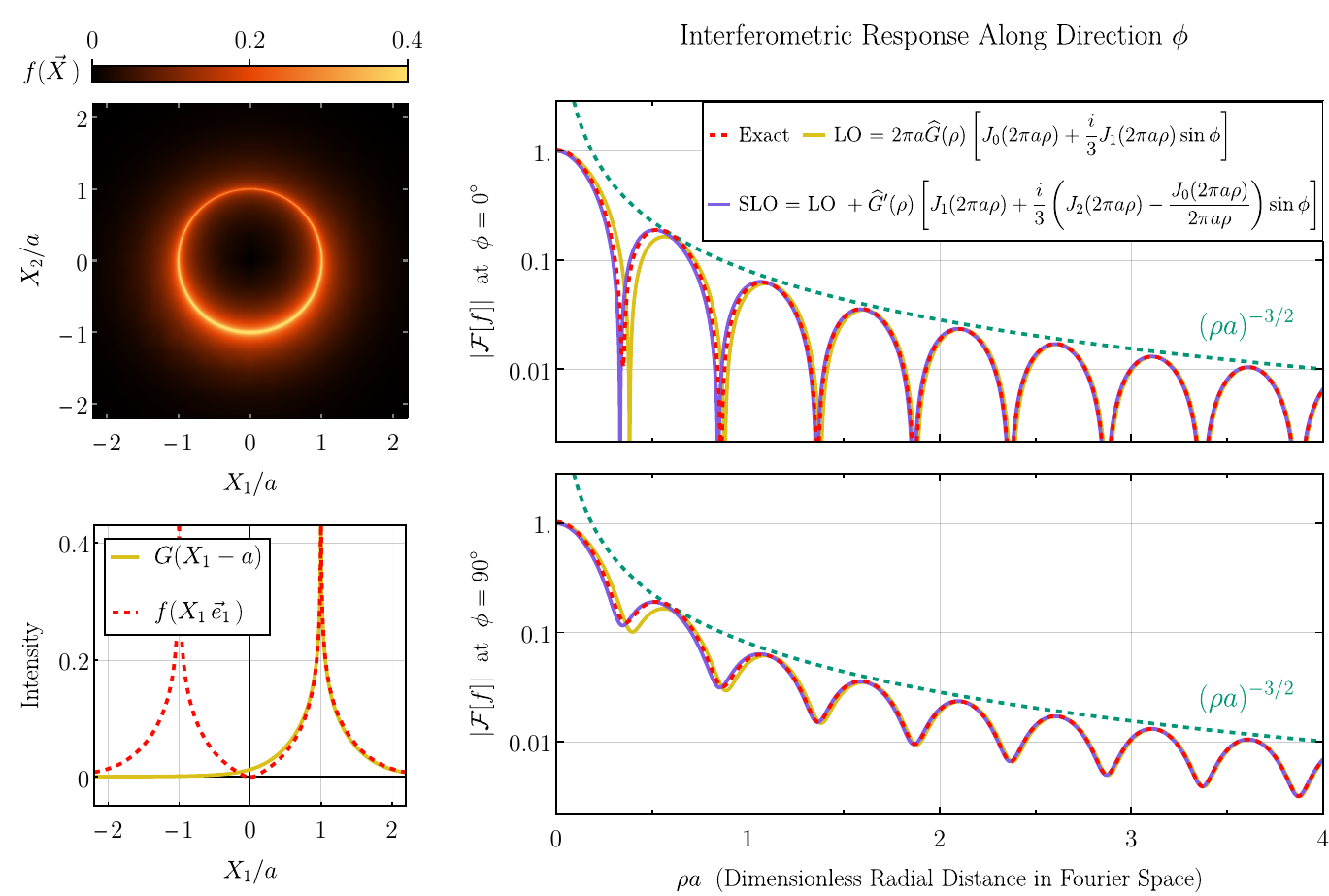}
    \caption{\textit{Top left: } Toy model for an image of a nearly circular photon ring from a spherical accretion flow onto a black hole with mild $m=1$ asymmetry \cref{eq: intro f}. \textit{Bottom left: } Cross section of the image along the $X_1$ axis ($\Vec{e}_1$ being the unit $X_1$-coordinate vector) compared to the peak-approximating function $G(X_1-a)$ given by \cref{eq: intro g0 g1 g-1 approx}. \textit{Top and bottom right: } Amplitude of the image Fourier transform $|\mathcal{F}[f]|(\rho,\phi)$ (i.e., the interferometric visibility) along the $\phi=0^{\circ}$ (top) and $\phi=90^{\circ}$ (bottom) directions. Curves show the exact Fourier response (computed numerically; red dashed), the leading order focal expansion (LO; yellow solid), and the subleading order focal expansion (SLO; blue sold), with $\widehat{G}(\rho)$ in these approximations given by \cref{eq: intro Ghat}. These approximations both reproduce the correct asymptotic decay rate of $(\rho a)^{-3/2}$ and perform well across all wavenumbers. The small discrepancy at $\rho=0$ for both LO and SLO orders stems from our imperfect representation of the ring profile $f(r,\theta)$ by the approximation $G(r-a)$, which we have chosen for analytic simplicity. At the origin in visibility space this corresponds to inaccurate total image flux. This would be remedied if one used in the focal expansion the exact 1D Fourier transform of the radial profile.}
    \label{fig: intro plot}
\end{figure*}

Before a Fourier transform can be taken, numerically or otherwise, we must specify the functional behavior on the full 2D plane. There are two problems with taking \cref{eq: intro log approx} at face value. First, the function would not be smooth at $r=0$, with discontinuity in the first derivative that could lead to spurious large-argument asymptotics of the Fourier transform. Second, the large-radius tail is both negative and divergent, compromising the interpretation of the function as a physical image on the sky. We deal with these issues through the following toy model:
\begin{align}\label{eq: intro f}
    \begin{split}
    &f(r,\theta)=\\
    &\frac{1}{4\pi^2\sqrt{2}}\left(1-\frac{1}{3}\sin\theta\right)e^{-\frac{1}{5}\left(\frac{a}{r}-1\right)} \log \left[1+\frac{a^4}{(r-a)^4}\right].
    \end{split}
\end{align}
The asymptotic issue has been dealt with by adding a constant inside the logarithm while the smoothness problem has been solved by forcing all derivatives at the origin to vanish via the $e^{-a/5r}$ factor. We have also added non-trivial angular dependence for illustrative purposes and normalized the function so that the Fourier transform at the origin will come out to be approximately unity. The image generated by \cref{eq: intro f} is shown in the upper left corner of fig.~(\ref{fig: intro plot}). We now seek to find the 2D Fourier transform $\mathcal{F}[f]$ of \cref{eq: intro f}.

The first step of our procedure is decomposing $f(r,\theta)$ into angular modes as in \cref{eq: ang mode decomposition introduction}. We have
\begin{align}\label{eq: intro g0 g1 g-1}
\begin{split}
        g_0(r)=&-6 i g_{1}(r)=6i g_{-1}(r)\\
        =&\frac{1}{4\pi^2 \sqrt{2}}e^{-\frac{1}{5}\left(\frac{a}{r}-1\right)} \log \left[1+\frac{a^4}{(r-a)^4}\right],
\end{split}
\end{align}
all other modes vanishing. Note that as $r$ represents physically the radial distance in the 2D plane, \cref{eq: intro g0 g1 g-1} is only defined for $r\geq 0$. The next step in our procedure is to extend this to a function on the full real line that matches \cref{eq: intro g0 g1 g-1} as closely as possible for $r\geq 0$ and is relatively small for $r<0$. Analytic simplicity is an important consideration here. In our present artificial scenario the choice is simple:
\begin{align}\label{eq: intro g0 g1 g-1 approx}
\begin{split}
        g_0(r)=&-6i g_{1}(r)=6i g_{-1}(r)\\
        \approx&\frac{1}{4\pi^2 \sqrt{2}}\log \left[1+\frac{a^4}{(r-a)^4}\right]\equiv G(r-a),
\end{split}
\end{align}
where in the last line we have defined $G(r)$ as our approximating function valid on the full real line. It is important for our procedure to center $G$ on the peak and treat $a$ as a translation parameter. A comparison of $G(r-a)$ to the cross section of $f(r,\theta)$ along the $X_1$ axis is shown in the bottom left corner of fig.~(\ref{fig: intro plot}).

Next, we calculate the \textit{one-dimensional} Fourier transform of $G(r)$. In this case one can show (see section~\ref{sec: log rings against all odds})
\begin{align}\label{eq: intro Ghat}
    \widehat{G}(p)=\frac{1}{2^{3/2}\pi^2 |p|}\left[1- e^{-\sqrt{2}\pi a |p| }\cos\left(\sqrt{2}\pi a p\right)\right].
\end{align}
Having done this, the final step is writing down the approximation to $\mathcal{F}[f]$. Denote again the spherical coordinates in Fourier space as $(\rho,\phi)$. We need to deal with each angular mode separately, the final answer can be divided into contributions from the $n=0$ and $|n|=1$ modes as
\begin{align}\label{eq: intro F f ang}
    \mathcal{F}\left[f\right](\rho,\phi)=2\pi \mathcal{J}_0(\rho)+4\pi \sin\phi \mathcal{J}_1(\rho)
\end{align}
where the functions $\mathcal{J}_0,\mathcal{J}_1$ cannot be found exactly -- this is where the focal approximation is made. The leading and subleading contributions to these quantities in our scheme are\footnote{The separation into leading/subleading orders can be slightly different if the function $G(r)$ is not even around the origin, see further sections for details.}
\begin{align}
    \begin{split}\label{eq: intro J0 ap}
        \mathcal{J}_0&(\rho)\approx \\
        &a \widehat{G}(\rho) J_0(2\pi a \rho)\qquad\quad \big\}\;\;\text{Leading Order (LO)}\\
        +&\frac{1}{2\pi}\widehat{G}'(\rho) J_1(2\pi a \rho) \qquad \big\}\;\;\text{Subleading Order (SLO),}
    \end{split}
\end{align}
\begin{align}
    \begin{split}\label{eq: intro J1 ap}
        \mathcal{J}_1&(\rho)\approx\\ 
        &a \widehat{G}(\rho) J_1(2\pi a \rho) \qquad\qquad\qquad\qquad\quad\;\;\; \big\}\;\;\text{LO}\\
        +&\frac{1}{2\pi}\widehat{G}'(\rho)\left[J_2(2\pi a \rho)-\frac{J_0(2\pi a\rho)}{2\pi a \rho}\right] \qquad \big\}\;\;\text{SLO.}
    \end{split}
\end{align}
These two equations are true for a generic even function $G(r)$; in the present case, we simply plug in \cref{eq: intro Ghat}. Once $\widehat{G}$ is known, higher focal orders are easy to find, requiring only the calculation of increasingly higher derivatives of it. We will analyze higher orders in the main body of the paper, stopping at SLO for now\footnote{Note also that we have not accounted for the error coming from the approximation \cref{eq: intro g0 g1 g-1 approx}. We will investigate the potential errors introduced by this extension step in further sections.}. 

Putting \cref{eq: intro F f ang,eq: intro J0 ap,eq: intro J1 ap} together, the desired Fourier transform can be split into leading and subleading contributions
\begin{align}
    \mathcal{F}[f] \approx \mathcal{F}_{\text{LO}}[f]+\mathcal{F}_{\text{SLO}}[f]
\end{align}
where
\begin{align}
    \begin{split}
        \mathcal{F}_{\text{LO}}[f]&(\rho,\phi)=\\
        &2\pi a\, \widehat{G}(\rho)\left[J_0(2 \pi a\rho)+\frac{i}{3} J_1(2 \pi a\rho) \sin \phi\right],  
    \end{split}
\end{align}
\begin{align}
    \begin{split}
        &\mathcal{F}_{\text{SLO}}[f](\rho,\phi)=\\
        &\widehat{G}'(\rho)\left[J_1(2\pi a \rho)+\frac{i}{3}\left(J_2(2\pi a \rho)-\frac{J_0(2\pi a \rho)}{2\pi a\rho}\right)\sin\phi\right].
    \end{split}
\end{align}
A comparison between the exact, leading, and subleading orders of our approximation is shown on the right side of fig.~(\ref{fig: intro plot}). The top plot shows the amplitude of the Fourier transform along the $\phi=0^{\circ}$ direction in the Fourier plane, while the bottom plot shows it along $\phi=90^{\circ}$. In both cases we see the LO focal expansion performs well across the full range plotted, with SLO improving on the already small discrepancy. Note the disagreement between numerics and our scheme is mostly confined to a small region between the small and large argument regimes\footnote{There is a discrepancy at the origin due to the approximation step \cref{eq: intro g0 g1 g-1 approx} but it is small enough to not be noticeable on these plots.}. Note in particular we obtain the correct large argument asymptotics (the range of our comparison plots is somewhat limited in the name of visual clarity, but plotting the functions further did not appear to reveal any sudden disagreement).

To summarize, our procedure allows us to perform a type of dimensional reduction, providing a remarkably accurate and robust approximation that reduces the two dimensional Fourier transform to expressions in terms of a one dimensional Fourier transform. The caveat is we must do it for each angular mode separately and must also go through an open-ended procedure of approximating the peak profile of our ring by a function on the full real line, all done in a way amenable to analytic calculation. While these problems are somewhat limiting, in situations when they can be overcome our method provides a robust and analytic approximation scheme.

In particular, the focal expansion is well suited to constructing models of interferometric responses to black hole images, which in full generality we expect to consist of a series of nearly-concentric `photon rings' 
\cite{PhotonRingReview}, instead of just the one ring of fig.~(\ref{fig: intro plot}). The focal expansion could then be applied to each ring separately to yield a robust model of the total response. We hope to explore the application of our method to this more realistic scenario in future work, restricting ourselves for the remainder of the present paper to deriving and exploring the focal expansion on its own in total generality.

\subsection{Notation \label{sec: notation}}
Before we proceed, a few words about notation. In all that follows we take the Fourier transform of a function\footnote{All of our procedures are linear so to consider complex functions one must simply apply the methods outlined below to the real and imaginary parts separately.} $f(\Vec{X}): \mathbb{R}^N \rightarrow \mathbb{R}$ to be defined by
\begin{equation} \label{Fourier transform definition}
    \mathcal{F}\left[f(\Vec{X})\right](\Vec{P})\equiv \int_{\mathbb{R}^N}f(\Vec{X}) e^{-2\pi i \Vec{P}\cdot \Vec{X}}d^n\Vec{X}
\end{equation}
where $\Vec{P},\Vec{X}\in \mathbb{R}^N$ and $\Vec{P}\cdot \Vec{X}$ is the standard Cartesian dot product. Sometimes we drop the argument of $f$ when it is clear from context, so that  $\mathcal{F}[f(\Vec{X})](\Vec{P})$ and $\mathcal{F}[f](\Vec{P})$ are both given by \cref{Fourier transform definition}. Note that with our normalization the inverse Fourier transform can be obtained by simply reversing the sign of the exponent in the equation above, meaning
\begin{equation}\label{eq: inverse is minus}
    \mathcal{F}^{-1}[f](\Vec{P})=\mathcal{F}[f](-\Vec{P}).
\end{equation}

We will exclusively denote the one dimensional Fourier transform with a hat: $\mathcal{F}[f]\equiv \widehat{f}$. In the case of two dimensions $N=2$ we also exclusively reserve $(r,\theta)$ and $(\rho, \phi)$ for the spherical coordinates of the $\Vec{X}$ and $\Vec{P}$ spaces respectively so that
\begin{align}
    r e^{i\theta}=&X_1+i X_2,\\
    \rho e^{i\phi}=&P_1+i P_2.
\end{align}

The even and odd parts of a function $f(z)$ are denoted and calculated as
\begin{align}\label{eq: f evenodd decomp}
    f^{\pm}(z)\equiv& \frac{f(z)\pm f(-z)}{2},
\end{align}
the plus and minus standing for even and odd parts respectively. We see in particular that any function can be uniquely decomposed into even and odd parts via $f(z)=f^{+}(z)+f^{-}(z)$. Note also that the operation of taking even/odd parts of a function commutes with taking the Fourier transform -- we will use this fact repeatedly. The superscripts $\pm$ will also be used to denote special operators $\left(\partial\right)^{\pm}$ we introduce in section~\ref{sec: focusing operators} but this use should not be confused with the even/odd meaning \cref{eq: f evenodd decomp}.

Finally, we will make use of rising and falling factorials defined and denoted for integer $n>0$ as
\begin{align}
    (a)^{\uparrow}_n\equiv & \prod_{m=0}^{n-1}(a+m),\\
    (a)^{\downarrow}_n\equiv & \prod_{m=0}^{n-1}(a-m)\label{eq: falling factorial}
\end{align}
respectively, and both equal to $1$ for the empty-product case $n=0$. We also use the double factorial $n!!$ with the standard definition (factorial product only over integers with the same parity as $n$), including the fact that $(-1)!!\equiv 1$.

\section{Derivation \label{sec: The Method}}

\subsection{Preliminaries \label{sec: preliminaries}}

For the remainder of the main body of this note we focus on the 2D Fourier transform, delegating the higher dimensional generalizations to Appendix~\ref{app: higher dim}. Using the spherical coordinates $(r,\theta)$ we can expand any function $f(\Vec{X})$ into a $\theta$-Fourier \textit{series} 
\begin{equation} \label{f theta fourier series}
     f(\Vec{X})=f(r,\theta)=\sum_{n\in \mathbb{Z}}g_n(r) e^{i n \theta},
\end{equation}
where\footnote{Since our ultimate interest lies in ring-like shapes one might wonder why we don't exploit this fact from the start, as in \cite{Gralla} or \cite{AlexPaper}. The problem here is that one cannot unambiguously assign a ring of some variable thickness and shape to an image. The generality of \cref{f theta fourier series} saves us from having to introduce such artificial complexities but comes at a cost of making non-circular shapes less natural -- see section~\ref{sec: making shapes}.}
\begin{equation}
    g_n(r)\equiv \frac{1}{2\pi}\int_0^{2\pi}f(r,\theta)e^{-i n \theta}d\theta. 
\end{equation}
We now use  the Anger-Jacobi expansion \cite{BesselWatson}
\begin{equation}\label{eq: Anger-Jacobi}
    e^{i z \cos \theta}=\sum_{m\in \mathbb{Z}}i^m J_m(z)e^{i m \theta},
\end{equation}
where $J_m$ is the $m$-th order Bessel function of the first kind and $J_{-m}=(-1)^m J_m$. We will make repeated use of \cref{eq: Anger-Jacobi}, one can understand it concisely as the statement that $i^m J_m(z)$ is the $m$-th coefficient of the $\theta$-Fourier series of $e^{iz\cos \theta}$, with $z$ a constant. Taking \cref{eq: Anger-Jacobi,Fourier transform definition,f theta fourier series} together we get
\begin{equation} \label{f hat decomposed into sums of integrals}
    \mathcal{F}[f](\rho, \phi)=2\pi\sum_{n\in\mathbb{Z}}e^{i n (\phi-\pi/2)}\int_0^{\infty}r g_n(r) J_n(2\pi r \rho)\, dr
\end{equation}
 and so we see that once we separate our function into angular modes the process of taking the Fourier transform is reduced to performing weighted integrals of Bessel functions. By itself this decomposition is not new, indeed, the integrals in \cref{f hat decomposed into sums of integrals} are commonly known as Hankel transforms \cite{NIST}. The principal question we are interested in is this: what can be said about the general behavior of such integrals? For a generic $g_n(r)$ it is unlikely the integral can be performed in closed form, the standard approach would then be to look for an asymptotic approximation of some kind (see for instance \cite{BenderOrszag}). We will instead take a novel approach motivated by the following set of qualitative observations.

The leading-order term in $J_n(2\pi r \rho)$ as $\rho\rightarrow \infty$ is \cite{BesselWatson}
\begin{align}\label{eq: Bessel asymptote}
    \begin{split}
        J_n(2 \pi r \rho)\;\Big|_{\rho \gg 1}\approx\frac{1}{\pi \sqrt{r \rho}}&\cos\left(-2\pi r \rho+n\pi/2+\pi/4\right),
    \end{split}
\end{align}
provided $r>0$. Now, assume $g_n(r)$ is sufficiently tightly peaked around some radius $r=a >0$ so that to leading $1/\rho$ order the integral in \cref{f hat decomposed into sums of integrals} becomes 
\begin{align} \label{eq: first approx}
    \begin{split}
        \int_0^{\infty}&r g_n(r) J_n(2\pi r \rho)\, \text{d}r\;\Big|_{\rho \gg 1}\\
        &\approx \frac{1}{\pi}\sqrt{\frac{a}{\rho}}\,\text{Re}\left[e^{i \pi (n/2+1/4)} \int_0^{\infty}g_n(r)e^{-2\pi i r \rho}dr\right].
    \end{split}
\end{align}
The last integral is almost the Fourier transform of $g_n(r)$, only the integration limits are incorrect. To remedy this, define an extension $G_n(r-a)$ of $g_n(r)$ to all of $\mathbb{R}$, shifted to make our peak centered at the origin. We will have more to say about choosing an extension in Section~\ref{section: peak at origin} but for now note that one might pick, for instance, a trivial extension  with $G_n(r)$ vanishing for $r<-a$ (although we will often make a less trivial choice for the sake of analytic tractability, e.g., fig.~\ref{fig: remainder example} in Section~\ref{section: peak at origin}). We then have
\begin{align} \label{eq: second approx}
    \begin{split}
        \int_0^{\infty}&r g_n(r) J_n(2\pi r \rho)\, \text{d}r\;\Big|_{\rho \gg 1}\\
        &\approx \frac{1}{\pi}\sqrt{\frac{a}{\rho}}\,\text{Re}\left[e^{i \pi (n/2+1/4)} \int_{\mathbb{R}}G_n(r-a)e^{-2\pi i r \rho}dr\right],
    \end{split}
\end{align}
where we assume the effect of extending $g_n$ to $G$ is negligible compared to the leading behavior.
Provided $G_n(r)$ is an even function so that $\widehat{G}_n(\rho)$ is real (we will analyze the general case later), \cref{eq: second approx} becomes approximately 
\begin{align} \label{eq: third approx}
    \begin{split}
        \int_0^{\infty}&r g_n(r) J_n(2\pi r \rho)\, \text{d}r\;\Big|_{\rho \gg 1}\\
            &\approx \frac{1}{\pi}\sqrt{\frac{a}{\rho}}\,\text{Re}\left[e^{i \pi (-2 a \rho+n/2+1/4)} \widehat{G}_n(\rho)\right]. 
    \end{split}
\end{align}
Applying the Bessel asymptotic \cref{eq: Bessel asymptote} backwards one can rewrite \cref{eq: third approx} as
\begin{align}\label{eq: app large rho}
    \int_0^{\infty}&r g_n(r) J_n(2\pi r \rho)\, \text{d}r\;\Big|_{\rho \gg 1}\approx a \widehat{G}_n(\rho) J_n(2\pi a \rho).
\end{align}
This simple result turns out to be more correct than one would expect, to see why we now look at the case $\rho\rightarrow 0$. In this limit the Bessel function varies little over the extent of the peak and so we can write
\begin{align} \label{eq: fourth approx}
    \begin{split}
        \int_0^{\infty}&r g_n(r) J_n(2\pi r \rho)\, \text{d}r\;\Big|_{\rho \ll 1}\\
            &\approx \int_{\mathbb{R}}r G_n(r-a) J_n(2\pi r \rho)\,\text{d}r\\
            &\approx a J_n(2\pi a \rho) \int_{\mathbb{R}}G_n(r-a)\,\text{d}r,
    \end{split}
\end{align}
where we assumed $G$ is tightly peaked at $r=a$ and $\rho$ is small so that the factor of $r J_n(2\pi r \rho)$ is approximately constant over the peak. An unweighted integral over the peak profile is simply $\widehat{G}_n(0)$ which, provided $\widehat{G}_n(\rho)$ has a convergent Taylor series at the origin, one can write as
$\widehat{G}_n(0)=\widehat{G}_n(\rho)+O(\rho)$. This means we may write \cref{eq: fourth approx} as
\begin{align} \label{eq: app small rho}
    \int_0^{\infty}&r g_n(r) J_n(2\pi r \rho)\, \text{d}r\;\Big|_{\rho \ll 1}\approx a \widehat{G}_n(\rho) J_n(2\pi a \rho)+O(\rho).
\end{align}

It appears that one can combine the $\rho \rightarrow 0$ and $\rho \rightarrow \infty$ limits given by \cref{eq: app large rho,eq: app small rho} and simply write
\begin{align} \label{eq: fifth approx}
    \begin{split}
        \int_0^{\infty}&r g_n(r) J_n(2\pi r \rho)\, \text{d}r\approx a\, \widehat{G}_n(\rho) J_n(2\pi a \rho),
    \end{split}
\end{align}
understanding the approximation to be \textit{globally} valid for all $\rho$, assuming nothing unexpected happens in the intermediate region between the small and large argument limits. Note that this global validity is afforded to us thanks to the appearance of $\widehat{G}_n$ which, being a Fourier transform, contains both large and small scale information about the image. We have in essence exchanged a Hankel transform for an approximate expression involving a 1D Fourier transform.

A glaring shortcoming of \cref{eq: fifth approx} is the fact that near the origin it is correct only to the leading, constant Taylor order.
This clearly motivates a strategy: 
we will to develop a general expansion based around the 1D Fourier transform of $G_n$ with leading term given by \cref{eq: fifth approx}, and with subleading terms used to match higher $\rho=0$ Taylor orders. If enforced at all orders, the `pincer maneuver' of combining the two limits \cref{eq: app large rho,eq: app small rho} into a single approximation \cref{eq: fifth approx} would then ideally yield an approximation without a regime of validity limited to small or large arguments and constrain the error in the intermediate regime between these two limits. Extending this universal behavior to subleading perturbative orders, however, is significantly more difficult and requires the development of several new analytic tools.

\subsection{Focusing Operators\label{sec: focusing operators}}

For most of the remainder of this section we generalize the Bessel integrals we have been so far considering to a generic linear integral transform of a function $f$ shifted by $a$:
\begin{equation} \label{eq: Ika generic}
    \mathcal{I}_{K,a}\left[f\right](\rho)\equiv\int_{\mathbb{R}}f(x-a) K(2\pi x \rho)\,\text{d}x,
\end{equation}
where we use the lowercase $x$ as a generic one-dimensional integration variable to differentiate it from both the uppercase $\Vec{X}$ describing the position on the image plane and the radial coordinate $r=||\Vec{X}||$.
We assume that whatever extension procedure necessary to define $f$ on the full real line has already been performed. Nothing about the kernel $K$ is assumed except that it has a well behaved Taylor series at the origin. 

We are interested in obtaining an approximation where $\widehat{f}$ plays a central role. Let the lowercase $p$ denote the frequency domain coordinate dual to our generic $x$, unrelated to the previously introduced $\Vec{P}$ and $\rho=||\Vec{P}||$ that correspond the 2D Fourier space of our image.  We first rewrite \cref{eq: Ika generic} by replacing $K$ with 
\begin{align}
    K(x)=\left[\mathcal{F}^{-1}\circ \mathcal{F}\right]\left[K\right](x)=\mathcal{F}[\widehat{K}(-p)](x) 
\end{align}
(where we used \cref{eq: inverse is minus}) and performing the $x$ integral -- this is essentially just the Fourier convolution theorem. Doing so yields 
\begin{equation} \label{eq: Fourier convolution trick}
\begin{split}
        \mathcal{I}_{K,a}\left[f\right](\rho)=&\int_{\mathbb{R}}f(x-a) K(2\pi x \rho)\,\text{d}x\\
        =&\int_{\mathbb{R}}\widehat{f}(2\pi p \rho) \widehat{K}(-p) e^{-i 2 \pi p (2\pi a \rho)}\text{d}p.
\end{split}
\end{equation}
Using \cref{eq: Fourier convolution trick} we can equivalently interpret our integral as a linear functional of $\hat{f}$ instead of $f$
\begin{align} \label{eq: linear functional alt}
    \widehat{\mathcal{I}}_{K,a}[\widehat{f}\,](\rho)\equiv \mathcal{I}_{K,a}\left[f\right](\rho)
\end{align}
where we will distinguish between the two alternative interpretations using a hat. Now note that we seem to get an expansion akin to what we desired by Taylor expanding  $\widehat{f}(2\pi p \rho)$ inside the integral around $p=1/2\pi$:
\begin{align} \label{eq: first attempt}
    \begin{split}
        \mathcal{I}_{K,a}\left[f\right](\rho)=&\sum_{n=0}^{\infty}\frac{(2\pi \rho)^n}{n!}\widehat{f}^{\,(n)}(\rho)\\
        &\times\int_{\mathbb{R}} (p-1/2\pi)^n\widehat{K}(-p) e^{-i 2 \pi p (2\pi a \rho)}\,\text{d}p\\
        =&\widehat{f}(\rho)K(2\pi a \rho)+\cdots.
    \end{split}
\end{align}
The first term is encouraging -- indeed, if $f(x-a)=x g(x-a)$ and $K=J_n$ it contains precisely our desired leading term \cref{eq: fifth approx} -- but we are not so lucky with all subsequent terms in the sum. The problem is the appearance of the bare factors of $\rho^n$ in the non-leading terms of \cref{eq: first attempt}. Being polynomial, they are badly behaved for large $\rho$. 

Although the above expansion fails globally, it motivates the path forward; we must avoid all polynomial factors of $\rho$ in our final expression as these are inherently local to the origin. We will instead make use of a special family of differential operators -- which we name \textit{focusing operators}\footnote{We refer to these as focusing operators because they progressively isolate higher-order local information for a given function, but we largely picked the term for convenience and ease of use.} -- and arrange the expansion around the action of these operators on $\widehat{f}(\rho)$ and $K(2\pi a \rho)$. We name the resulting series a \textit{focal expansion}. To fix the coefficients of this series we will Taylor expand it at $\rho=0$ and match it to a Taylor expansion of $\mathcal{I}_{K,a}\left[f\right]$ obtained directly from \cref{eq: Fourier convolution trick}.
While this last matching step seems inherently local to the origin, surprisingly it seems to capture global behavior of the integral when $K=J_n$ (and, potentially, for a wider class of oscillatory kernels) as we will see when we consider examples of our approach in section~\ref{sec: examples}. We encourage a practically minded reader not interested in the abstract details of the matching procedure to skip to the summary of our method for the case $K=J_n$ in section \ref{subsec: summary}. 

We proceed by by first introducing the operators. Assuming $n\geq 0$ we define the \textit{n-th even and odd focusing operators} respectively as\footnote{Operators of this type are known to be relevant in Bessel function analysis -- see \cite{BesselWatson,NIST} -- and more broadly in studies of differential equations upon rewriting them using the Cauchy-Euler operator $z \partial$. We are, however, unaware of any prior work using them to create series expansions of integrals the way we do here.}
\begin{align} \label{eq: focusing +}
    \left(\partial\right)_n^+&\equiv \left(\partial-\frac{n-1}{z}\right)\left(\partial-\frac{n-2}{z}\right)\cdots \left(\partial-\frac{1}{z}\right)\partial, \\
    \left(\partial\right)_n^-&\equiv \left(\partial-\frac{n }{z}\right)\left(\partial-\frac{n-1}{z}\right)\cdots \left(\partial-\frac{1}{z}\right) \label{eq: focusing -},
\end{align}
where $z$ is a generic variable (later we will need it to stand in for either of our $x$ or $p$) and $\partial = d/dz$; we suppress $z$ while writing $\partial$ as the argument over which derivation is performed will always be clear from context. The notation is meant to call to mind the falling factorial \cref{eq: falling factorial} due to the visually similar product structure.  For $n=0$ we take the standard empty product convention so that $\left(\partial\right)^{\pm}_0=1$. When applying these to a function $f(z)$ we will often refer to $\left(\partial\right)^{\pm}_n f$ as the \textit{n-th even/odd focus of f} or that we are \textit{focusing f to n-th even/odd order}. Note also that  $\left(\partial\right)^+_{n+1}=\left(\partial\right)^-_n\, \partial$. An alternative version with the operator products expanded can be shown to be
\begin{align} \label{eq: focal op explicit +}
    \left(\partial\right)^+_n&=\sum_{k=0}^{n-1}\frac{(n-1+k)!}{(n-k-1)! k!}\frac{1}{(-2 z)^k}\partial^{n-k},\\
        \label{eq: focal op explicit -}\left(\partial\right)^-_n&=\sum_{k=0}^{n}\frac{(n+k)!}{(n-k)! k!}\frac{1}{(-2 z)^k}\partial^{n-k},
\end{align}
with \cref{eq: focal op explicit +} valid only for $n\geq 1$. Note that for large $z$, unless the derivatives of $f(z)$ decay slower than their parent function, \cref{eq: focal op explicit -,eq: focal op explicit +} show the $n$-th foci of $f(z)$ are both approximately equal to $f^{(n)}(z)$. This means that we do not suffer any additional power-law divergences at large argument not already present in $f(z)$ or its derivatives when calculating the foci.

To understand what motivates the form and naming of \cref{eq: focusing -,eq: focusing +}, we apply $\left(\partial\right)^-_2$ to an odd power series:
\begin{align} \label{eq: explicit example of cancelation}
    \begin{split}
        \left(\partial\right)^-_2 & \left(a_0 z+a_1 z^3+ a_2 z^5+\cdots\right)\\
        =&\left(\partial-\frac{2}{z}\right)\left(0\cdot a_0+2 a_1 z^2+4 a_2z^4+\cdots\right)\\
        =&\,0\cdot a_1 z+8a_2z^3+\cdots.
    \end{split}
\end{align}
Every factor of $\left(\partial - n/z\right)$ simply lets us cancel the remaining leading term of the power series, the $n$ being a combinatorial factor resulting from chaining multiple of these operator factors in a row. This behavior will continue to hold in general as long as we apply the odd/even operators to odd/even power series respectively. First note that 
\begin{align}
    \begin{split}\label{eq: even foc power}
        \left(\partial\right)^+_n z^k=& \frac{k!!}{(k-2n)!!}z^{k-n}\\
    =&k(k-2)\cdots (k-2(n-1))\;z^{k-n},
    \end{split}\\
    \nonumber  \text{\phantom{test}}\\ 
    \begin{split} \label{eq: odd foc power}
        \left(\partial\right)^-_n z^k=& \frac{(k-1)!!}{(k-2n-1)!!}z^{k-n}\\
    =&(k-1)(k-3)\cdots (k-2n+1)\;z^{k-n},
    \end{split}
\end{align}
both of which can be proved by induction in $n$. In particular \cref{eq: even foc power,eq: odd foc power} imply that focusing a non-negative power of $z$ of the same parity as the operator cannot lead to negative powers of $z$ -- the higher order foci will simply vanish:
\begin{align} \label{eq: foc low vanish}
    \left(\partial\right)^+_n z^{2k}\Big|_{n>k}=\left(\partial\right)^-_nz^{2k+1}\Big|_{n>k}=0
\end{align}
so that the lowest powers one can obtain when focusing (with appropriate parity) $z^{2k}$ and $z^{2k+1}$ are $z^k$ and $z^{k+1}$ respectively. This will not be the case for powers of  opposite parity to the operator, $\left(\partial\right)^+_2 z=1/z$ for instance, and is precisely the reason we introduced separate even and odd operators in the first place. Even functions should be focused at even order and odd functions at odd order to avoid introducing a divergence at the origin. Bringing together \cref{eq: even foc power,eq: odd foc power,eq: foc low vanish} we see that
\begin{align} \label{eq: action on even power series}
    \left(\partial\right)^+_n\sum_{k=0}^{\infty}a_k z^{2 k}&=(2n)!! \,a_n z^n +O(z^{n+2}),\\
    \label{eq: action on odd power series}\left(\partial\right)^-_n\sum_{k=0}^{\infty}a_k z^{2 k+1}&=(2n)!! \,a_n z^{n+1} +O(z^{n+3}),
\end{align}
where in both cases the higher order terms only depend on $\{a_k\}_{k>n}$ -- all dependence on $\{a_k\}_{k<n}$ has been removed in the process of focusing. The key takeaway from \cref{eq: action on even power series,eq: action on odd power series} is that focusing operators isolate progressively higher orders of both the coefficients $a_k$ and raw powers of $z$ near the origin, all done in a manner that is well behaved for large arguments. These facts will prove crucial for approximating our integral, \cref{eq: Fourier convolution trick}.

We saw how raw powers of $z$ behave when focused but, as we shall soon see, we will also need to describe how foci of an arbitrary function $h(z)$ change when multiplied by a power. The relevant expressions are
\begin{align}\label{eq: product of power and function 1}
    \left(\partial\right)^{\pm}_n z^{2k} h(z)=&\sum_{j=0}^k (n)^{\downarrow}_j 2^j \binom{k}{j}z^{2k-j} \left(\partial\right)^{\pm}_{n-j} h(z),\\
        \label{eq: product of power and function 2}\left(\partial\right)^{\pm}_n z^{2k\mp 1} h(z)=&\sum_{j=0}^k(n)^{\downarrow}_j 2^j \binom{k}{j} z^{2k-j\mp 1} \left(\partial\right)^{\mp}_{n-j} h(z),
\end{align}
both of which can be proven by induction in $n$. Note that multiplication by an even power leaves the parity of the foci of $h$ involved in the expression invariant, while multiplication by an odd power flips it.

\subsection{The Focal Expansion} \label{subsec: The Focal Expansion}

Before we utilize focusing operators we first Taylor-expand the integral \cref{eq: Fourier convolution trick} at small $\rho$. Doing so will suggest how a focal expansion is to be assembled in the first place by matching Taylor orders near the origin. We begin by assuming our kernel $K$ is a purely even function -- it will turn out the general case can be reduced to that of even kernels. Assume $\widehat{f}$ and $K$ to have Taylor series near the origin given by
\begin{align} \label{eq: Taylor exp of f and K}
    \widehat{f}(z)=&\sum_{k=0}^{\infty}c_k z^k, \quad
    K(z)=\sum_{k=0}^{\infty}\alpha_k z^{2k},
\end{align}
where the $c_k$ must be understood as free parameters and the coefficients of the focal expansion cannot depend on them -- we seek a series which, once found, is valid for any $\widehat{f}$ without the need to recalculate any coefficients. By contrast, we assume the kernel is fixed so that every $\alpha_k$ is a known constant on which the focal coefficients may depend. With that said, from \cref{eq: Fourier convolution trick,eq: Ika generic,eq: Taylor exp of f and K} we obtain
\begin{align}
\begin{split}
        \mathcal{I}_{K,a}\left[f\right](\rho)=&\sum_{k=0}^{\infty}c_k (2\pi \rho)^k \mathcal{F}\left[p^k \widehat{K}(-p)\right](2\pi a\rho)\\
        =&\sum_{k=0}^{\infty}c_k (i \rho)^k  K^{(k)}(2\pi a\rho),
\end{split}
\end{align}
where in the last line we also used the identity $\mathcal{F}\left[x^n h(x)\right](p)=(i/2\pi)^n \widehat{h}^{(n)}(p)$ and, once again, \cref{eq: inverse is minus}. Finally, using the Taylor expansion of the kernel we have
\begin{align} \label{eq: Ika exp}
        &\mathcal{I}_{K,a}\left[f\right](\rho)=\nonumber\\
        &\quad\sum_{k\geq 0}\;\sum_{m\geq k/2} c_k(2\pi a\rho)^{2m} \left[ \alpha_m\,(2m)^{\downarrow}_k\left(\frac{i}{2\pi a}\right)^k\right].
\end{align}
The term in the square brackets should be understood as a constant for fixed $k$ and $m$ while the prefactor $c_k(2\pi a\rho)^{2m}$ represents the functional behavior we need to match with the focal expansion. There are two independent dimensions at play here: we must match the $c_k$ as well as the $\rho$ behavior. 

This is where the focusing operators come in. Note that using \cref{eq: even foc power,eq: odd foc power,eq: Taylor exp of f and K} we have
\begin{align}
\begin{split} \label{eq: focus of f+}
        \left(\left(\partial\right)^+_k \widehat{f}^+\right)(\rho)=&\sum_{l=k}^{\infty}c_{2l}\; \rho^{2l-k}2^k(l)^{\downarrow}_k\\
    =&c_{2k}\;\rho^k\; (2k)!!+\cdots,
\end{split}\\
&\text{\phantom{filler}}\nonumber\\
    \begin{split} \label{eq: focus of f-}
        \left(\left(\partial\right)^-_k \widehat{f}^-\right)(\rho)=&\sum_{l=k}^{\infty}c_{2l+1}\; \rho^{2l+1-k}2^k(l)^{\downarrow}_k\\
    =&c_{2k+1}\;\rho^{k+1}\; (2k)!!+\cdots,
\end{split}\\
&\text{\phantom{filler}}\nonumber\\
    \begin{split}\label{eq: focus of K}
        \left(\left(\partial\right)^+_k K\right)(2\pi a \rho)=&\sum_{l=k}^{\infty}\alpha_{l}\; (2\pi a\rho)^{2l-k}2^k(l)^{\downarrow}_k\\
    =&\alpha_{k}\;(2\pi a\rho)^{k}\; (2k)!!+\cdots,
\end{split}
\end{align}
where we also used the fact that $(2l)!!/(2l-2k)!!=2^k(l)^{\downarrow}_k$.  Matching the $c_k$'s can be done naturally by focusing $\widehat{f}$ and we can then fix the $\rho$ behavior at that $c_k$ order through multiplication by an appropriate series in foci of $K$. For instance, say we want to match the $c_0$ order of \cref{eq: Ika exp} : 
\begin{align} \label{eq: c0}
    c_0\sum_{m=0}^{\infty}\alpha_m (2\pi a \rho)^{2m} .
\end{align}
Inspired by the multiplicative character of \cref{eq: fifth approx} we see this can be done via a term of the form
\begin{align} \label{eq: c0fix}
\left(\left(\partial\right)^+_0\widehat{f}^+\right)(\rho)\sum_{n=0}^{\infty}A^+_{0,n}\left(\left(\partial\right)^+_{2n}K\right)(2 \pi a \rho),
\end{align}
provided the coefficients $A^+_{0,n}$ are chosen so that the $\rho$-Taylor expansion of \cref{eq: c0fix} matches that of \cref{eq: c0}.  Note that this is possible since $\left(\partial\right)^+_{2n}K(2 \pi a \rho)$ starts at order $\rho^{2n}$ and so the matching is iteratively well defined; $A^+_{0,0}$ fixes the $\rho^0$ order, $A^+_{0,1}$ fixes the $\rho^2$ order, etc. Every term `overshoots' at higher $\rho$ orders, but that can be accounted for in subsequent steps. In a similar vein, \cref{eq: c0fix} contains dependence on higher order $c_{2k}$'s beyond just $c_0$ but we can fix that later with higher order foci of $\widehat{f}$. For instance, to fit the $c_2$ order we must account for \cref{eq: Ika exp} but also cancel the spurious terms we introduced in \cref{eq: c0fix}, together these contributions are 
\begin{align} \label{eq: c2}
\begin{split}
        -c_2\sum_{n=0}^{\infty}&\Big[ \frac{n\left(n-1/2\right)}{\pi^2 a^2}\alpha_n(2\pi a\rho)^{2n}\\
        &+\rho^2 A^+_{0,n}\left(\left(\partial\right)^+_{2n}K\right)(2 \pi a \rho)\Big].
\end{split}
\end{align}
From \cref{eq: focus of f+,eq: focus of K} we see that we can fit \cref{eq: c2} using
\begin{align} \left(\left(\partial\right)^+_1\widehat{f}^+\right)(\rho)\sum_{n=0}^{\infty}A^+_{1,n}\left(\left(\partial\right)^+_{2n+1}K\right)(2 \pi a \rho),
\end{align}
where the $A^+_{1,n}$ are once again picked to match the $\rho$-Taylor expansion of \cref{eq: c2}. The key fact here is that $\left(\partial\right)^+_1\widehat{f}^+$ has no dependence on $c_0$ so we don't disturb the lower order we had already fixed. Again, we overshoot at higher $c_{2k}$ orders but can account for that once we reach them in the matching procedure. Extending this reasoning to all $c_k$ orders of even and odd parity we can write the following ansatz:
\begin{align}\label{eq: Focal Expansion}
\begin{split}
        \mathcal{I}_{K,a}&\left[f\right](\rho)=\widehat{\mathcal{I}}_{K,a}[\widehat{f}\,](\rho)=\\
        &\sum_{k,n=0}^{\infty}A^+_{k,n} \left(\left(\partial\right)^+_k\widehat{f}^+\right)(\rho) \left(\left(\partial\right)^+_{2n+k}K\right)(2 \pi a \rho)\\
        +&\sum_{k,n=0}^{\infty}A^-_{k,n} \left(\left(\partial\right)^-_k\widehat{f}^-\right)(\rho) \left(\left(\partial\right)^+_{2n+1+k}K\right)(2 \pi a \rho),
\end{split}
\end{align}
where the first line matches all the $c_{2k}$ and the second all the $c_{2k+1}$ orders. We will refer to \cref{eq: Focal Expansion} as \textit{the focal expansion} of $\mathcal{I}_{K,a}\left[f\right](\rho)$, this is the central tool that we have developed.

Before finding the coefficients in \cref{eq: Focal Expansion} we briefly discuss the case of kernels with non-definite parity. First note that we can always decompose $K=K^++K^-$ and write
\begin{align}
    \mathcal{I}_{K,a}\left[f\right](\rho)=\mathcal{I}_{K^+,a}\left[f\right](\rho)+\mathcal{I}_{K^-,a}\left[f\right](\rho),
\end{align}
meaning all we need is to develop a way to deal with purely odd kernels. Note, however, that any odd kernel $K^-(z)$ with a Taylor series at the origin defines an even function with a Taylor series at the origin via $\frac{K^-(z)}{z}$. This means that we may write
\begin{align}
    \begin{split}
    \mathcal{I}_{K^-(z),a}&\left[f(x)\right](\rho)\\
    &=2\pi \rho\,\mathcal{I}_{\frac{K^-(z)}{z},a}\big[(x+a) f(x)\big](\rho)\\
    &=2\pi \rho\,\widehat{\mathcal{I}}_{\frac{K^-(z)}{z},a}\left[\frac{i}{2\pi}\widehat{f}\,'+ a \widehat{f}\;\right](\rho),
    \end{split}
\end{align}
where in the last step we have used \cref{eq: linear functional alt}. We have thus reduced the problem to the already analyzed case of even kernels. To summarize, for a generic kernel $K(z)$ the focal expansion can be found by writing
\begin{align}
    \begin{split}
        \mathcal{I}_{K(z), a}&\,[f](\rho)=\widehat{\mathcal{I}}_{K^+(z), a}\,[\widehat{f}\,](\rho)\\
        &+i\rho \,\widehat{\mathcal{I}}_{\frac{K^-(z)}{z}, a}\,[\widehat{f}\,'\,](\rho)+2\pi\rho a\,\widehat{\mathcal{I}}_{\frac{K^-(z)}{z}, a}\,[\widehat{f}\,](\rho)
    \end{split}
\end{align}
and applying \cref{eq: Focal Expansion} to each integral separately.

We now discuss finding the coefficients $A_{k,n}^{\pm}$. One must substitute the explicit expressions \cref{eq: focus of f+,eq: focus of f-,eq: focus of K} into \cref{eq: Focal Expansion} and equate the coefficients at each $c_k(2\pi a \rho)^{2m}$ order to the known expansion \cref{eq: Ika exp}. We omit the tedious details and instead list the final recurrence relations:
\begin{align}
        (-1)^k \alpha_m &(2m)^{\downarrow}_{2k}=\sum^{m-k}_{n=0} \sum_{l=0}^k(2\pi a)^l A_{l,n}^+ \alpha_{l+n+m-k}\nonumber\\
   \label{eq: rec A+} &\times 2^{2(n+l)}(k)^{\downarrow}_l (l+n+m-k)^{\downarrow}_{2n+l},\\
&\text{\phantom{filler}}\nonumber\\
        (-1)^k i \alpha_m &(2m)^{\downarrow}_{2k+1}=\sum^{m-k-1}_{n=0} \sum_{l=0}^k(2\pi a)^l A_{l,n}^- \alpha_{l+n+m-k}\nonumber\\
   \label{eq: rec A-} &\times 2^{2(n+l)+1}(k)^{\downarrow}_l (l+n+m-k)^{\downarrow}_{2n+l+1},
\end{align}
where the first recurrence is non-vanishing only for $m\geq k \geq 0$ and the second only for $m>k\geq 0$ (this is a result of the $m$ lower bound in \cref{eq: Ika exp}). These recurrence relations can be solved iteratively. 
Thinking of $A^+_{l,n}$ as an infinite matrix, we see from \cref{eq: rec A+} that to calculate $A^+_{k,m-k}$ we must first know the values of all \textit{other} coefficients contained in the finite $k$-by-$(m-k)$  submatrix stretching from $A^+_{0,0}$ to the desired $A^+_{k,m-k}$. The coefficient $A^+_{0,0}$ is unique in that it doesn't depend on any others, from \cref{eq: rec A+} it can be seen to be $1$ as long as $\alpha_0$ is non-vanishing. Thus, there are different `paths' one can take to find the desired $A^+_{k,m-k}$; one option would be to go row by row in the submatrix, calculating $A^+_{0,0},\cdots,A^+_{0,m-k}$ one after another, then calculating the row below, and so on. Completely analogous logic follows for finding $A^-_{l,n}$ using \cref{eq: rec A-}. This iterative procedure also makes clear that for each kernel $K$ the coefficients $A^{\pm}_{l,n}$ , if they exist\footnote{We run into non-existence issues if any of the $\alpha$'s vanish as that sets to 0 a term we need for a particular iterative step, see Appendix~\ref{app: solving the general focal recurrence} for more details.}, are uniquely fixed. As an example, here are the first several focal coefficients:
\begin{align}
    \begin{split} \label{eq: A iterative sol example}
        A^+_{0,0}&=1,\quad  A^+_{0,1}=0, \quad
        A^+_{1,0}=\frac{-(\alpha_0+2\alpha_1)}{8\, a \pi\, \alpha_1}, \\
        &\text{\phantom{filler}}\\
        A^-_{0,0}&=i,\quad  A^-_{0,1}=0, \quad
        A^-_{1,0}=\frac{-i(\alpha_1+12\alpha_2)}{16\, a \pi\, \alpha_2}.
    \end{split}
\end{align}
One can keep going indefinitely, at least algebraically -- we do not claim the focal expansion \cref{eq: Focal Expansion} converges for all possible kernels. See Appendix \ref{app: counterexamples} for two explicit counterexamples with poor convergence behavior.

While finding a finite subset of focal coefficients can be easily performed algorithmically, finding closed form analytic solutions for all of the $A^{\pm}_{l,n}$ presents a more formidable challenge. For a generic $K$ we delegate results of this type to Appendix \ref{app: solving the general focal recurrence} since, as we show now, the problem simplifies dramatically for the case that interests us the most.

\subsection{Bessel Function Kernels \label{sec: Bessel function Kernels}}

In the special case of $K=J_n$ we introduce the notation
\begin{align} \label{eq: definition of mathcalJ}
    \mathcal{I}_{J_n, a}\left[f\right](\rho)\equiv \mathcal{J}_{n,a}\left[f\right](\rho).
\end{align}
We start by investigating $K=J_0$. The Taylor expansion of $J_0(z)$ is \cite{BesselWatson}
\begin{align}
    J_0(z)=\sum_{k=0}^{\infty}\frac{(-1)^k}{4^k k!^2}z^{2k}
\end{align}
and so, in our earlier notation of \cref{eq: Taylor exp of f and K}, we have $\alpha_k=(-1)^k/4^k\, k!^2$. Evaluating the first few focal coefficients iteratively we get what seems to be (for a different derivation see appendix \ref{app: solving the general focal recurrence})
\begin{align} \label{eq: A Bessel}
    A_{l,n}^{+}=-i A_{l,n}^-=\frac{\delta_{n,0}}{(4 \pi a)^l}\frac{(2l-1)!!}{l!},
\end{align}
where $\delta$ is the Kronecker delta. Eq. (\ref{eq: A Bessel}) can indeed be seen to solve the recurrences \cref{eq: rec A+,eq: rec A-}, both of which reduce to the hypergeometric identity 
\begin{align} \label{eq: strange hyp geo}
    \frac{(2m-1)!!}{2^k k!\, m!\, (2m-2k-1)!!}=&\sum_{l=0}^k \frac{(-1)^l (2l-1)!!}{2^l l!\, (k-l)!\, (l+m-k)!}
\end{align}
that can be proved via, for instance, Sister Celine's method \cite{AB}. By uniqueness of the solutions to \cref{eq: rec A-,eq: rec A+} whenever none of the $\alpha_i$'s vanish we then see that, indeed, the guess \cref{eq: A Bessel} gives the focal coefficients for $K=J_0$. 

There is another significant simplification that occurs for this kernel. Bessel functions obey the following recurrence relations \cite{BesselWatson}:
\begin{align}\label{eq: Bessel recurrence 1/x}
    \frac{2n}{z}J_n(z)=&J_{n-1}(z)+J_{n+1}(z),\\
    \label{eq: Bessel recurrence d/dx}2 \partial J_n(z)=&J_{n-1}(z)-J_{n+1}(z),
\end{align}
using which one can prove that focusing $J_0$ simply results in higher order Bessel functions:
\begin{align}\label{eq: foci of J0}
    \left(\partial\right)^+_n J_0=(-1)^n J_n.
\end{align}
Putting \cref{eq: A Bessel,eq: foci of J0} together we see that for this kernel the focal expansion \cref{eq: Focal Expansion} takes the drastically simplified form
\begin{align}
    \begin{split} \label{eq: J0 Focal Expansion}
        &\mathcal{J}_{0,a}\left[f\right](\rho)=\sum_{k=0}^{\infty}\frac{(2k-1)!! }{k!}\frac{1}{(-4\pi a)^k}\\\times&\left[(\left(\partial\right)^+_k\widehat{f}^+)(\rho)J_k(2\pi a \rho)-i (\left(\partial\right)^-_k\widehat{f}^-)(\rho)J_{k+1}(2\pi a \rho)\right] 
    \end{split}
\end{align}
with each sum corresponding to matching $\rho$ powers at a given $c_k$ order in \cref{eq: Focal Expansion} vanishing beyond the leading term. 

When trying to apply the logic that led to \cref{eq: J0 Focal Expansion} to higher-order Bessel function kernels we run into an immediate problem: the Taylor series of $J_m(z)$ starts at order $z^m$ which means that for $m>0$ at least one of the $\alpha_k$ of \cref{eq: Taylor exp of f and K} vanish. This invalidates the iterative procedure of the previous section (see Appendix \ref{app: solving the general focal recurrence} for more details). There is, however, another way. With \cref{eq: J0 Focal Expansion} we can calculate focal expansions for higher order Bessel function kernels indirectly. One begins by noting that \cref{eq: Bessel recurrence d/dx} Fourier transformed and multiplied by $i^n$ becomes precisely the recurrence relation for Chebyshev polynomials of the first kind $T_n$, with the same initial conditions save for a multiplicative factor of $\widehat{J}_0$.  This means we have \cite{IntegralTransforms}
\begin{align} \label{eq: Jn F transform}
    \widehat{J}_n(p)=(-i)^n T_n(2\pi p) \widehat{J}_0(p),
\end{align}
where the explicit formulas for the polynomials are \cite{abramowitz1965handbook,NIST} $T_0=1$ and 
\begin{align} \label{eq: Tn explicit}
    T_n(2\pi p)=&\frac{n}{2}\sum_{j=0}^{\lfloor n/2 \rfloor}(-1)^j\frac{(n-j-1)!}{j! (n-2j)!}(4\pi p)^{n-2j}
\end{align}
for $n\geq 1$ with $\lfloor x \rfloor$ denoting the floor function. 
Using \cref{eq: Tn explicit,eq: Jn F transform,eq: Fourier convolution trick,eq: linear functional alt} we immediately see (in keeping with the notation of \cref{eq: Fourier convolution trick} $p$ here represents the variable being integrated over):
\begin{align} \label{eq: arbitrary n prelim}
    \begin{split}
        \mathcal{J}_{n,a}\left[f\right](\rho)=&\widehat{\mathcal{J}}_{n,a}\left[\widehat{f}(2\pi p \rho)\right](\rho)\\
    =&(-i)^n\widehat{\mathcal{J}}_{0,a}\left[\widehat{f}(2\pi p \rho)T_n(-2\pi p)\right](\rho)\\
    =&\frac{n}{2}\sum_{j=0}^{\lfloor n/2 \rfloor}\left(\frac{2 i}{\rho}\right)^{n-2j}\frac{(n-j-1)!}{j! (n-2j)!}\\
    &\quad\times\widehat{\mathcal{J}}_{0,a}\left[\widehat{f}(2\pi p \rho)(2\pi p \rho)^{n-2j}\right](\rho).
    \end{split}
\end{align}
If we expand the last factor in the above sum, mainly
\begin{align}\label{eq: the focal term}
    \widehat{\mathcal{J}}_{0,a}\left[\widehat{f}(2\pi p \rho)(2\pi p \rho)^{n-2j}\right](\rho)
\end{align}
into a focal series using \cref{eq: J0 Focal Expansion} we will need to evaluate the following foci\footnote{Note that in going to the focal expansion the argument $2\pi p \rho$ of $\widehat{f}$ turns into just $\rho$ -- compare \cref{eq: J0 Focal Expansion} with \cref{eq: Fourier convolution trick}.}, depending on the parity of $n$:
\begin{align} \label{eq: foc comp 1}
\left(\partial\right)^{\pm}_{k} \left(z^{2\Delta} \widehat{f}(z)\right)^{\pm}\Big|_{z=\rho}=&\left(\partial\right)^{\pm}_{k} z^{2\Delta} \widehat{f}^{\pm}(z)\Big|_{z=\rho}\\
\label{eq: foc comp 2}\left(\partial\right)^{\pm}_{k} \left(z^{2\Delta+1} \widehat{f}(z)\right)^{\pm}\Big|_{z=\rho}=&\left(\partial\right)^{\pm}_{k} z^{2\Delta+1} \widehat{f}^{\mp}(z)\Big|_{z=\rho}
\end{align}
where $\Delta \equiv \lfloor n/2 \rfloor-j$ and where we wrote the odd and even parts of the composite functions explicitly via the odd and even parts of $\widehat{f}$. We can expand out \cref{eq: foc comp 1,eq: foc comp 2} using the focal operator properties \cref{eq: product of power and function 1,eq: product of power and function 2} derived earlier. Doing so in the focal expansion of \cref{eq: the focal term}, plugging the resulting series back into \cref{eq: arbitrary n prelim}, and finally arranging all of the terms in increasing $\left(\partial\right)^{\pm}_k \widehat{f}$ order\footnote{In the process one is forced to evaluate a non-trivial hypergeometric sum, like in \cref{eq: strange hyp geo} this can be done using Sister Celine's method \cite{AB}.} we obtain the focal expansions for generic even and odd order Bessel function kernels:
\begin{align}
    \begin{split}\label{eq: focal expansion even bessel}
        \mathcal{J}_{2n,a}&\left[f\right](\rho)=\sum_{k=0}^{\infty}\frac{(-1)^n}{(-4 \pi a)^k}\Bigg{\{ }\\
    &\left(\left(\partial\right)^+_k \widehat{f}^+\right)(\rho)\sum_{l=0}^{n}\Bigg[\frac{(2k+2l-1)!!}{k!}\frac{(-4)^l}{1+\Omega_{l,n}}\\
    &\quad\quad\quad\quad \quad\quad\quad\quad \times \binom{n+l}{2l}\frac{J_{k+l}(2\pi a \rho)}{(2\pi a\rho)^l}\Bigg]\\
    -i&\left(\left(\partial\right)^-_k \widehat{f}^-\right)(\rho)\sum_{l=0}^{n}\Bigg[\frac{(2k+2l-1)!!}{k!}\frac{(-4)^l}{1+\Omega_{l,n}}\\
    &\quad\quad\quad\quad \quad\quad\quad\quad \times \binom{n+l}{2l}\frac{J_{k+l+1}(2\pi a \rho)}{(2\pi a\rho)^l}\Bigg]\Bigg{\}}
    \end{split}
\end{align}
where we defined
\begin{align}\label{eq: Omega}
    \Omega_{l,n}\equiv\begin{cases}
        0, &n=0,\\
        \frac{l}{n}, &n\geq 1,
    \end{cases}
\end{align}
and
\begin{align}\label{eq: focal expansion odd bessel}
    \begin{split}
        \mathcal{J}_{2n+1,a}&\left[f\right](\rho)=\sum_{k=0}^{\infty}\frac{(-1)^n}{(-4 \pi a)^k}\Bigg{\{ }\\
    &\left(\left(\partial\right)^+_k \widehat{f}^+\right)(\rho)\sum_{l=0}^{n}\Bigg[\frac{(2k+2l-1)!!}{k!}(-4)^l\\
    &\quad\quad\quad\quad \quad\quad\quad\quad \times \binom{n+l}{2l}\frac{J_{k+l+1}(2\pi a \rho)}{(2\pi a\rho)^l}\Bigg]\\
    +i&\left(\left(\partial\right)^-_k \widehat{f}^-\right)(\rho)\sum_{l=0}^{n+1}\Bigg[\frac{(2k+2l-1)!!}{k!}\frac{(-4)^l}{1+\Omega'_{l,n}}\\
    &\quad\quad\quad\quad \quad\quad\quad\quad \times \binom{n+l+1}{2l}\frac{J_{k+l}(2\pi a \rho)}{(2\pi a\rho)^l}\Bigg]\Bigg{\}}
    \end{split}
\end{align}
where we defined
\begin{align}\label{eq: Omega prime}
    \Omega'_{l,n}\equiv\begin{cases}
        3\delta_{l,1}, &n=0,\\
        \frac{l(4n+2l+1)}{2n(n+1)+l}, &n\geq 1.
    \end{cases}
\end{align}

Despite appearance, the factors of $1/\rho$ in \cref{eq: focal expansion even bessel,eq: focal expansion odd bessel} do not lead to divergences at the origin since 
\begin{align}\label{eq: bessel order}
    J_{m+l}(z)/z^l = O(z^m)
\end{align}
for small arguments. However, what \cref{eq: bessel order} also shows is that every term in the sums over $l$ in \cref{eq: focal expansion even bessel,eq: focal expansion odd bessel} at fixed $k$ comes in at the same $\rho^k$ or $\rho^{k+1}$ order near the origin. This means the behavior for small $\rho$ is obscured due to a possible cancellations between many different terms. In contrast, the large $\rho$ behavior of each of these sums will be clearly dominated by the $l=0$ term. One can remedy this and make both small and large $\rho$ behavior of the $l$ sums manifest -- as this is a somewhat cosmetic fix with a non-trivial derivation we delegate the details to appendix \ref{app: alt form of bessel kernels} and simply copy the final result here for convenience:
\begin{align}\label{eq: focal expansion alt master main_body}
    \begin{split}
        \mathcal{J}_{N,a}&\left[f\right](\rho)=\sum_{k=0}^{\infty}\frac{(2k-1)!!}{k! (-4\pi a)^k}\Bigg\{\\
        &\left(\left(\partial\right)^+_k \widehat{f}^+\right)(\rho)\Bigg[\sum_{m=0}^{\min(\lfloor N/2\rfloor,k)}\frac{J_{N+k-m}(2\pi a \rho)}{(\pi a \rho)^m}\\
        &\text{\phantom{aaaaaaaaaaaaaaa}}\times\binom{N}{N-2m}(k)^{\downarrow}_m\Bigg]\\
        -i&\left(\left(\partial\right)^-_k \widehat{f}^-\right)(\rho)\Bigg[\sum_{m=0}^{\min(\lceil N/2 \rceil,k+1)}\frac{J_{N+k-m+1}(2\pi a\ \rho)}{(\pi a \rho)^m}\\
        &\text{\phantom{aaaaaaaaaaa}}\times\binom{N+1}{N-2m+1}(k+1)^{\downarrow}_mQ_{N,k}(m)\Bigg]\Bigg\}
    \end{split}
\end{align}
where $N$ is an arbitrary non-negative integer, $\lceil x \rceil$ is the ceiling function, and we have defined
\begin{align}\label{eq: Q def main_body}
    Q_{N,k}(m)\equiv 1-m \frac{N+2(k+1)}{(k+1)(N+1)}.
\end{align}
Note that this procedure had the unexpected benefit of unifying the even and odd expressions. 

Equation~(\ref{eq: focal expansion alt master main_body}) is the main theoretical result of the present work. It extends the qualitative ``pincer maneuver" logic of section \ref{sec: preliminaries} into a proper, all-orders approximation scheme that preserves the key property of being accurate over the full range of arguments. As an immediate sanity check, observe that if $f(r,\theta)=\delta (r-a)$ then $\widehat{f}^+=1$, $\widehat{f}^-=0$ and so the whole focal expansion collapses to just the even $k=0$ term, mainly, $J_N(2\pi a \rho)$ as expected.

Note that the leading factor of $1/(4 \pi a)^k$ in  \cref{eq: focal expansion alt master main_body} does not violate dimensional analysis, the units cancel with those coming from the foci of $\hat{f}$ which are composed of derivatives and factors of $1/\rho$. This motivates a rough heuristic: if $f(x)$ has some dimensionful measure of width $w$ we anticipate that $(\partial)^{\pm}_k\hat{f}$ will contribute a factor of $w^k$ which turns the leading factor of $1/(4 \pi a)^k$ into the dimensionless ratio $(w/4 \pi a)^k$. This is a crude estimate -- one can expect factors of $w$ to arise naturally from derivatives but the dimensionful $1/\rho$ factors in the foci of $\hat{f}$ complicate the situation. 

Nevertheless, this reasoning allows us to draw some modest conclusions about the convergence properties of \cref{eq: focal expansion alt master main_body}. We will not investigate the convergence conditions fully rigorously, however, note that if we assume the foci of $\widehat{f}$ don't grow significantly with the order $k$  the summands in \cref{eq: focal expansion alt master main_body} will grow roughly as
\begin{align} \label{eq: leading large k behavior}
    \frac{(2k-1)!!}{k!}\left(\frac{-w/a}{4 \pi}\right)^k
\end{align}
where we ignored any large $k$ behavior of the Bessel functions -- this is conservative as one can show the maximum of $J_m$ falls like $1/m^{1/3}$ for large $m$ \cite{abramowitz1965handbook}. If we were considering a series with coefficients given by \cref{eq: leading large k behavior} it would converge absolutely to
\begin{align}
    \frac{1}{\sqrt{1+w/2\pi a}}
\end{align}
for all $w/a<2\pi$. 
This gives us some hope as to the convergence of \cref{eq: focal expansion alt master main_body}; at the very least it is not obviously asymptotic (i.e., a  useful approximation but formally factorially divergent) in character. We see the convergence of the focal expansion will likely depend on the factor $w/(2 \pi a)$ being small but, as we shall see in the next section, we can in principle force it to be smaller if need be.

In the case that our Bessel function integral comes from a Fourier-transform, like in \cref{f hat decomposed into sums of integrals}, the integrand will contain an extra multiplicative factor of $r$ so that 
\begin{align}\label{eq: extra x}
    f(x-a)=x g(x-a)
\end{align}
where $g(x-a)$ is now the radial profile (extended to the full real line) of one of the angular modes of our 2D image function.  This means $f(x)=x g(x)+a g(x)$ and so $\widehat{f}(\rho)=a\widehat{g}(\rho)+i/2\pi \widehat{g}'(\rho)$. The focal expansion is linear in the function being integrated over which means (using the notation of \cref{eq: linear functional alt})
\begin{align}
    \mathcal{J}_{N,a}\left[x g(x-a)\right](\rho)=a\widehat{\mathcal{J}}_{N,a}\left[\widehat{g}\right](\rho)+\frac{i}{2\pi}\widehat{\mathcal{J}}\left[\widehat{g}'\right](\rho).
\end{align}
We see that to obtain the relevant series one must sum two separate focal expansions. One must be careful here when it comes to orders, as discussed above $\widehat{g}'$ will have an extra factor of $w$ meaning that the $(k+1)$th term in the focal expansion of $\widehat{\mathcal{J}}\left[\widehat{g}\right]$ comes in at the same order as the $k$th term in the focal expansion of $\widehat{\mathcal{J}}\left[\widehat{g}'\right]$. In particular, we see only the $\widehat{\mathcal{J}}\left[\widehat{g}\right]$ term contributes to leading $O((w/a)^0)$ order. Note also that, as a result, the leading focal order can also be obtained by simply setting the extra factor of $x$ in \cref{eq: extra x} to $a$.

Before moving on, we address the problem of choosing the appropriate shift amount $a$ of the profile function $f(x-a)$ in our integral \cref{eq: Ika generic}. As it stands it is an unconstrained a free parameter; once we have chosen an integral and fixed the kernel the integrand is also fixed, but our assignment of $f$ and $a$ is not. Any shift $a\rightarrow a+\delta $ must be accompanied by redefining our function  $f(x)\rightarrow f(x+\delta)$. On the whole this maneuver does not change, of course, the actual value of \cref{eq: Ika generic} but its effect on the focal expansion \cref{eq: focal expansion alt master main_body} seems more troubling. We see that $a$ directly controls the frequency of oscillations for large $\rho$, which is certainly fixed and not a free parameter. 

To resolve this apparent contradiction, consider the leading terms of \cref{eq: focal expansion alt master main_body} at large $\rho$, using \cref{eq: f evenodd decomp,eq: focal op explicit +,eq: focal op explicit -,eq: Bessel asymptote} they simplify to 
\begin{align}
\begin{split}\label{eq: JN large rho}
        &\mathcal{J}_{N,a}[f](\rho)\Big|_{\rho \gg 1}\approx \sum_{k=0}^{\infty}\frac{i^{N+k} (2k-1)!!}{k! (-4\pi a)^k 2\pi \sqrt{a\rho}}\\
    &\times\left[e^{\frac{i\pi}{4}} \hat{f}^{(k)}(\rho)e^{-i2\pi a \rho}+(-1)^{N+k} e^{-\frac{i\pi}{4}}\hat{f}^{(k)}(-\rho)e^{i2\pi a \rho}\right].
\end{split}
\end{align}
Now, as mentioned above changing $a\rightarrow a+\delta$ also requires changing $f(x)\rightarrow f(x+\delta)$ if we want the integral itself to remain fixed. But the argument shift on $f$ corresponds to multiplication by the phase $e^{i2\pi \delta \rho}$ in Fourier space. This means that
\begin{align}\label{eq: fhatk change}
    \widehat{f}^{(k)}(\rho)\rightarrow e^{i2\pi \delta \rho}\sum_{j=0}^{k}\binom{k}{j}(i2\pi \delta)^{k-j}\widehat{f}^{(j)}(\rho).
\end{align}
Because the phase in \cref{eq: fhatk change} factors out front we see that, on the whole, the frequency $2\pi a$ of the oscillations in \cref{eq: JN large rho} remains totally unchanged under the shift $a\rightarrow a+\delta$. In contrast, it is apparent that under this shift the envelope of the oscillations will undergo changes at each focal order, all of which will presumably cancel at each other when resummed.

It is still advisable to choose $a$ to correspond to the center, in some sense, of the weight of the profile function so that $f(x-a)$ is highly peaked at $x=a$ and small elsewhere. The benefits of this are apparent in a situation where, say, the peak profile is odd or even with respect to its center -- choosing $a$ to coincide with this center will then set $\hat{f}^{+}$ or $\hat{f}^{-}$ respectively to zero. If we chose a different, non-optimal value of $a$ we would be forced to deal with non-trivial cancellations between the odd and even parts of the sum. In general, we suspect a careful choice of $a$ can improve the convergence speed of the focal expansion. However, the above considerations show that choosing some particular $a$ is not essential. For complicated peak profiles there might be no unique canonical choice of $a$ but the focal expansion can survive an imperfect choice without having its oscillations go out-of-phase with the true answer for large arguments. In short: making a good $a$ choice can be useful, but is not essential to get the correct large argument frequency and phase behavior -- it will only affect the convergence of the envelope function. 

\subsection{\label{section: peak at origin}Remainders and Central Peaks}

We now address more formally the issues of extending a function on $[0,\infty)$ to a function on the full real line and that of approximating integrals with weight concentrated near or at the origin. The salience of the latter is now apparent: if our function is peaked at the origin and, as discussed at the end of the previous section, we are to choose the shift parameter $a$ close to the peak, then the naive choice would be $a=0$. That, however, renders the focal expansions \cref{eq: focal expansion even bessel,eq: focal expansion odd bessel} manifestly ill-defined. 

We will soon see these two problems are connected, but let us address the extension one first. To recap, we are concerned with a situation where our integral is \textit{not} of the form \cref{eq: Ika generic} but instead
\begin{align} \label{eq: Ika* generic}
    \mathcal{I}_{K,a}^*\left[f\right](\rho)\equiv\int_0^{\infty}f(x-a)K(2\pi x \rho)\,\text{d}x,
\end{align}
where we will use the star to indicate a restricted integration range.
We seek to extend $f$ to the full real line via a function $F$ so that the integral to be evaluated is of the form \cref{eq: Ika generic} which can be focally expanded as \cref{eq: Focal Expansion}. But this step is subtle -- we need an extension whose Fourier transform is analytically tractable, otherwise we could just trivially set 
\begin{align} \label{eq: Ftrivial}
    F_{\text{trivial}}(x)\equiv 
    \begin{cases}
        f(x), &x\geq -a,\\
        0, &x<-a,
    \end{cases}
\end{align}
so that $\mathcal{I}^*_{K,a}[f]=\mathcal{I}_{K,a}[F]$ exactly. The problem is that in many cases we will not be able to write $\widehat{F}_{\text{trivial}}$ in closed form. In general, $F$ will need to be a compromise that accurately captures the majority of the weight of the profile $f$ and that has an analytically tractable 1D Fourier transform. We will see explicit examples of this in sections \ref{sec: gaussian ring} and \ref{sec: log ring}.

\begin{figure}
    \includegraphics[width=1.03\linewidth]{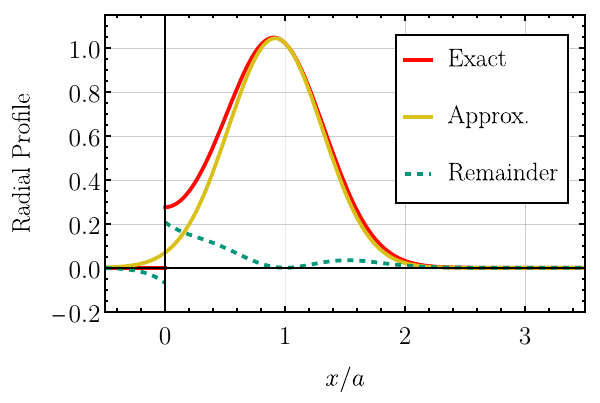}
    \caption{Two possible extensions of a radial function to the full real line. The solid red line corresponds to the trivial extension $F_{\text{trivial}}(x-a)$ (as in \cref{eq: Ftrivial}) which reproduces the value of the integral \cref{eq: Ika* generic} exactly. A possible compromise approximation $F(x-a)$ is shown in solid yellow and the difference $(F_{\text{trivial}}-F)(x-a)$ in dashed green. These particular functions are taken from Section~\ref{sec: gaussian ring}: the exact function is \cref{eq: gaussian general gn} while the approximation is \cref{eq: gaussian gn general approx} with $N=1$, $\sigma=0.4 a$, $n=0$, and $\gamma_0=1$.}
    \label{fig: remainder example}
\end{figure}

For the generic case, the most we can really say here is that if the definition of $F_{\text{trivial}}(x)$ in \cref{eq: Ftrivial} leads to a discontinuity at $x=-a$ then the compromise extension $F(x)$ will likely need to smooth it out --  for an example see fig.~(\ref{fig: remainder example}) where a Gaussian tail does the job.
We see from this example that our procedure will tend to make the remainder peaked (in this case discontinuously) at the origin. If we obtain a remainder not peaked at the origin we could in principle repeat this procedure until we \textit{are} faced with integrating over a function peaked at the origin. In any case, we see that for an accurate assessment of generic peak profiles we need to develop a way of dealing with functions that are peaked at the origin for which the focal expansion breaks down.\footnote{Note that we expect this to be a practical problem mostly for the $n=0$ angular mode of the decomposition \cref{f theta fourier series}. This is because, assuming our parent function $f$ is continuous, only the $n=0$ mode can be non-vanishing at the origin.}

Our strategy for dealing with integrals of this type is to remove weight from the origin in an analytically tractable manner and then apply the focal expansion to the remaining weight, which will by construction be peaked away from the origin. This can be done with the help of the following identities \cite{BesselWatson}:
\begin{align} \label{eq: exp fit}
    \int_0^{\infty}&e^{-c_1 x}J_n(c_2 x)\,\text{d}x=\frac{\left(\sqrt{c_1^2+c_2^2}-c_1\right)^n}{c_2^n \sqrt{c_1^2+c_2^2}},\\
    \int_0^{\infty}&e^{-c_1 x^2}J_n(c_2 x)\,\text{d}x=\sqrt{\frac{\pi}{4 c_1}}e^{-c_2^2/8c_1}I_{n/2}\left(\frac{c_2^2}{8 c_1}\right), \label{eq: gaussian fit}
\end{align}
where $c_1, c_2$ are positive constants and $I_{n/2}$ is a modified Bessel function of the first kind of (possibly fractional) order $n/2$. The procedure is now as follows: if the peak of our function $f(x-a)$ at $x=0$ has vanishing derivative, we fit a Gaussian $A e^{-c_1 x^2}$ to it, otherwise we fit the exponential $A e^{-c_1 x}$. We then subtract said fit from $f(x-a)$, leaving us with a function that is peaked away from the origin and to which we may apply the focal expansion without worry. We can account for the subtracted weight analytically using \cref{eq: exp fit,eq: gaussian fit}. For a discontinuous peak like the remainder in fig.~\ref{fig: remainder example} we would do this separately for both the $x>0$ and the $x<0$ parts (the latter after a change of integration variables $x\rightarrow-x$ and noting $J_n(-z)=(-1)^n J_n(z)$). We will see one explicit example of this procedure in section~\ref{section: peak at origin}.

As mentioned in the previous section, the focal expansion is, roughly speaking, a perturbative expansion in the parameter $(w/4\pi a)$ where $w$ is the width and $a$ the position of the peak. It is plausible then that a simple leading-order fit described above, despite moving the peak away strictly from the origin, will not result in $a$ big enough to make the expansion well behaved. In that situation we can analytically fit higher orders of the Taylor expansion of $f(x-a)$ near $x=0$ using functions of the type $A x^k e^{-c_1 x}$. This would hopefully allow us to move even more weight away from the origin, the corresponding analytic integrals we would need can be obtained simply by taking a derivative over $c_1$ of \cref{eq: exp fit}.

\subsection{Summary} \label{subsec: summary}

We now summarize the main results of Section~\ref{sec: The Method}. These approximation procedures are the main results of the present work; their most important property is that they are \textit{uniformly valid}, meaning they represent the true function well over a full range of arguments, not just in the small or large argument limits (there are caveats for Fourier transforms of delta-function rings -- see section~\ref{sec: delta function rings}). We stress that this is the main non-trivial property of the focal expansion: despite fixing its coefficients via simple Taylor matching at the origin, the expansions' unusual structure renders this local matching globally meaningful. We now describe the procedure of approximating a generic Bessel $J_n$ function integral:\\

\textit{Consider an integral of the form
\begin{align}\label{eq: summary int}
    \mathcal{I}_1\equiv\int_{\varepsilon}^{\infty}f(r) J_n(2\pi r \rho)\,\text{d}r.
\end{align}
with the lower limit $\varepsilon$ taken to be either $0$ or $-\infty$. To calculate the focal expansion of \cref{eq: summary int} analytically one needs to first transform $f$ via a series of semi-qualitative steps:}
\begin{enumerate}
    \item Determine whether $f(r)$ is significantly peaked at the origin (as compared to its values for $|r|>0$). If not, skip to step 4, replacing $g$ with $f$.
    \item If $f'(0)=0$ fit a Gaussian $A e^{-B r^2}$ to $f(r)$ at the origin, otherwise fit an exponential $A e^{-B r}$, where $A,B$ are constants.
    \item Subtract the fit function from $f(r)$ and call the result $g(r)$. The subtraction can be accounted for analytically as the integral over the fit function can be expressed in closed form via \cref{eq: exp fit,eq: gaussian fit}.
    \item It remains to calculate the integral over $g(r)$. First determine an approximate radius $r=a>0$ at which $g(r)$ is peaked. The choice can be rough and qualitative without fundamentally affecting the accuracy of the subsequent expansion. \textit{Note: if $\varepsilon=-\infty$ it may happen after the weight-removal step above that there are two separate peaks, one on each side of the origin. These have to be manually separated into two functions, each peaked in only one region. See section~\ref{section: peak at origin} for an example.}
    \item  Choose an approximating function $G(r)$ such that $G(r-a)\approx g(r)$ for $r>0$ but which is defined on the full real line, that last requirement will be trivially satisfied if $\varepsilon=-\infty$. The choice should be guided by analytic simplicity and introducing as little extra weight away from the $r=a$ peak as possible. In many practical situations the value of $a$ will be directly suggested by an explicit phase factor $e^{-2\pi i a \rho}$ in $\widehat{g}(\rho)$. We refer to $a$ as the \textit{shift parameter}. 
\end{enumerate}
Once $G$ and $a$ have been chosen, we switch to quantitatively analyzing the integral 
\begin{align}\label{eq: summary int2}
    \mathcal{I}_2\equiv\int_{\mathbb{R}}G(r-a) J_n(2\pi r \rho)\,\text{d}r.
\end{align}
We continue:
\begin{enumerate}
    \setcounter{enumi}{5}
    \item Calculate the 1D Fourier transform $\widehat{G}(\rho)$ and separate it out into even and odd parts $\widehat{G}^{\pm}(\rho)$.
    \item The integral \cref{eq: summary int2} can be written as \textit{focal expansion}. This is a uniformly valid expansion so for most practical purposes the leading order (LO) should suffice, it is given by $G_n(r)$:
    \begin{align}\label{eq: summary LO}
        \begin{split}
            \widehat{G}_n^+(\rho)&J_{n}(2\pi a \rho)\\
        -i \widehat{G}_n^-(\rho)&\Bigg[J_{n+1}(2\pi a \rho)- \frac{n J_{n}(2\pi a \rho)}{2\pi a \rho}\Bigg].
        \end{split}
    \end{align}
    \item If more accuracy is needed, the full extent of the focal expansion is most succinctly given by \cref{eq: focal expansion alt master main_body} (where the definitions \cref{eq: Q def main_body,eq: focusing +,eq: focusing -,eq: definition of mathcalJ,eq: Ika generic} are used) upon replacing $f\rightarrow G$. Roughly speaking, if $w$ is some measure of width of the peak $G(r)$ then the focal expansion is a perturbative series in $(w/4\pi a)$.
    \item By construction $\mathcal{I}_1\approx \mathcal{I}_2$ and so we have obtained our approximation. If more accuracy is needed at this last step, one may recursively apply the previous steps to the error term $G(r-a)-g(r)$, where one takes $g(r)=0$ for $r<0$, and splits the full integral over $\mathbb{R}$ into two separate $r>0$, $r<0$ terms expressible as \cref{eq: summary int}.
\end{enumerate}

We can immediately apply this core approximating technique to  2D Fourier transforms:\\

\textit{Consider a 2D plane with polar coordinates $(r,\theta)$ and a function $f(r,\theta)$ on it. We wish to calculate its 2D Fourier transform $\mathcal{F}[f](\rho,\phi)$, with $(\rho,\phi)$ being the polar coordinates in Fourier space. We proceed as follows:}
\begin{enumerate}
    \item Perform a Fourier \textit{series} of $f$ in $\theta$ at each radius $r$ to obtain
    \begin{align}\label{eq: summary f series}
        f(r,\theta)=\sum_{n\in\mathbb{Z}}g_n(r)e^{in\theta}
    \end{align}
    with the functions $g_n$ given in \cref{f theta fourier series}.
    \item The 2D Fourier transform of $f$ can then be written as 
    \begin{align} \label{eq: summary ang decomp of F[f]}
        \mathcal{F}[f](\rho, \phi)=2\pi\sum_{n\in\mathbb{Z}}e^{i n (\phi-\pi/2)}\mathcal{J}_n(\rho)
    \end{align}
    with $\mathcal{J}_n$ given by
    \begin{align}\label{eq: summary intj}
        \mathcal{J}_n(\rho)=\int_0^{\infty}r g_n(r) J_n(2\pi r \rho)\, dr.
    \end{align}
    \item We now proceed to apply the approximation method described below \cref{eq: summary int} to each of the integrals $\mathcal{J}_n$ from the previous step\footnote{In practice one might have to settle for a finite angular resolution and truncate after a finite number of angular modes.}. Here we have a choice: either we apply the first several steps to $r g_n(r)$ as a whole, or to $g_n(r)$ alone. If we choose the former, we follow all instructions in the previous procedure and we are done. We call the final result \textit{the focal expansion} of $\mathcal{F}[f]$.
    \item Assume instead one chooses to deal with each $g_n(r)$ alone (that is, removing weight at the origin, extending to the full real line, etc.). Call the extension to the full real line, complete with shift parameter $a$, $G_n(r-a)$. The function relevant to the focal expansion is then $r G_n(r-a)$, undoing the shift we get $(r+a)G_n(r)$. This means the relevant 1D Fourier transform is
    \begin{align}\label{eq: summary effect of r}
        a\left( \widehat{G}_n(\rho)+\frac{i}{2 \pi a}\widehat{G}'_n(\rho)\right),
    \end{align}
    and this is the function to be used in the focal expansion\footnote{Note that if $G_n$ has a definite parity (so that the peak is even/odd with respect to its center) then the odd/even components of \cref{eq: summary effect of r} will cleanly correspond to the two distinct terms that constitute it.}. Because the derivative term brings with it an extra factor of $w/a$ (where $w$ is a measure of the width of the peak in $G_n$) the leading $O((w/a)^0)$ order is given by simply replacing $\widehat{G}$ in \cref{eq: summary LO} by $a \widehat{G}_n$.
    \item If more accuracy is needed once can, as with the previous procedure, take successive focal orders using \cref{eq: focal expansion alt master main_body}. We once again call the final result \textit{the focal expansion} of $\mathcal{F}[f]$. The small caveat here is that one will need to sum two focal expansions for two separate functions: one for $a\widehat{G}_n$ and the other for $\frac{i}{2\pi}\widehat{G}_n'$. As mentioned above a derivative brings with it an extra factor of $w/a$ and so the term of order $O((w/a)^k)$ in the full expansion for \cref{eq: summary effect of r} consists of a sum of the $k$th focal term in the expansion for $a \widehat{G}_n$ alone and the $(k-1)$th focal term in the expansion of $\frac{i}{2\pi}\widehat{G}'_n$ alone (the latter taken to vanish if $k=0$). 
\end{enumerate}

Intuitively, we may think of the focal expansion as a cousin to the Fourier convolution theorem, albeit an unusual one that separates the angular and radial dimensions. The Fourier convolution theorem holds that, in any dimension, we have
\begin{align}\label{eq: convolution theorem}
    \mathcal{F}\left[f_1*f_2\right]=\mathcal{F}[f_1]\mathcal{F}[f_2]
\end{align}
where $*$ represents convolution and the right hand side is simply a product. Now, say that the function $f(r,\theta)$ is circular ring of radius $a$ with a radial profile $G(r)$ all around. The Fourier transform of a delta function ring is $J_0(2\pi a \rho)$ (see Section~\ref{sec: delta function rings}) so a tempting -- but incorrect -- heuristic would be to treat the ring image as a 1D radial convolution. 
This argument is of course a totally incorrect application of the convolution theorem, for one the functions being convolved in \cref{eq: convolution theorem} are both 2D and so are their Fourier transforms, while we have to deal with the 1D Fourier transform of the profile. Nevertheless one can intuitively think of the focal expansion as the rigorous reflection of this naive want of making 2D functions play by 1D convolution rules: it separates out the radial direction (represented by the 1D Fourier transforms of the profiles) from the angular (represented by the Bessel functions and the angular mode label $n$). The price we have to pay is that a simple product of two functions no longer suffices -- we must also introduce special operators and an infinite sum.

\section{Examples \label{sec: examples}}

In this section, we benchmark and stress-test the focal expansion on several representative families of ring-like profiles. We start with \emph{annulus rings} (section~\ref{sec: annulus rings}), for which both the exact Fourier transform and the focal series can be obtained analytically, providing a controlled validation of the expansion. We then turn to \emph{Gaussian rings} (section~\ref{sec: gaussian ring}), which probe two practical choices required by the method---the extension of the radial profiles to the full real line and the choice of shift parameter. Next we consider \emph{logarithmic rings} (section~\ref{sec: log ring}) motivated by gravitational lensing in a black hole's photon ring; here the relevant Hankel-type integrals are evaluated numerically and the examples probe the behavior of the focal expansion in the presence of integrable divergences and slow tails. We also include a \emph{centrally peaked} example to test the prescription of section~\ref{section: peak at origin} for removing weight at the origin. Finally, we discuss \emph{delta-function rings} of zero thickness (section~\ref{sec: delta function rings}) with non-circular shapes, where the focal expansion remains accurate for individual angular modes but the full Fourier transform can involve nontrivial resummations over modes.

\subsection{Annulus Rings \label{sec: annulus rings}}

\subsubsection{General Calculations \label{sec: annulus general calc}}

We first analyze rings whose radial profile is given by a rectangular function multiplying a polynomial so that
\begin{align} \label{eq: f for annulus ring}
    f(r,\theta)=\Pi\left(\frac{r-a}{w}\right)\sum_{n\in \mathbb{Z}}e^{i n \theta}\gamma_n p_n(r-a)
\end{align}
where the sum over $n$ represents angular modes, $p_n$ are polynomials (not to be confused with the $p$ that we have used in the previous section as an integration variable), and
\begin{align}
    \Pi(x)\equiv 
    \begin{cases}
        1, &|x|\leq 1/2,\\
        0, &|x|>1/2.
    \end{cases}
\end{align}
Equation~(\ref{eq: f for annulus ring}) has finite radial support of width $w$ and for $w< 2a$, which we assume from now on, represents a radially and angularly modulated annulus.

This family of functions is particularly useful in probing the validity of the focal expansion as here we can calculate both the true $\mathcal{F}[f]$ and every order of our approximation fully analytically. We begin by describing how to calculate the former. The angular modes $g_n(r)$ of $f(r,\theta)$ have the radial dependence given by
\begin{align}\label{eq: gn annulus general}
    g_n(r)=\gamma_n p_n(r-a)\Pi\left(\frac{r-a}{w}\right).
\end{align}
Per Section~(\ref{subsec: summary}) we see we must evaluate the integrals
\begin{align} \label{eq: integral annulus general}
    \mathcal{J}_n\equiv \gamma_n \int_{a-w/2}^{a+w/2}\text{d}r\, r \,p_n(r-a) J_n(2\pi r \rho)
\end{align}
where we again assumed $a> w/2$. Here the recurrence relations \cref{eq: Bessel recurrence 1/x,eq: Bessel recurrence d/dx} come to help. Taking their difference and multiplying by $z^k$ we get
\begin{align}
    z^k J_{n+1}(z)=(n+k)z^{k-1}J_n(z)-\partial\left(z^k J_n(z)\right).
\end{align}
This can be used to iteratively reduce any integral over $z^k J_n(z)$ to one over a bare Bessel function, one can show the end result is
\begin{align} \label{eq: integral zk Jn}
    \begin{split}
        \int&\text{d}z\, z^k J_n(z)=2^k \left(\frac{n+k-1}{2}\right)^{\downarrow}_k \int\text{d}z\, J_{n-k}(z)\\
        -&\sum_{j=0}^{k-1}2^j  \left(\frac{n+k-1}{2}\right)^{\downarrow}_j z^{k-j}J_{n-j-1}(z)+\text{Const.}
    \end{split}
\end{align}
As for the remaining integral, the recurrence \cref{eq: Bessel recurrence d/dx} can once again be used to obtain a closed form solution, albeit at a small price of having to introduce just one extra special function:
\hypergeometricsetup{
  fences=brack,
  separator={,},
  divider=bar,
}
\begin{align}
    \int& \text{d}z\, J_{2n+1}(z)= J_0(z)-2\sum_{j=0}^n J_{2j}(z)+\text{Const.} \label{eq: integral J2n+1}\\
    \int& \text{d}z\, J_{2n}(z)=-2\sum_{j=0}^{n-1}J_{2j+1}+\int \text{d}z\, J_0(z)\label{eq: integral J2n}\\
    \int& \text{d}z\, J_0(z)=\pFq{1}{2}{1/2}{1,3/2}{-\frac{z^2}{4}}\;z\;+\text{Const.}\label{eq: integral J0}
\end{align}
where we wrote the last integral using a particular generalized hypergeometric function. Once we have \cref{eq: integral J0,eq: integral J2n,eq: integral J2n+1,eq: integral zk Jn} calculating any of the integrals given by \cref{eq: integral annulus general} is straightforward and so we omit the details.

Having found the exact result, we move to the focal expansion. Extending \cref{eq: gn annulus general} to the full real line is trivial as its support lies entirely within $r>0$ and we may then simply take the extension to vanish for $r\leq 0$ without incurring any error. The 1D Fourier transform of this extension is then
\begin{align}
    \widehat{g}_n^{\text{ext.}}(\rho)=e^{-2\pi i a\rho}\gamma_n \;p_n\left(\frac{i}{2\pi}\frac{d}{d\rho}\right)\frac{\sin(\pi w \rho)}{\pi \rho}
\end{align}
 and the phase factor in front means the natural choice of the shift parameter is simply $a$ (retroactively justifying the notation). Using the notation of Section~(\ref{subsec: summary}) we then have
\begin{align}
    G_n(r)=&\gamma_n p_n(r)\Pi\left(r/w\right),\\
    \widehat{G}_n(\rho)=&\gamma_n \;p_n\left(\frac{i}{2\pi}\frac{d}{d\rho}\right)\frac{\sin(\pi w \rho)}{\pi \rho},\label{eq: Gn annulus general}
\end{align}
where in the last step we used the fact that the Fourier transform of a rectangular step is the sinc function; note that the polynomial factor becomes a differential operator acting on sinc. Once we have \cref{eq: Gn annulus general} we calculate \cref{eq: summary effect of r}, take the even and odd components, and plug them into the focal expansion \cref{eq: focal expansion alt master main_body} to obtain our approximation.

Since we have the exact analytic results \cref{eq: integral zk Jn,eq: integral J0,eq: integral J2n,eq: integral J2n+1} for the family of annulus rings one might dismiss this set of examples as a simple diagnostic tool for the focal expansion. The diagnostic aspect is certainly true -- with the unconventional nature of our approach a robust way to check its accuracy to arbitrary orders is crucial -- but even here the focal expansion has something to offer for the pragmatically minded. Instead of a nebulous sum of a large number of Bessel functions as provided by the exact formulas, the leading order focal expansion gives us a superposition of only two terms, each with a base oscillation at frequency $2\pi a$ set by two Bessel functions, which is modulated in amplitude by the Fourier transform of the radial cross-section of the ring (if the radial function is even with respect to $r=a$ this modulation becomes a trivial multiplication). In this way the focal expansion cleanly separates the radial and angular structures.

\begin{figure*}
    \centering
    \includegraphics[width=\linewidth]{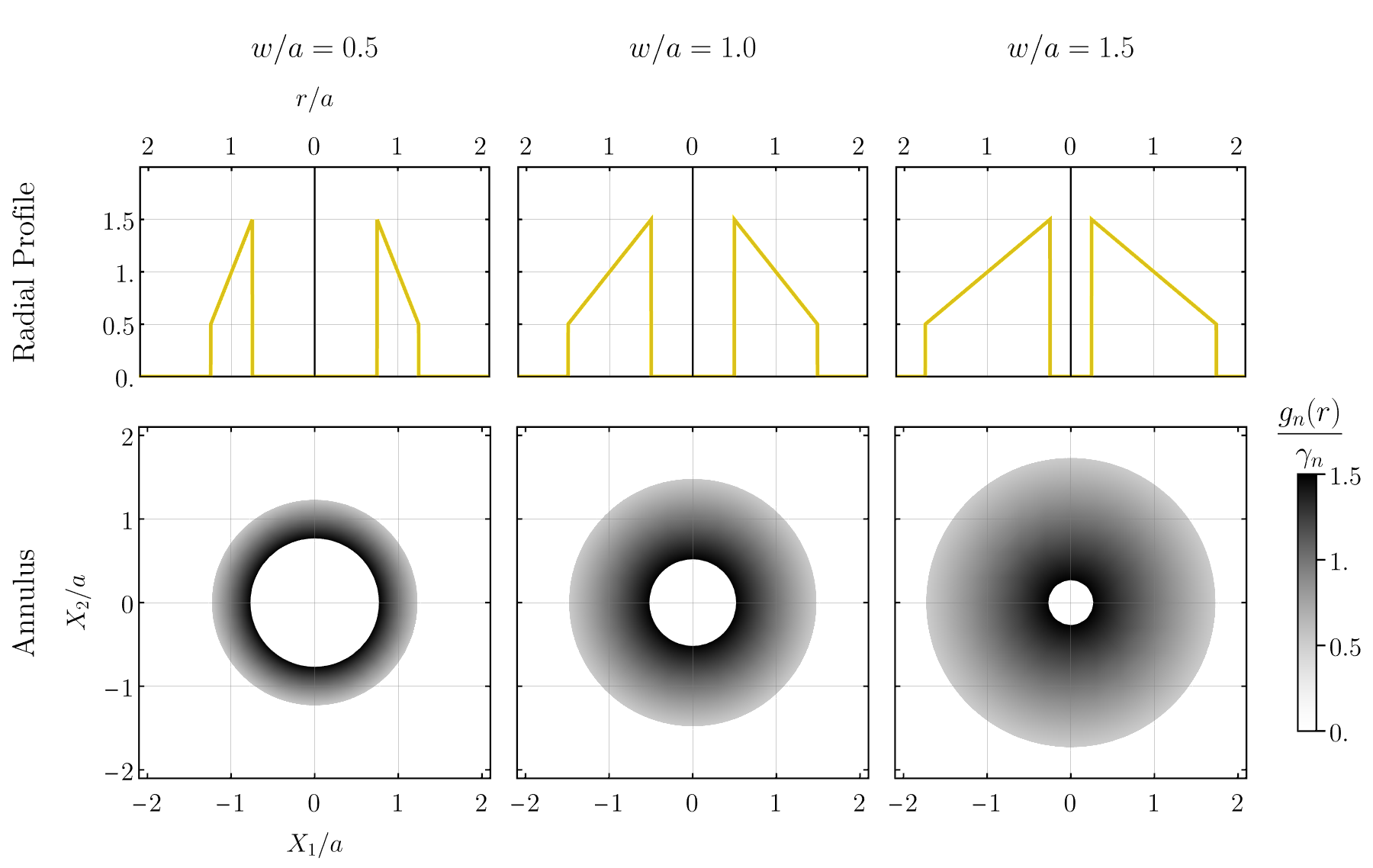}
    \caption{Example radial profiles for an annulus ring, with polynomial modulation given by \cref{eq: annulus 1 poly} and for several different values of the width $w$. These plots should not be confused with the function $f(r,\theta)$ itself which is to be Fourier transformed, they represent the radial dependence of the individual angular mode functions $g_n(r)$ which can then be summed with non-trivial angular dependence to give a non-spherically symmetric $f(r,\theta)$.}
    \label{fig: annulus density 1}
\end{figure*}

\subsubsection{Example 1: Simple Annulus Rings \label{sec: annulus example 1}}

To showcase the flexibility of the focal expansion we will skip over the case of trivial polynomial dependence $p_n=1$ (corresponding to a flat ``tophat") and jump straight to 
\begin{align}\label{eq: annulus 1 poly}
    p_n(r-a)=1-\frac{(r-a)}{w},
\end{align}
where we assume every angular mode, however many we ultimately choose to have, has the same radial behavior (we will relax this assumption in the next example). A few example radial profiles of the ring angular modes are shown in fig.~(\ref{fig: annulus density 1}).

We begin by analyzing the $n=0$ angular mode.
The exact value of integral \cref{eq: integral annulus general} is quite cumbersome but we state it here for the sake of comparison. It is given by 
\begin{align}\label{eq: annulus 1 exact}
    \begin{split}
        \mathcal{J}_0=&\gamma_0\bigg[\frac{(1+a/w)}{(2\pi\rho)^2}z J_1(z)-\frac{1}{w(2\pi\rho)^3}\bigg((2+z^2)J_1(z)\\
        -&zJ_2(z)-\pFq{1}{2}{1/2}{1,3/2}{-\frac{z^2}{4}}\;z\bigg)\bigg]\bigg|^{z=2\pi\rho(a+w/2)}_{z=2\pi\rho(a-w/2)}
    \end{split}
\end{align}
where we are meant to evaluate the difference between our expression at $z=2\pi\rho(a+w/2)$ and $z=2\pi\rho(a-w/2)$. For the focal expansion, on the other hand, we have\footnote{Without a loss of generality we may restrict ourselves to non-vanishing angular modes for which $\gamma_n\neq 0$.}
\begin{align}
    \frac{1}{\gamma_n}\widehat{G}_n(\rho)=\frac{\sin (\pi w \rho)}{\pi\rho}\left(1+\frac{i}{2\pi w \rho}\right)-\frac{i \cos(\pi w \rho)}{2\pi \rho}
\end{align}
and so the leading focal order is 
\begin{align}\label{eq: annulus 1 LO}
    \begin{split}
        \frac{1}{a^2\gamma_0}\mathcal{J}_0&\big|_{\text{Focal, LO}}=J_0(2\pi a \rho)\frac{\sin(\pi w \rho)}{\pi \rho a} \\
        +&\frac{J_1(2\pi a \rho)}{2\pi \rho a}\left(\frac{\sin(\pi w \rho)}{\pi w \rho}-\cos(\pi w\rho)\right)
    \end{split}
\end{align}
with the subleading focal correction given by

\begin{figure*}         \includegraphics[width=\linewidth]{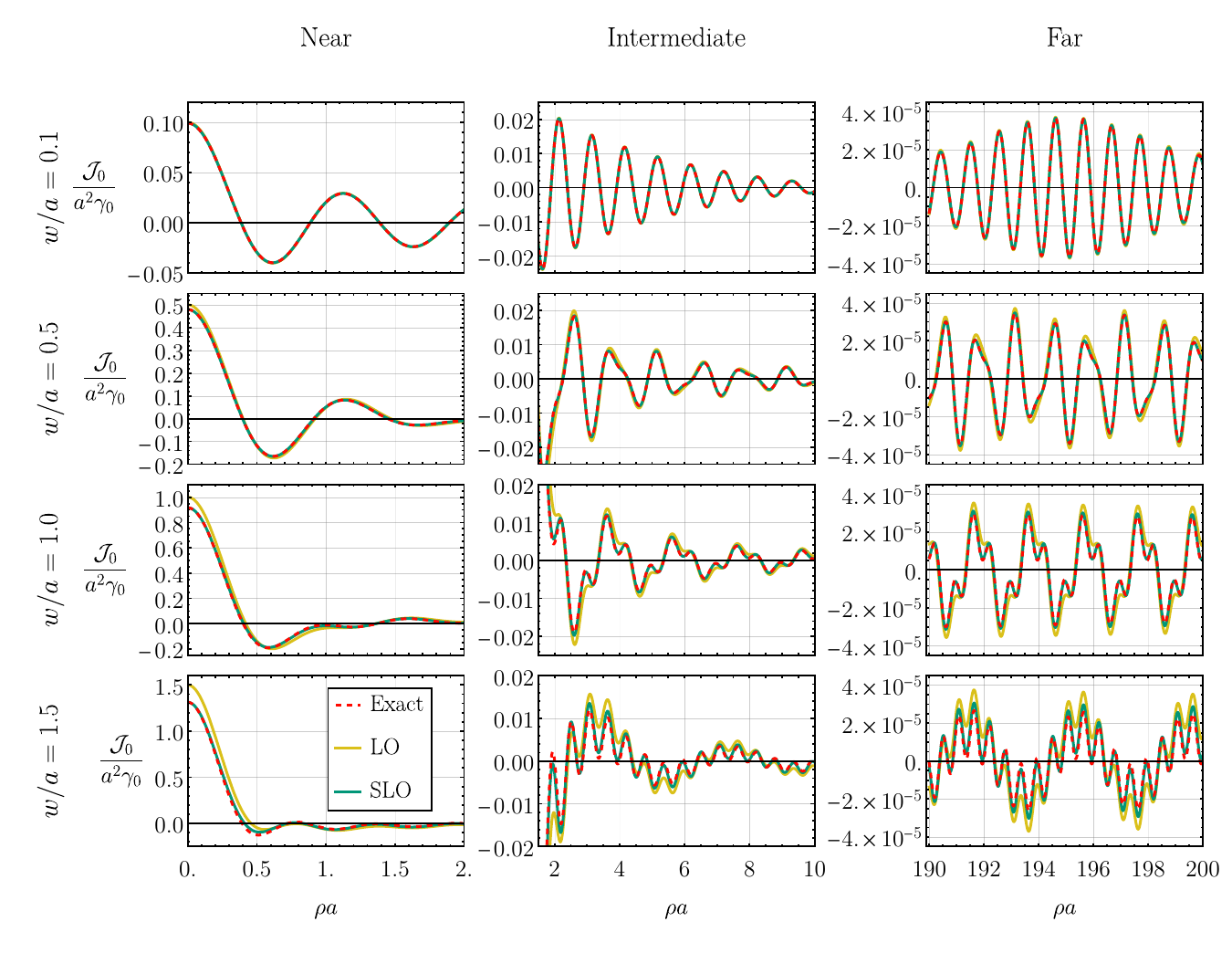}
    \caption{The Fourier transforms of several angularly symmetric annulus rings, described via $\mathcal{J}_0$ from \cref{eq: integral annulus general}. The last three rows correspond to the rings in fig.~(\ref{fig: annulus density 1}). Curves show exact results using \cref{eq: annulus 1 exact} (red, dashed line) compared to those calculated using the focal expansion \cref{eq: annulus 1 LO,eq: annulus 1 SLO} to leading (yellow) and subleading (green) order over three regimes of $\rho$ values and for four ring widths. The range in the far-$\rho$ regime represents the generic large-$\rho$ behavior in each case. We see that LO provides a reasonable and SLO an excellent approximation to the exact values across all of the plots, with discrepancies worse for larger widths.}
    \label{fig: annulus J0 width comparison}
\end{figure*}

\begin{align}\label{eq: annulus 1 SLO}
    \begin{split}
        \frac{1}{a^2\gamma_0}&\mathcal{J}_0\big|_{\text{Focal, SLO}}=\\
        -\frac{w}{a}&\frac{J_0(2\pi\rho a)}{2\pi \rho a}\left[\sin(\pi \rho w)\left(\frac{1}{2}-\frac{1}{(\pi \rho w)^2}\right)+\frac{\cos(\pi\rho w)}{\pi \rho w}\right]\\
        +\frac{w}{a}&\frac{J_1(2\pi\rho a)}{4\pi \rho a}\left[\cos(\pi\rho w)-\frac{\sin(\pi\rho w)}{\pi\rho w}\right]\\
        -\frac{w}{a}&\frac{J_2(2\pi\rho a)}{8\pi \rho a}\left[\sin(\pi\rho w)\left(1-\frac{3}{(\pi\rho w)^2}\right)+\frac{3\cos(\pi\rho w)}{\pi\rho w}\right]
    \end{split}
\end{align}
where we have elected to split orders based on powers of $w/a$ as described in section \ref{subsec: summary}. 
The main advantage of the focal approach over the exact result \cref{eq: annulus 1 exact} is the clear separation of the dependence on angular structure -- the fact we're looking at something ring shaped and so expect oscillations with frequency $2\pi a$ -- from the radial structure -- the effects of finite thickness captured in the frequency $2\pi w$. In \cref{eq: annulus 1 exact} these are tangled together inside Bessel and hypergeometric function arguments in a cumbersome and physically opaque expression. By contrast, the leading focal order \cref{eq: annulus 1 LO} is short and captures finite width effects as an amplitude modulation of frequency $2\pi w$ on top of standard Bessel function oscillations at frequency $2\pi a$. The subleading focal order \cref{eq: annulus 1 SLO} is more cumbersome but still carries this amplitude-modulation structure.

\begin{figure*}
    \centering
    \includegraphics[width=\linewidth]{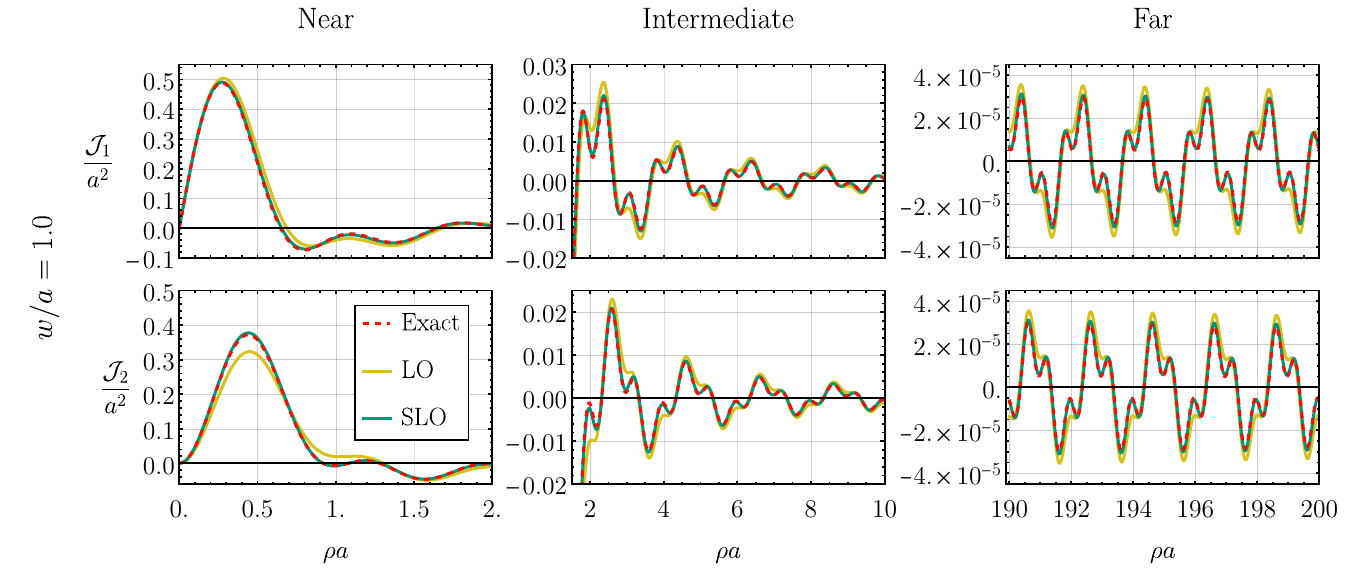}
    \caption{The $n=1,2$ angular modes of the Fourier transform of an annulus ring. The plots depict the values of $\mathcal{J}_1$ and $\mathcal{J}_2$ from \cref{eq: integral annulus general} calculated exactly (red, dashed line) compared to those calculated using the focal expansion to leading (yellow) and subleading (green) order for the single ring width $w/a=1$. Much like in fig.~(\ref{fig: annulus J0 width comparison}) we see that LO provides a reasonable and SLO an excellent approximation to the exact values across all of the plots, with discrepancies worse for larger widths.}
    \label{fig: annulus J1 J2}
\end{figure*}

Moving forward we will refer to leading order as LO, subleading order as SLO, sub-subleading order as S$^2$LO, and so on. Sometimes these will refer to contributions of individual focal orders like in \cref{eq: annulus 1 LO,eq: annulus 1 SLO} but we will also use them to denote the sum of focal terms \textit{up to} a given order. The distinction will be clear from context.

The exact, LO, and SLO results for $\mathcal{J}_0$ at $w/a$ values taken from fig.~(\ref{fig: annulus density 1}) are shown in fig.~(\ref{fig: annulus J0 width comparison}). We have split the plots into near ($\rho a \leq 2$) intermediate ($2\leq \rho a \leq 10$) and far ($190\leq \rho a \leq 200$) regimes for clarity. Overall, we see that the leading focal order (yellow line) provides a good overall estimate of $\mathcal{J}_0$ for all three widths, in all argument regimes\footnote{The discrepancy at the origin for LO does not represent a breakdown of our Taylor-order matching reasoning of Section~(\ref{subsec: The Focal Expansion}) and is an artifact of splitting the expansion based on powers of $w/a$, as described in section~(\ref{subsec: summary})}. Going to SLO fixes these shortcomings -- the SLO focal approximation (green line) is almost indistinguishable from the exact results (red, dashed line). Note in particular that the focal approximation obtains the correct phase and frequencies at large arguments -- the range of the far regime is representative of the behavior.  The approximations are decrease in accuracy with increasing $w/a$, as expected.  

The general behavior we see in figure~(\ref{fig: annulus J0 width comparison}) persists for higher angular orders: figure~(\ref{fig: annulus J1 J2}) below showcases the performance of focal expansion for $\mathcal{J}_1$ and $\mathcal{J}_2$ for the $w/a=1.0$ case. Once again, the LO provides a good and SLO excellent approximation across the full range of arguments. 

\begin{figure}
    \centering
    \includegraphics[width=\linewidth]{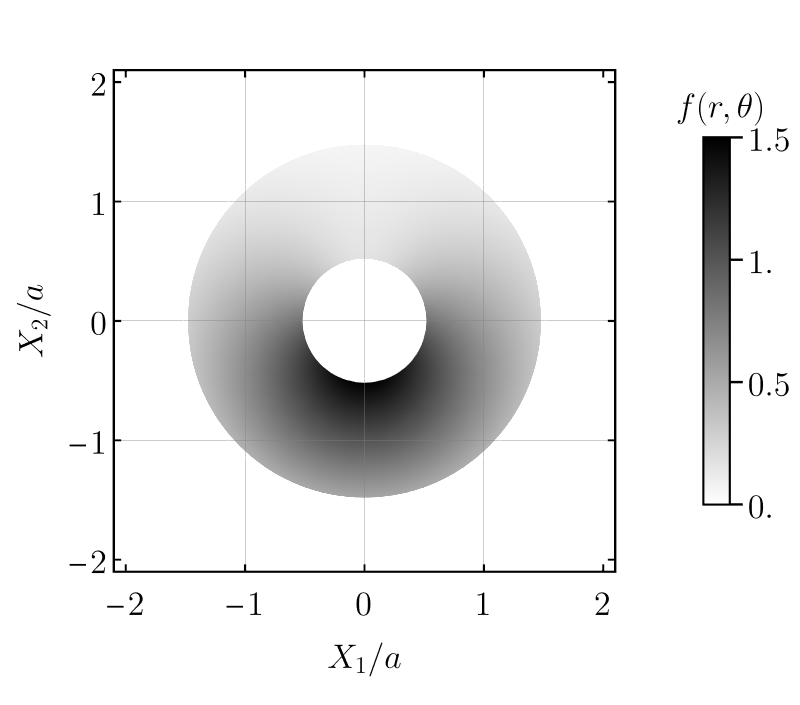}
    \caption{The density plot of an annulus ring with a rudimentary angular dependence. The function $f(r,\theta)$ is given by \cref{eq: annulus 2 fuzzy ellipse f}.}
    \label{fig: annulus first angular example}
\end{figure}

\begin{figure*}
    \centering
    \includegraphics[width=\linewidth]{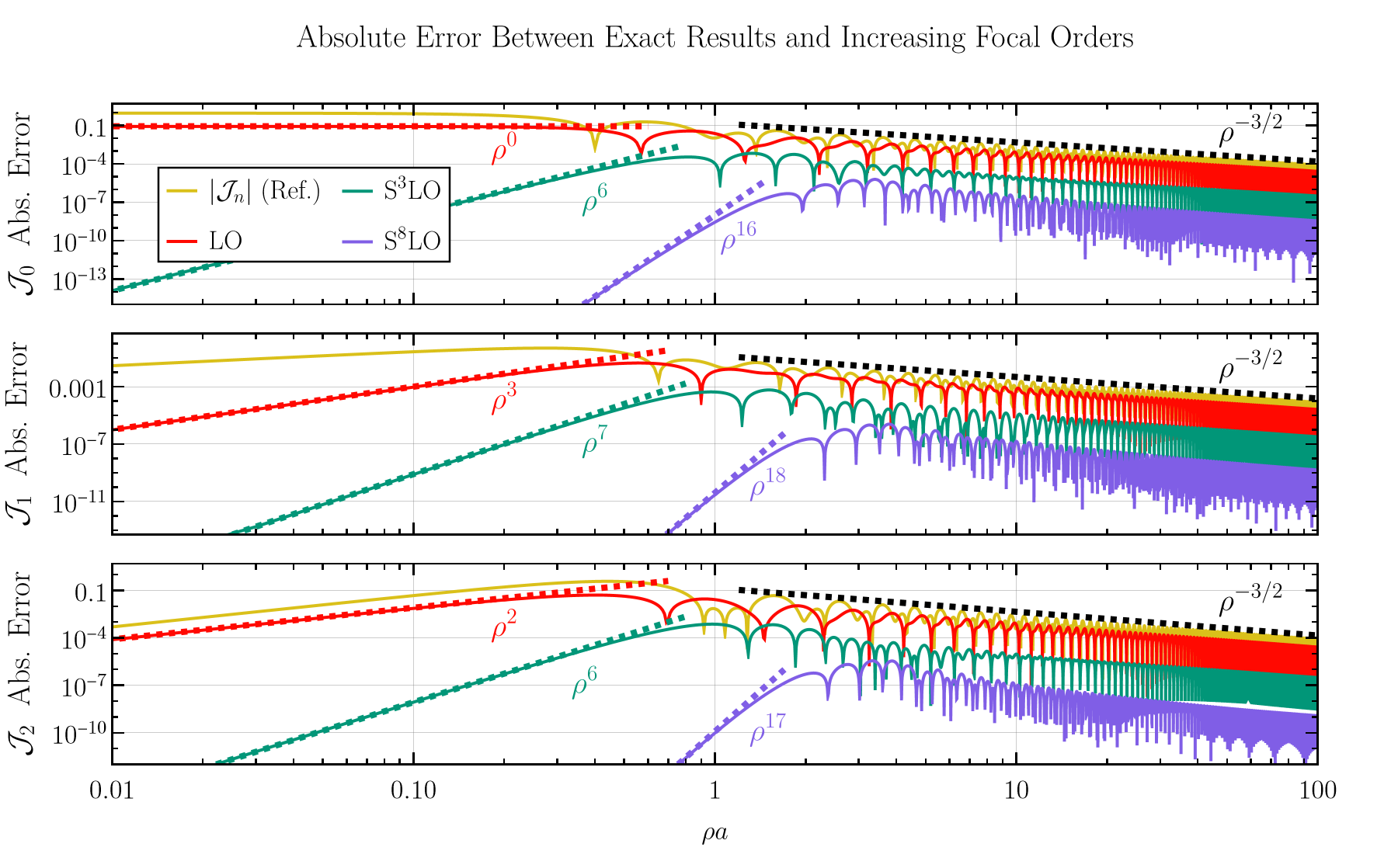}
    \caption{
    Log-log plots showing the absolute error of the focal expansion applied to an annulus ring, for several approximation orders. Specifically, we show the error \cref{eq: annulus 1 abs errors} between the exact expressions for $\mathcal{J}_0,\mathcal{J}_1, \mathcal{J}_2$ of section~\ref{sec: annulus example 1} and LO, S$^3$LO, S$^8$LO of their respective focal expansions, calculated for $w/a=1$ and with the $p_n$ of \cref{eq: f for annulus ring} given by \cref{eq: annulus 1 poly}. The full values of $|\mathcal{J}_0|, |\mathcal{J}_1|, |\mathcal{J}_2|$ were also added to the plots in yellow for reference. We see successive focal orders improve performance of the approximation over the full range of arguments, with no evidence of finite $\rho$ range of validity. The exact subleading orders (3 and 8) were chosen on the basis of visual clarity. The dashed lines indicate power laws of the form $A \rho ^{\alpha}$, with values of $\alpha$ stated nearby in the same color as the line. For $\mathcal{J}_0$ these are exactly in accordance with the focal matching performed in section~\ref{subsec: The Focal Expansion} (the $n$th focal order fits the function near the origin up to $\rho^{2 n}$). Since the focal expansion for Bessel kernels of order greater than 0 was found indirectly we don't have a priori values for these power laws in the cases of $\mathcal{J}_1$ and $\mathcal{J}_2$, and so for the bottom two plots all dashed lines were determined empirically from the data.}
    \label{fig: annulus 1 errors}
\end{figure*}
To understand better how performance of the focal expansion improves with successive orders we investigate the absolute errors, 
\begin{align}\label{eq: annulus 1 abs errors}
    \Big|\mathcal{J}_n(\rho)-\mathcal{J}_n|_{\text{Focal, S}^k\text{LO}}(\rho)\Big|,
\end{align}
for $n=0,1,2$ and several increasing orders $k$ in the representative case $w/a=1$. The results are shown in fig.~(\ref{fig: annulus 1 errors}) where $k=0,3,8$ were chosen for the sake of clarity and to avoiding too much overlap of the functions plotted. Note that both axes are logarithmic. The most striking feature of fig.~(\ref{fig: annulus 1 errors}) is the fact that successive focal orders increase the accuracy \textit{everywhere} -- there is no evidence of a limited regime of validity in $\rho$ space. This overall improvement is expected based on our derivation. First, note the red, green, and blue dashed lines which match the behavior near the origin to simple $A \rho^k$ power laws. We clearly see polynomial improvement at higher orders as expected from our Taylor series matching of section~\ref{subsec: The Focal Expansion}. Second, note that for large arguments the improvement manifests multiplicatively: all orders have the same $\rho^{-3/2}$ decay (marked by the black dashed line) suppressed by smaller and smaller multiplicative constants as the focal order increases (as multiplication by a small factor on a log-log plot corresponds visually to translating the plot downwards). This is exactly what we would expect if the large $\rho$ series was made up of terms suppressed by increasing powers of $w/2\pi a$, as we argued for in section~\ref{sec: Bessel function Kernels}. Both of these behaviors -- polynomial improvement near the origin and multiplicative improvement asymptotically -- should be generic properties of focal expansions if our derivation is to be trusted. Not all of this behavior is clear, however, as close inspection of fig.~(\ref{fig: annulus 1 errors}) will show the successive orders don't quite seem to be suppressed by the expected factors of $1/2\pi$ (since here $w/a=1$) at large $\rho$. In fact, we found the order-by-order improvement seems to oscillate between a factor of $1/4$ and just under $1$. It is clear more effort is needed to understand the convergence behavior of focal expansions but we defer all such investigations to future work, satisfied for the time being with the encouraging numerical results we have shown.

\begin{figure*}
    \includegraphics[width=\linewidth]{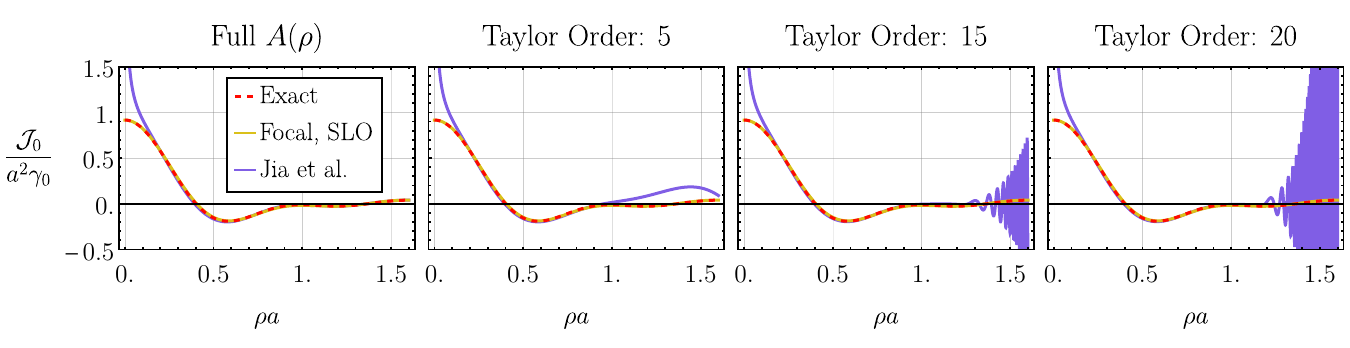}
    \caption{Comparison of the method from \cite{AlexPaper}, both the full and Taylor-expanded versions described by the authors, to the exact as well as focal results for $\mathcal{J}_0$ in the case of $f(r,\theta)=\gamma_0\left(2-\frac{r}{a}\right)\,\Pi\left(\frac{r}{a}-1\right)$ (corresponding to the middle column of fig.~(\ref{fig: annulus density 1})). The first column shows the non-Taylor expanded case, while the remaining columns show the Taylor-expanded versions recommended by the authors of \cite{AlexPaper}. We see that both the full and expanded versions of the method of \cite{AlexPaper} break down near the origin. In addition, while the full version performs extremely well above around $\rho a =0.2$, the Taylor expanded version seems to break down a little above $\rho a=1$. Note in particular that, comparing the last two columns, it appears that the breakdown point moves slightly closer to the origin for large Taylor orders, possibly indicating a finite radius of convergence. The authors of \cite{AlexPaper} experimented with expanding the phase and absolute value of \cref{eq: Alex Au} to different orders, but we elected to remain agnostic and expanded both to the same Taylor order when plotting.}
    \label{fig: Alex comp}
\end{figure*}

Before moving on to the next example, let us briefly see how our results for the individual $n=0,1,2$ angular orders can be used to describe annulus functions with non-trivial angular dependence. Consider a simple example function with the angular dependence 
\begin{align}\label{eq: annulus 2 fuzzy ellipse f}
    f(r,\theta)=\left(2-\frac{r}{a}\right)\Pi\left(\frac{r}{a}-1\right)(2.5-2\sin\theta)
\end{align}
shown below in fig.~(\ref{fig: annulus first angular example}).
The Fourier transform is then given by (noting that $\mathcal{J}_{-1}=-\mathcal{J}_1$)
\begin{align}
    \mathcal{F}\left[f\right](\rho,\phi)=2\pi\left[2.5 \mathcal{J}_0(\rho)+2 i \sin \phi \mathcal{J}_1(\rho)\right],
\end{align}
with the real part given entirely by the $n=0$ angular mode and the imaginary part given entirely by the $n=1$ angular mode. Since here $w/a=1$ we see there is no need to make new plots to check the accuracy of the focal expansion: the real part is shown by the second row of fig.~(\ref{fig: annulus J0 width comparison}) while the imaginary part is shown by the first row of fig.~(\ref{fig: annulus J1 J2}) (the former rescaled by a constant and the latter by a constant and $\sin \phi$). Naturally the same conclusions we drew before hold here too, LO is a good while SLO an excellent approximation. This situation, however, exemplifies an important catch -- the focal expansion can only be applied to individual angular modes, not to the whole function $f$ itself. Here this is a somewhat pedantic comment as the radial behavior of each mode is identical, but this observation becomes more salient in the next example we consider.

Before moving on, however, let us compare our method to existing techniques. In particular, we compare to the prescription of \cite{AlexPaper}. We focus on the spherically symmetric case, so that only $\mathcal{J}_0$ is non-vanishing, and limit ourselves to $a/w=1$ for brevity. According to the authors, for large $\rho$ we should approximate our present example by
\begin{align}\label{eq: Alex full}
    \frac{1}{a^2\gamma_0}\mathcal{J}_0\approx \frac{A(\rho)}{2 \pi a \sqrt{\rho}}e^{i \frac{\pi}{4}-2\pi i r_{\text{COL}}\rho}+\text{c.c.}
\end{align}
with c.c denoting the complex conjugate and where\footnote{The authors of \cite{AlexPaper} reserve $A$ for the absolute value of this integral, we are simplifying the notation.}
\begin{align}\label{eq: Alex Au}
    \frac{A(\rho)}{\sqrt{a}}\equiv \int_0^1\text{d}q\, e^{-2\pi i \rho (a/2+q a-r_{\text{COL}})} \left(\frac{3}{2}-q\right)\sqrt{\frac{1}{2}+q}.
\end{align}
The constant $r_{\text{COL}}$ being determined by evaluating integrals similar to the one above without the oscillating phase, in the present case it turns out to be
\begin{align}
    \frac{r_{\text{COL}}}{a}=\frac{3 \left(117 \sqrt{3}-23\right)}{14 \left(33 \sqrt{3}-17\right)}.
\end{align}
Here the authors give us two options, we may attempt to evaluate \cref{eq: Alex Au} directly, or Taylor expand the phase and evaluate a finite number of the resulting integrals\footnote{We are simplifying here for the sake of brevity, the authors Taylor expand not the complex phase in \cref{eq: Alex Au} itself but the absolute value and phase of $A(\rho)$ separately. We follow their lead when plotting below. See \cite{AlexPaper} for details.}. The presence of a square root in \cref{eq: Alex Au} is a generic feature of the approach of \cite{AlexPaper} and it unfortunately makes analytic evaluation of $A(\rho)$ difficult. Even in the present simple example doing so necessitates the introduction of additional special functions (the Fresnel $S, C$ functions to be exact), we omit writing down the full expression for brevity and simply plot the comparison in the first column of fig.~(\ref{fig: Alex comp}). We only plot the near-origin region as the non-Taylor expanded version of \cref{eq: Alex full} is visually indistinguishable from the exact result beyond it.  If we instead opt to expand the phase, the necessary integrals are somewhat easier to evaluate but the resulting series appears to have a finite radius of convergence (compare the last two columns in fig.~(\ref{fig: Alex comp})), located somewhere inside the near-origin region.

The focal expansion appears more analytically tractable when compared directly to the above method, but this might not hold universally. Certainly, if we allow ourselves to write the final result with a remaining 1D oscillatory integral over the radial cross-section of our image it is preferable for that 1D integral to not contain extra multiplicative factors of square roots etc. This, combined with its uniform validity, suggests the focal approach to be more useful, at least for spherically symmetric rings. That last comment must be kept in mind: if the function to be Fourier transformed is not spherically symmetric the focal expansion requires us to first decompose it into spherical modes, which is not required in the method of \cite{AlexPaper}, and the calculations necessary to do so might prove non-tractable. In addition, while the alternative approach of \cite{AlexPaper} becomes invalid for small $\rho$, it \textit{does} give the full leading asymptotic order for large $\rho$: if we were to calculate successive orders in this framework (the authors have not done so but it seems possible in principle) we would expect to see power-law improvements with increasing orders, instead of the weaker, constant factor power suppression of the focal expansion seen in fig.~(\ref{fig: annulus 1 errors}). 

To summarize our comparison, while the focal expansion appears to us to be more analytically tractable than the method of \cite{AlexPaper} employed without the extra Taylor expansion step, both have their advantages and could be useful in different contexts. The usefulness of the Taylor-expanded method of \cite{AlexPaper} seems more dubious -- both due to the fact that the presence of polynomial factors of $\rho$ renders the asymptotics false and the fact that the resulting power series appears to have a finite radius of convergence in some cases (as seen in fig.~(\ref{fig: Alex comp})).

\subsubsection{Example 2: Non-Circular Rings \label{sec: making shapes}}

\begin{figure}
    \centering
    \includegraphics[width=\linewidth]{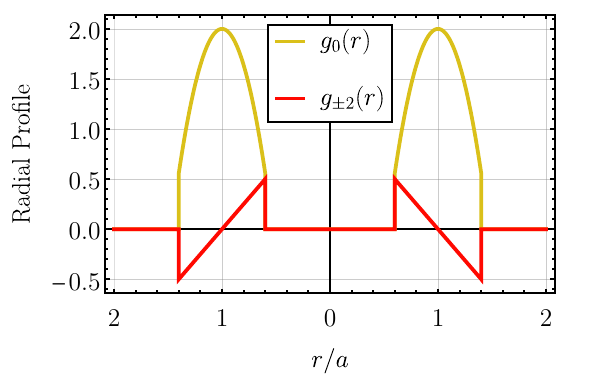}
    \caption{The radial profiles of the non-vanishing angular modes of the fuzzy ellipse $f(r,\theta)$ given by \cref{eq: annulus 2 g0g2} and shown in fig.~(\ref{fig: annulus 2 density}).}
    \label{fig: annulus 2 radial}
\end{figure}

\begin{figure}
    \centering
    \includegraphics[width=\linewidth]{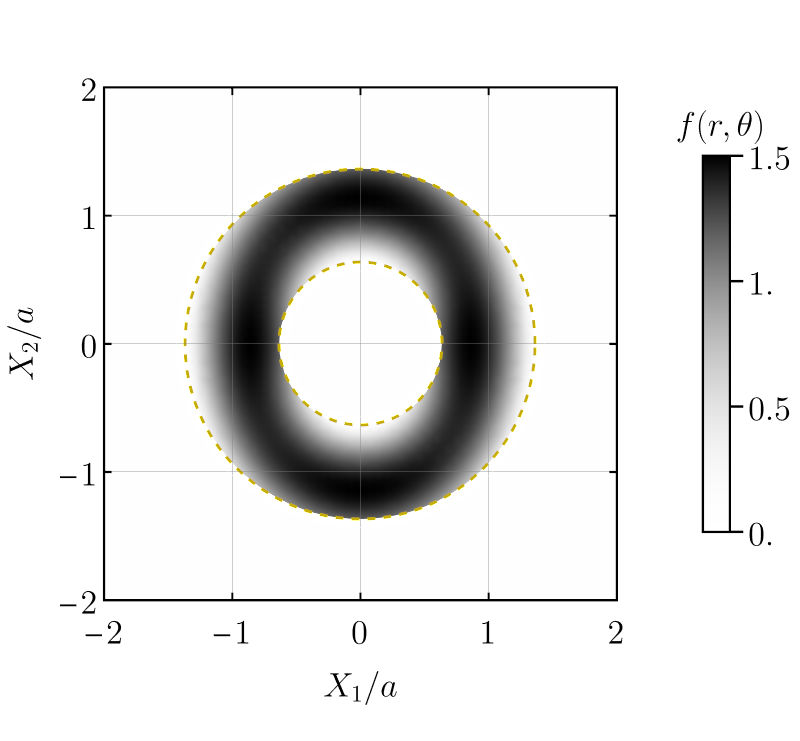}
    \caption{The density plot of the fuzzy ellipse formed using annulus rings, with $f(r,\theta)$ given by \cref{eq: annulus 2 f}. The dashed circles denote the boundaries of the support of each angular mode. Note that we were able to make a non-circular shape (a fuzzy ellipse with major axis coinciding with the $y$ axis) appear with just two angular modes, but the key was choosing the radial profiles of those two modes to not be identical. }
    \label{fig: annulus 2 density}
\end{figure}

\begin{figure*}
    \centering
    \includegraphics[width=\linewidth]{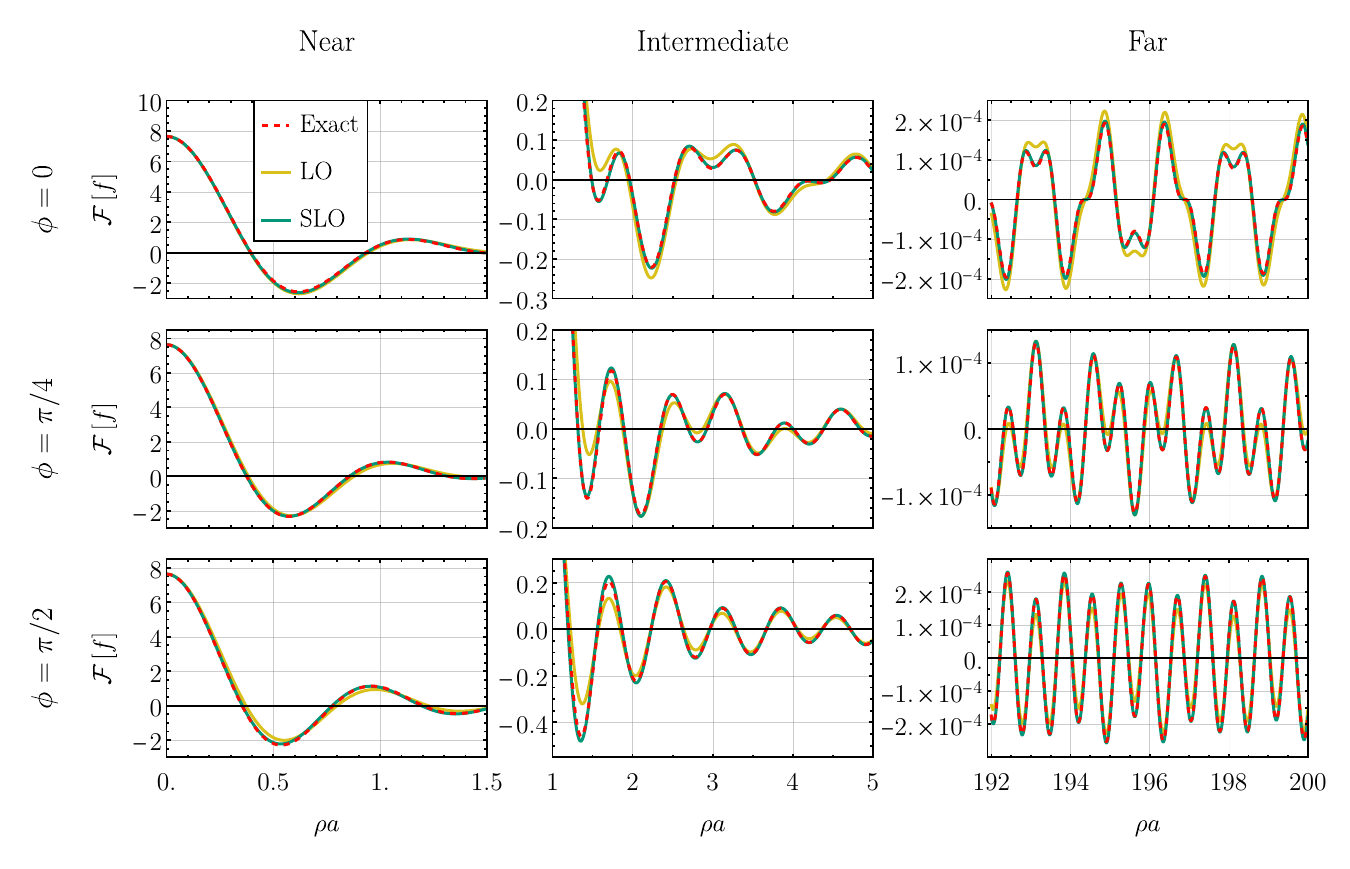}
    \caption{Comparison of the exact Fourier transform of the fig.~(\ref{fig: annulus 2 density}) fuzzy ellipse to the LO and SLO focal expansion, at three representative angles in the Fourier plane ($0^{\circ}$, $45^{\circ}$, and $90^{\circ}$ measured counterclockwise from the horizontal axis). We plotted each angle for three $\rho$ regimes: near ($0\leq \rho a\leq 1.5)$, intermediate ($1.5\leq \rho a \leq 5$), and far ($192\leq \rho a \leq 200)$. The range for the far regime is representative for phase agreement at large argument. We see the LO provides a good approximation across all plots with only minor errors around the maxima while the SLO is nearly indistinguishable from the true values. In particular, the focal approximation gives the correct phases and frequencies at large arguments.}
    \label{fig: annulus 2 angular comparison}
\end{figure*}
\begin{figure*}
    \centering
    \includegraphics[width=\linewidth]{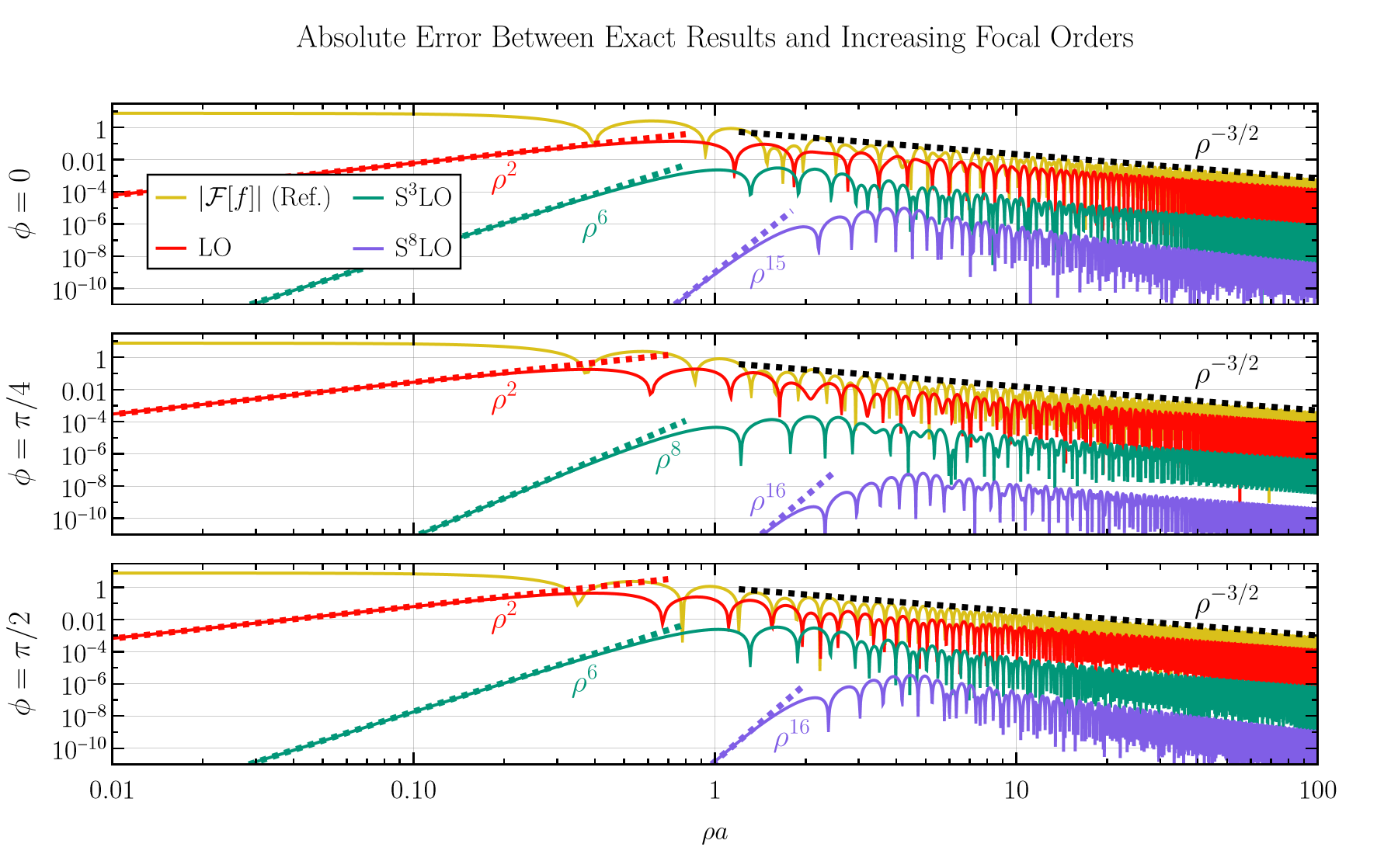}
    \caption{Log-log plots showing the absolute error of the focal expansion applied to the fig.~(\ref{fig: annulus 2 density}) fuzzy ellipse for several angles in the Fourier plane and for several focal approximation orders. Specifically, we show the errors $\big|\mathcal{F}[f](\rho,\phi)-\mathcal{F}[f]|_{\text{Focal, S}^k\text{LO}}(\rho,\phi)\big|$ between the exact expressions and the LO, S$^3$LO, S$^8$LO orders of the corresponding focal expansion, calculated at the angles $0^{\circ}$, $45^{\circ}$, and $90^{\circ}$ in the Fourier plane (as measured anti-clockwise from the horizontal). The full value of $|\mathcal{F}[f]|$ was also added to the plots in yellow for reference. We see successive focal orders improve performance of the approximation over the full range of arguments, with no evidence of finite $\rho$ range of validity. The exact subleading orders (3 and 8) were chosen on the basis of visual clarity. The dashed lines indicate power laws of the form $A \rho ^{\alpha}$, with values of $\alpha$ stated nearby in the same color as the line. These power laws were determined empirically from the data.}
    \label{fig: annulus 2 error}
\end{figure*}

We will now showcase how different angular modes can be summed together to create the appearance of a non-circular shape. The key is relaxing the assumption of the previous example demanding all angular modes have the same radial dependence -- it is precisely differences in radial behavior between individual angular modes that give rise to rings of various non-circular shapes. As an example, consider a function for which the polynomial modulation $p_n$ of the rectangular pulse differ with $n$ as
\begin{align}\label{eq: annulus 2 g0g2}
    \begin{split}
        &g_0(r)=2\left(1 - 4.5 \left(\frac{r}{a}-1\right)^2\right) \Pi\left(1.25 \left( \frac{r}{a}-1\right)\right),\\
        &g_{\pm 2}(r)=1.25 \left(\frac{r}{a}-1\right)\Pi\left(1.25 \left(\frac{r}{a}-1\right)\right),
    \end{split}
\end{align}
and $g_n=0$ identically for $n\neq0,\pm 2$. The radial profiles of the non-vanishing angular modes are shown in fig.~(\ref{fig: annulus 2 radial}) while the full function
\begin{align}\label{eq: annulus 2 f}
    f(r,\theta)=g_0(r)+2g_2(r)\cos2\theta
\end{align}
is shown in fig.~(\ref{fig: annulus 2 density}). We see it appears as a fuzzy ellipse with the major axis aligned with the $X_2$ axis.

Using the methods described in section~\ref{sec: annulus general calc} one can easily calculate both the exact response and the focal approximations to it. We omit the details other than to note that the exact result can be seen to once again non-trivially mix the angular and radial structure, just like in \cref{eq: annulus 1 exact}, while the focal expansion separates them out multiplicatively, just like in \cref{eq: annulus 1 LO,eq: annulus 1 SLO}. Moving on to showing the end results, figure~(\ref{fig: annulus 2 angular comparison}) shows the exact Fourier response compared to the LO, and SLO of the focal expansion in the case of fuzzy ellipse fig.~(\ref{fig: annulus 2 density}) at three representative angles in the Fourier plane: $0^{\circ}$, $45^{\circ}$, and $90^{\circ}$. Once again, we see levels of agreement very similar to those seen in the plots of the previous section: the LO provides a good and the SLO an excellent approximation to the exact results. The situation is similar when we look at the log-log plots of the absolute errors at the three angles, shown in fig.~(\ref{fig: annulus 2 error}), which also very closely resemble their equivalents in the previous section.

To summarize, despite being framed mostly from the angle of spherical symmetry, the focal expansion can just as readily be used in non-spherically symmetric situations. The only difference is that non-circular shapes will in general have angular modes with differing radial profiles. In that sense, the focal expansion approach is robust and versatile, making no strict assumptions about the structure of $f(r,\theta)$, other than demanding any significant weight at $r=0$ be removed by procedures described in section~\ref{section: peak at origin} and that our function be reasonably approximated with a finite number of its angular modes\footnote{Although this last assumption can prove to be false in extreme cases like delta function rings with zero thickness, see section~\ref{sec: delta function rings}.}. The leading focal order is generally a satisfactory approximation for practical purposes which, because of the relative simplicity of its general analytic form \cref{eq: summary LO}, lets us interpret the angular modes of the Fourier transform as sums of Bessel functions with large frequency $2\pi a$ (capturing the angular information -- the ring part) modulated in amplitude by 1D Fourier transforms of the respective radial profiles (capturing the radial information -- the thickness). This suggests that the effects of radial structure on the behavior of the Fourier transform are most naturally understood by first separating the problem out into a countable set of angular modes. It raises the tantalizing possibility of extracting radial structure information from the Fourier transform -- like the details of the photon ring structure in the application case of the EHT -- systematically, with minimal bias, and in a way tractable to our finite analog minds.

\subsection{Gaussian Rings \label{sec: gaussian ring}}

\subsubsection{General Calculations}

\begin{figure*}
    \centering
    \includegraphics[width=\linewidth]{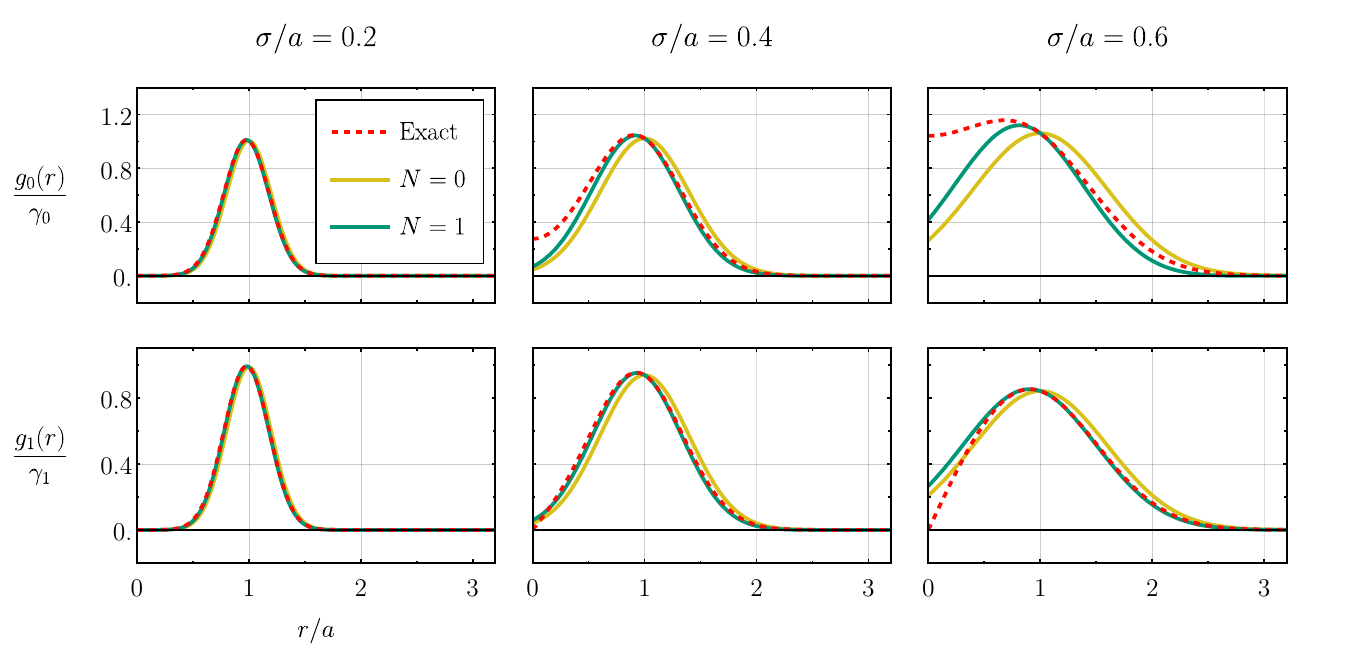}
    \caption{
    Radial profiles of the first two angular modes of a Gaussian ring at three representative values of the standard deviation $\sigma$, with both the exact function and two approximations plotted. The exact values (red, dashed) are given by \cref{eq: gaussian general gn}, while the approximations (solid yellow and green) are given by \cref{eq: gaussian gn general approx,eq: gaussian q poly} at Taylor orders $N=0$ and $N=1$. Note the approximation gets better for smaller $\sigma/a$, mirroring the behavior of the focal expansion. The $N=1$ approximation performs reasonably across all plots, incurring significant errors mostly in scenarios where the exact result has non-trivial weight at the origin. The $N=0$ performs worse: note in particular it does not predict the correct position of the peak, with the discrepancy growing worse with larger $\sigma$.}
    \label{fig: gaussian radial profile approx}
\end{figure*}

We open our analysis of functions with non-compact support by considering the family of Gaussian rings, defined as the convolution of a delta function ring of radius $a$ and a Gaussian:
\begin{align}\label{eq: Gaussian f general}
    \begin{split}
    f(r,\theta)=&\int_{0}^{\infty}|x|\,\text{d}|x| \int_{0}^{2\pi}\text{d}\varphi\, \gamma(\varphi) \delta(|x|-a)\\
    \times&\frac{1}{\sigma \sqrt{2\pi}}e^{-\frac{1}{2 \sigma^2}\left[(|x|\cos\varphi-r\cos\theta)^2+(|x|\sin\varphi-r\sin\theta)^2\right]}
    \end{split}
\end{align}
with $\sigma$ the standard deviation of the Gaussian and $\gamma(\theta)$ an arbitrary angular dependence with Fourier series
\begin{align}
    \gamma(\theta)=\sum_{n\in \mathbb{Z}}\gamma_n e^{i n \theta}.
\end{align}
Thanks to the convolution theorem we can calculate the Fourier transform of $f(r,\theta)$ right away: it is simply the product of a Gaussian and the Fourier transform of $\gamma(\theta)\delta(r-a)$, the latter of which can be evaluated with \cref{eq: Anger-Jacobi}. The final expression is
\begin{align}\label{eq: gaussian general F[f]}
\begin{split}
    \mathcal{F}\left[f\right]&(\rho,\phi)\\
    =&(2\pi)^{3/2} a\sigma \,e^{-2(\pi\sigma\rho)^2} \sum_{n\in\mathbb{Z}}\gamma_n(-i)^ne^{i n \phi} J_n(2\pi a \rho)
\end{split}
\end{align}
so that with our earlier notation 
\begin{align}
\mathcal{J}_n(\rho)=\gamma_n\sqrt{2\pi} a \sigma e^{-2(\pi\sigma\rho)^2}J_n(2\pi a \rho).    
\end{align}

The integral in \cref{eq: Gaussian f general} can also be evaluated in closed form. One must use a complexified version of \cref{eq: Anger-Jacobi} and the fact that $(-i)^n J_n(i z)=I_n(z)$ where $I_n$ is the modified Bessel function of the first kind \cite{BesselWatson}. In the end we get
\begin{align}\label{eq: Gaussian f general evaluated}
    f(r,\theta)=\frac{a\sqrt{2\pi}}{\sigma} e^{-(a^2+r^2)/2\sigma^2}\sum_{n\in\mathbb{Z}}e^{in\theta}\gamma_n I_n\left(\frac{ar}{\sigma^2}\right)
\end{align}
and so the angular modes are
\begin{align}\label{eq: gaussian general gn}
    g_n(r)=\gamma_n\frac{a\sqrt{2\pi}}{\sigma} e^{-(a^2+r^2)/2\sigma^2}I_n\left(\frac{ar}{\sigma^2}\right).
\end{align}

Unlike the annular rings of the previous section, the focal expansion here is non-trivial. The problem is that the trivial extension of \cref{eq: gaussian general gn} to the full real line (obtained by simply making the functions vanish for $r<0$) does not have a closed form 1D Fourier transform (as far as we were able to verify). To stay in the spirit of our investigation we must then employ some clever approximation scheme that captures the central aspects of the functions $g_n(r)$ while also making their 1D Fourier transforms tractable. Since we do have the exact result \cref{eq: gaussian general F[f]} we can use the family of Gaussian rings to probe the validity of some of these more qualitative steps the focal expansion calls for. In particular, we will investigate the effects of adding non-zero weight in the $r<0$ tails and choosing non-optimal shift parameters.

We begin by noting that the functions $I_n(z)$ have the asymptotic behavior
\begin{align}\label{eq: In asymptotic}
    I_n(z)\big|_{z\gg 1}\approx \frac{e^{z}}{\sqrt{2 \pi z}}
\end{align}
for large real $z$. Now, one can easily plot $g_n(r)$ for small values of $\sigma$ and convince themselves that it appears to be shaped like a Gaussian peaked at $r=a$, which would be true exactly if $I_n$ was composed of \textit{just} the exponential factor in \cref{eq: In asymptotic}. This motivates a strategy: first separate out the exponential as in 
\begin{align}
    I_n(ar/\sigma^2)=e^{ar/\sigma^2}\left(I_n(ar/\sigma^2) e^{-ar/\sigma^2}\right)
\end{align}
and then Taylor expand the term in parenthesis up to some finite order $N$ around $r=a$. In the end we find that
\begin{align}\label{eq: gaussian gn general approx}
    g_n(r)\approx \gamma_n \frac{a\sqrt{2\pi}}{\sigma}e^{-(r-a)^2/2\sigma^2}q_{n,N}(r-a)
\end{align}
where $N$ is the order of the Taylor approximation and $q_{n,N}$ is the resulting polynomial
\begin{align}\label{eq: gaussian q poly}
    q_{n,N}(z)=\sum_{j=0}^N \frac{z^j}{j!}\left[\left(\frac{d}{dr}\right)^j I_n(ar/\sigma^2)e^{-ar/\sigma^2}\right]\Bigg|_{r=a}.
\end{align}
Note that one could in principle calculate the Taylor coefficients in \cref{eq: gaussian q poly}  analytically, using the recurrence relation (note the sign difference when compared to \cref{eq: Bessel recurrence d/dx})
\begin{align}
    2\partial I_n(z)=I_{n+1}(z)+I_{n-1}(z).
\end{align}
We omit further details since, as it turns out, $N=0,1$ provide serviceable approximations for a reasonable range of $\sigma$ values already -- see fig.~(\ref{fig: gaussian radial profile approx}).

We still need to extend $g_n(r)$ to the full real line in order to use the focal expansion and here we are faced with a choice. If one wishes to make no further approximations the first instinct would be to take $g_n(r)=0$ for $r<0$, the Fourier integral will involve the error function but still be analytically doable. On the other hand, the form of \cref{eq: gaussian gn general approx} makes it tempting to simply keep the Gaussian $r<0$ tail without cutting it off, especially for values of $\sigma$ for which we expect its effects to be small (compare the height of the peak to the values at the origin in fig.~(\ref{fig: gaussian radial profile approx})). We will compare both approaches and refer to these functions as $g^{\text{ext. 1}}_n$ in the tails-on case and $g^{\text{ext. 2}}_n$ in the tails-off case. For an example of a direct comparison between the two see fig.~(\ref{fig: remainder example}) in section~\ref{section: peak at origin} above. Starting with the former approach, the 1D Fourier transform of the extension is
\begin{align}\label{eq: gaussian gener gn ext 1}
    \widehat{g}^{\text{ ext. 1}}_n(\rho)=e^{-i 2\pi\rho a} 2\pi a q_{n,N}\left(\frac{i}{2\pi}\frac{d}{d\rho}\right)e^{-2(\pi\sigma\rho)^2},
\end{align}
the explicit factor of $e^{-i2\pi \rho a}$ implies the natural choice of the shift parameter is just $a$ (retroactively justifying our notation). Using the notation of section~\ref{subsec: summary} we then have
\begin{align}\label{eq: Gaussian Gn general tails}
    \widehat{G}^{1}_n(\rho)=2\pi a\, q_{n,N}\left(\frac{i}{2\pi}\frac{d}{d\rho}\right)e^{-2(\pi\sigma\rho)^2}.
\end{align}

If we remove the tails we still get the factor of $e^{-i2\pi a \rho}$ in $\widehat{g}^{\text{ext. 2}}_n$, meaning the shift parameter is once again naturally chosen to be just $a$, but the final expression is more complicated:
\begin{align}\label{eq: Gaussian Gn general no tails}
    \begin{split}
            \widehat{G}^{2}_n(\rho)=&\pi a\, q_{n,N}\left(\frac{i}{2\pi}\frac{d}{d\rho}\right)\Bigg[e^{-2(\pi\sigma\rho)^2}\\
            \times&\left(1-\text{Erf}(i \sqrt{2}\pi\sigma \rho-a/\sqrt{2}\sigma)\right)\Bigg]
    \end{split}
\end{align}
with Erf being the error function. Note that if we take the limit $\sigma/a\rightarrow 0$ in $\widehat{G}^{2}_n$ while keeping $\rho$ constant we recover the expression for $\widehat{G}^{1}_n$. The derivatives in \cref{eq: Gaussian Gn general no tails} can in principle be evaluated analytically by noting $\text{erf}\,'(z)=2 e^{-z^2}/\sqrt{\pi}$. 

Having found these $\widehat{G}^{1}_n$ and $\widehat{G}^{2}_n$, we can plug them into the focal expansion and compare it to the exact result \cref{eq: gaussian general F[f]}. One might expect the tails-off extension $\widehat{g}_n^{\text{ ext. 2}}$ to be preferable, since it does not introduce any unphysical weight at negative $r$. However, as we show below, the resulting non-smoothness at $r=0$ can dominate the large-$\rho$ behavior and spoil the asymptotics. In practice we find that the smooth tails-on extension $\widehat{g}_n^{\text{ ext. 1}}$, while approximate, better preserves the correct large-$\rho$ decay in the parameter regime of interest.

\subsubsection{Example 1: Choosing Tails}

\begin{figure}
    \centering
    \includegraphics[width=\linewidth]{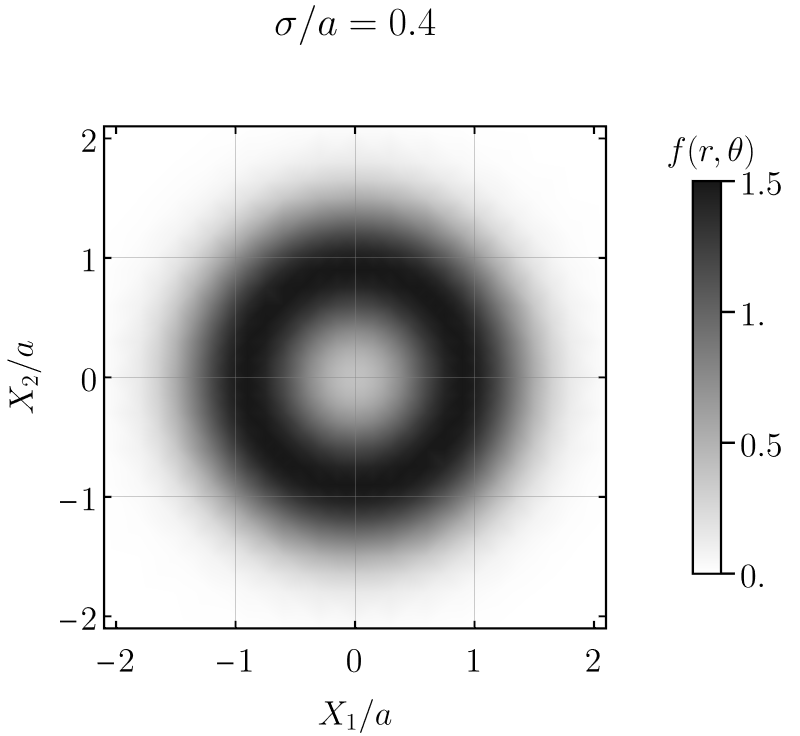}
    \caption{Density plot of an example Gaussian ring intensity with trivial angular dependence, the function $f(r,\theta)$ being given by \cref{eq: Gaussian f general evaluated} with $\sigma/a=0.4$, $\gamma_0=1$, and all other $\gamma_n$ vanishing.}
    \label{fig: gaussian density 1}
\end{figure}

\begin{figure*}
    \centering
    \includegraphics[width=\linewidth]{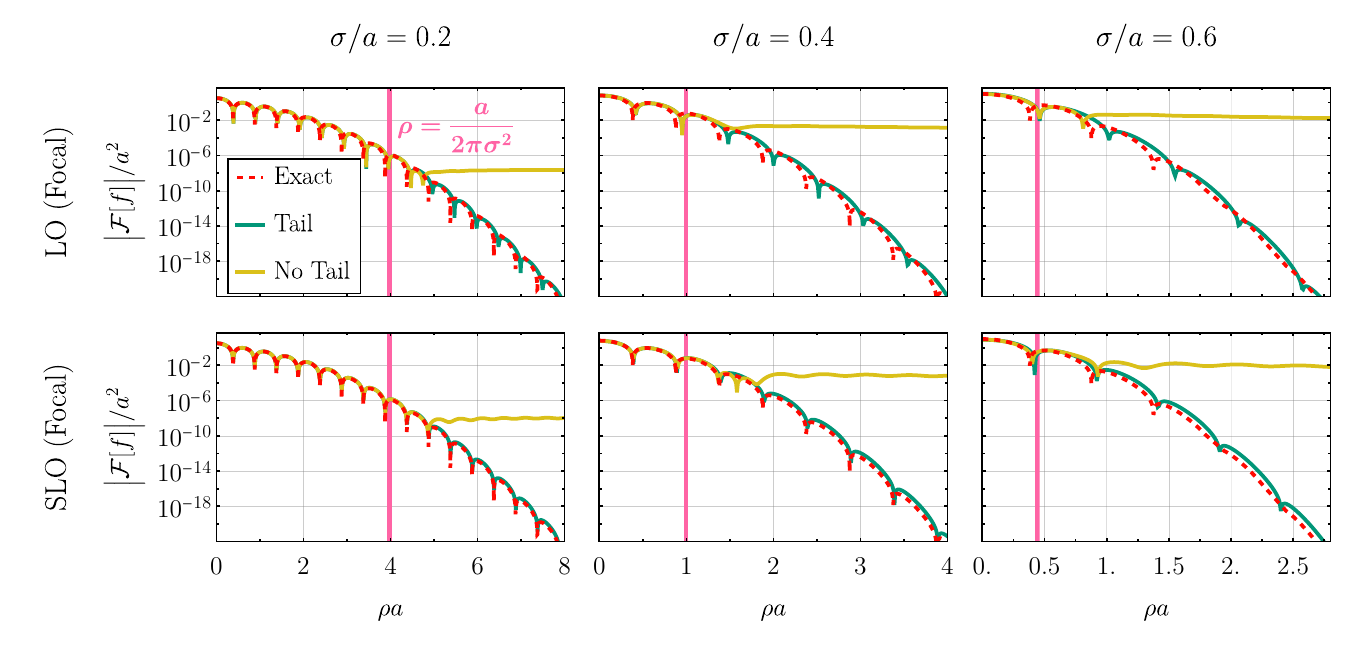}
    \caption{The comparison of the tails-on and tails-off versions of the focal expansion to the exact Fourier transform of a spherically symmetric Gaussian ring,
    with three representative standard deviations plotted (for an example density plot see fig.~(\ref{fig: gaussian density 1})). The tails-on and tails-off expressions are given by \cref{eq: Gaussian Gn general tails} and \cref{eq: Gaussian Gn general no tails} respectively with $N=1$. 
    Note all vertical axes are logarithmic and all horizontal ones linear. All plotted functions decay rapidly, hence the relatively small range of $\rho$ values. The first row contains result for the leading order focal expansion, while the second contains result for the subleading order focal expansion. In addition, vertical purple lines demarcate the point $\rho=a/2\pi \sigma^2$ at which we theoretically expect the tails-off approach to break down, we see these are slightly premature but generally correct. Once the tails-off approach breaks it decays like $1/\rho$ -- this appears nearly horizontal when compared to the exponential decay of the other functions (and, adding to the visual confusion in the case of $\sigma/a=0.2$, even undergoes a period of gentle increase before the $1/\rho$ decay kicks in). The tails-on approach does not suffer the same problems and continues to work well over the range plotted, especially for SLO and smaller $\sigma$ values.
    }
    \label{fig: Gaussian Tail-No Tail comparison}
\end{figure*}
\begin{figure*}
    \centering
    \includegraphics[width=\linewidth]{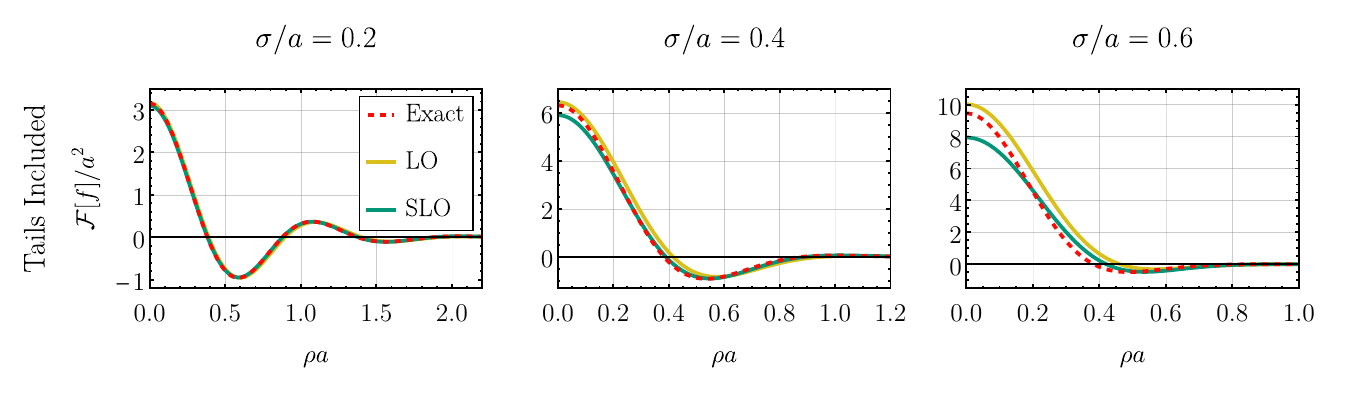}
    \caption{
    The comparison of the tails-on focal expansion at LO and SLO to the exact results for Gaussian rings with $\sigma/a=0.2, 0.4$, and $0.6$. The data is identical to fig.~(\ref{fig: Gaussian Tail-No Tail comparison}), with the only difference being the vertical axis is linear, not logarithmic, for viewing convenience. Note that for the largest $\sigma$ the change from LO to SLO leads to dubious improvements. This was to be expected -- for large $\sigma$ the errors from keeping the tails on become significant and they have to be accounted for manually, adding subleading focal order will not help. 
    }
    \label{fig: Gaussian Tails on no Log}
\end{figure*}

The issue becomes clear once we investigate \cref{eq: Gaussian Gn general no tails} in the large $\rho$ limit. We will focus on the case of $n=0$ as this is where the problems we are about to encounter are the most pronounced, this corresponds to a spherically symmetric (so the only non-vanishing $\gamma_n$ is $\gamma_0=1$) Gaussian-blurred ring -- see figure~(\ref{fig: gaussian density 1}) for an example with $\sigma/a=0.4$.

While we do recover \cref{eq: Gaussian Gn general tails} from \cref{eq: Gaussian Gn general no tails} if we take $\sigma/a\rightarrow 0$ and keep $\rho$ constant, the behavior is markedly different if these roles are reversed, with $\sigma$ constant and $\rho\rightarrow \infty$. One can show that for $|\text{Arg}z|<3\pi/4$ the leading asymptotic of the error function as $|z|\rightarrow\infty$ is \cite{abramowitz1965handbook} 
\begin{align}
    \text{Erf}z\big|_{|z|\gg 1}\approx 1-\frac{1}{z \sqrt{\pi}}e^{-z^2}.
\end{align}
Using this in \cref{eq: Gaussian Gn general no tails} we see that for large enough $\rho$ we actually \textit{lose} exponential suppression and instead get
\begin{align} \label{eq: gaussian G2 asymptotic}
   \widehat{G}_0^2(\rho)\big|_{\rho \gg a/2\pi \sigma^2}\approx \frac{a \,q_{0,N}(-a)}{i \sigma \rho \sqrt{2\pi}}e^{-a^2/2\sigma^2}e^{2\pi i\rho a}
\end{align}
This is a big problem as from the exact expression \cref{eq: gaussian general F[f]} for the Fourier transform we see the leading behavior must be given by exponentially suppressed Bessel functions. Instead, if we use the focal expansion with \cref{eq: Gaussian Gn general no tails} we will see the appearance of exponential suppression until around  $\rho\approx a/2\pi\sigma^2$, at which point the behavior will switch to the dramatically slower $1/\rho$ decay.

This does not, however, mean that the focal expansion fails to deliver on its promises -- we had simply used the wrong extension for $g_n(r)$. Looking at \cref{eq: Gaussian Gn general tails} we see that, had we chosen to keep the tails, we would have retained the crucial exponential decay. In fig.~(\ref{fig: Gaussian Tail-No Tail comparison}) we compare the exact expression against the focal expansion where we kept the $r<0$ tails on the radial profiles  and a focal expansion with the tails cut off. We indeed see the tails-on expression performs remarkably well (especially at SLO) while the tails-off expressions breaks down as predicted. This behavior goes against the naive expectation that the tails-off version is more correct on account of not having to use another approximation to obtain the function $G^2_n(r)$.

The major hint for what is going wrong comes from the structure of the asymptotic formula \cref{eq: gaussian G2 asymptotic}: first, note that $q_{0,\infty}(-a)=I_0(0)=1$ by \cref{eq: gaussian q poly}, and second, note the explicit phase $e^{2\pi i \rho a}$ nullifies the arguments we had used to choose the shift parameter to be $a$. The former already guides us to consider the point $r=0$, while the latter is more apparent if we use \cref{eq: gaussian G2 asymptotic} to write the asymptotic for $\widehat{g}_0^{\text{ext. 2}}$
\begin{align} \label{eq: Gaussian gn asymp}
   \widehat{g}_0^{\text{ext. 2}}(\rho)\big|_{\rho \gg a/2\pi \sigma^2}\approx \frac{a\, e^{-a^2/2\sigma^2}}{i \sigma \rho \sqrt{2\pi}}.
\end{align}
There is no explicit oscillatory phase in \cref{eq: Gaussian gn asymp} -- were we to take this at face value we would conclude the shift parameter vanishes, invalidating the focal expansion. Both of these issues point to $r=0$ as the culprit.

With this the reason for our discrepancy is obvious in retrospect. By cutting off the tails we have introduced a totally artificial discontinuity into the problem and discontinuities in position space\footnote{Either literal discontinuities of the function or those of its higher order derivatives. For instance, only the $n=0$ mode will have a discontinuity in the radial profile at the origin if we cut off the tails, for arbitrary $n$ the radial profile \cref{eq: gaussian general gn} will instead have discontinuities in its $n$th derivative.} lead to slower asymptotic decays in Fourier space. Whatever benefit we incurred from doing the 1D Fourier transforms of the radial profiles exactly is totally overpowered at large $\rho$ by this unnatural choice. It is still possible that summing all focal orders somehow resumms to the correct asymptotic behavior, but that approach is certainly no longer natural to the problem at hand.
Keeping all this in mind, the focal expansion still does a satisfactory job, as can be seen from fig.~(\ref{fig: Gaussian Tail-No Tail comparison}) and fig.~(\ref{fig: Gaussian Tails on no Log}), in the latter of which we plotted the tails-on case separately without logarithmic vertical axes to make comparing the functions easier.

To summarize: \textit{when extending the radial profiles $g_n(r)$ to the full real line one must not introduce artificial discontinuities and keep the extension at least as smooth as the parent function $f(r,\theta)$ if one wishes to get the correct asymptotic behavior out of the focal expansion.} Of course, in this particular case the issue is somewhat pedantic as, due to Gaussian fall-offs, these discrepancies appear at orders so small as to be mostly immaterial unless $\sigma/a$ is large (see fig.~(\ref{fig: Gaussian Tail-No Tail comparison})), at which point one should not be using the bare focal expansion anyway and instead first remove the weight at the origin as described in section~\ref{section: peak at origin}.

\subsubsection{Example 2: Non-Optimal Shift Parameter}

\begin{figure*}
    \centering
    \includegraphics[width=\linewidth]{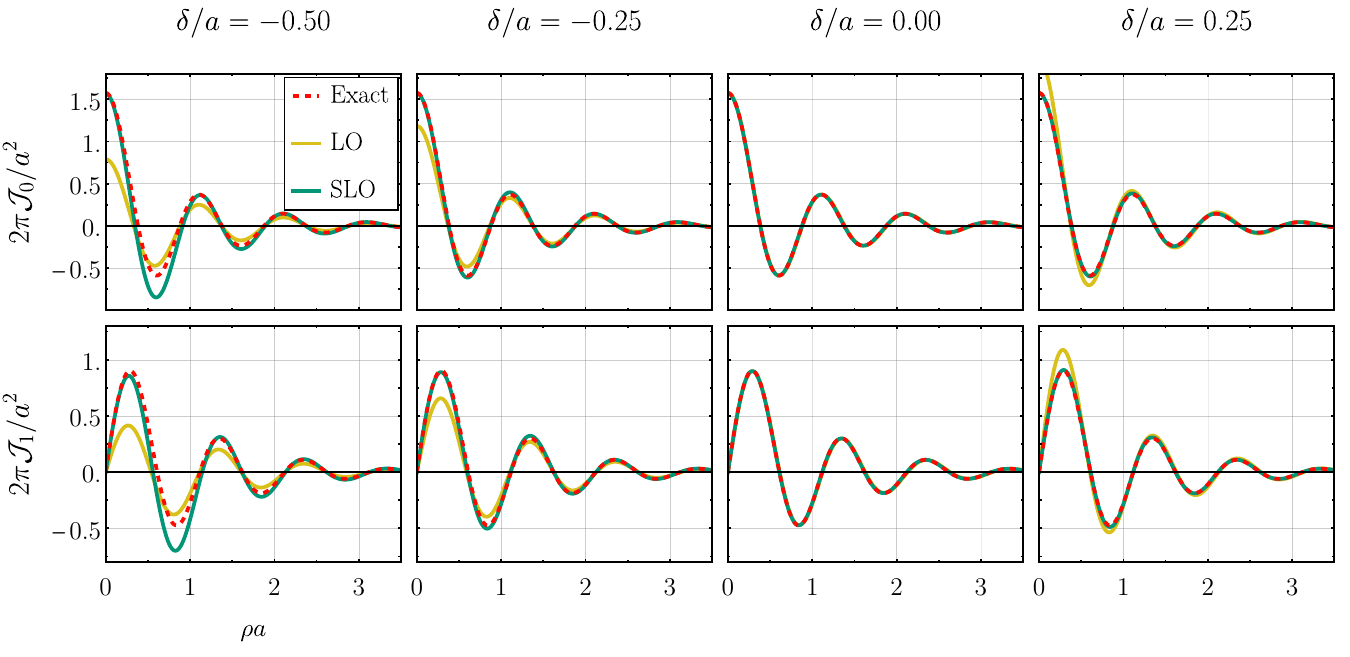}
    \caption{Examples of the behavior of the focal expansion when a non-optimal shift parameter is chosen -- in this case $a+\delta$. The first two angular mode integrals $\mathcal{J}_0,\mathcal{J}_1$ for a Gaussian ring with  $\sigma/a=0.1$ are plotted. Note that changing the shift parameter appears to have no effect on the phase or frequency, just the speed of convergence of the amplitude of the focal expansion.}
    \label{fig: Gaussian different a comp}
\end{figure*}
\begin{figure}
    \centering
    \includegraphics[width=\linewidth]{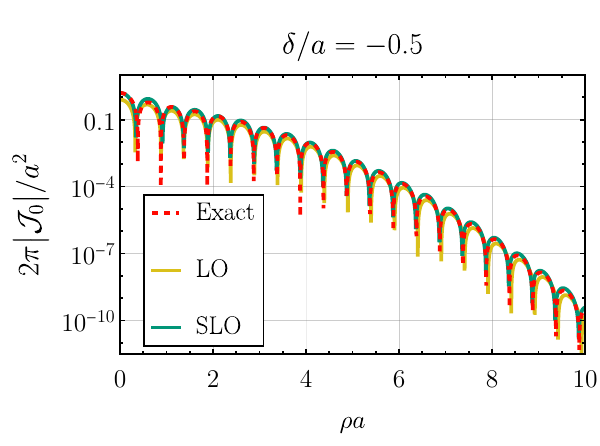}
    \caption{The data from the first plot of the first row of fig.~(\ref{fig: Gaussian different a comp}), chosen for the relatively large discrepancy between focal and exact results, plotted with the vertical axis logarithmic and over a larger range of arguments to make the frequency and phase behavior more apparent. Note we see no evidence of the focal expansion providing the incorrect phase or frequency.}
    \label{fig: Gaussian non optimal a log plot}
\end{figure}

There is a curiosity of the Gaussian ring we are yet to comment on. Note that the actual peak of the ring $f(r,\theta)$ is \textit{not} positioned at $r=a$. This is easiest to see from  fig.~(\ref{fig: gaussian radial profile approx}) where the yellow $N=0$ lines represent unmodulated Gaussians with centers at $r=a$ (see \cref{eq: gaussian gn general approx}). We see the unmodulated Gaussian peaks are positioned at slightly too large radii, while the $N=1$ expressions (green) are generally more faithful to the exact result.

The intuitive reason for this phenomena is simple: when convolving a Gaussian with a delta function circle there will be more overlap between the Gaussian tails facing inward as compared to those facing outwards from the circle due to the spherical geometry, and so we will see more weight distributed slightly closer to the origin. However, the frequencies in Fourier transform \cref{eq: gaussian general F[f]} are unchanged, depending only on the original radius $a$ (as is trivial from the Fourier convolution theorem). The focal expansion too picks this uncorrected radius (see the explicit oscillatory phase factor in \cref{eq: gaussian gener gn ext 1}). But here is where one might be worried: our prescription for the focal expansion in section~\ref{subsec: summary} instruct us to pick the shift parameter as close to the peak as possible, what then if we had picked the corrected radius instead of $a$? One might rightfully worry this would lead to discrepancies in the frequencies and as a result to the focal expansion drifting out of phase with the exact result for large $\rho$.

We did argue on theoretical grounds why this is \textit{not} a problem at the end of section~\ref{sec: Bessel function Kernels} (in short: picking a different shift parameter not only changes the frequencies of the Bessels but adds an extra oscillatory factor to the $\widehat{G}_n$'s and these two effects nullify each other at large arguments). It is worth seeing this play out in a realistic scenario, however, and because of their radius-frequency discrepancy the Gaussian rings are a good opportunity to do so. 

Assume one changes the shift parameter by $\delta$ so that it is now $a+\delta$. The effects of this change on the focal expansion of the angular mode integrals $\mathcal{J}_0,\mathcal{J}_1$ for a few values of $\delta/a$ and fixed $\sigma/a=0.1$ are shown below in fig.~(\ref{fig: Gaussian different a comp}). In addition, we chose the poorest performing case for $\mathcal{J}_0$ and plotted it in fig.~(\ref{fig: Gaussian non optimal a log plot}) with the vertical axis logarithmic to make discerning the frequency at larger arguments easier. As advertised, there appears to be no issues with frequency in any of the plots, only the amplitude has more difficulty converging to the exact result as we move away from $\delta=0$. Note in particular the SLO improving upon the performance of LO, showing that imperfect choice for the shift parameter can be overcome with higher focal orders. 
If one continues to look at more negative values of $\delta$ the focal expansion becomes significantly worse the closer we get to $\delta/a=-1$. This is because at that point the new shift parameter is zero and as a result the focal expansion is ill defined\footnote{It does seem, however, to work in the neighborhood of that point. It would be interesting to see if this can be used to glean any insight into the convergence properties of the focal expansion -- we leave this for future work.}.

\begin{figure}
    \centering
    \includegraphics[width=\linewidth]{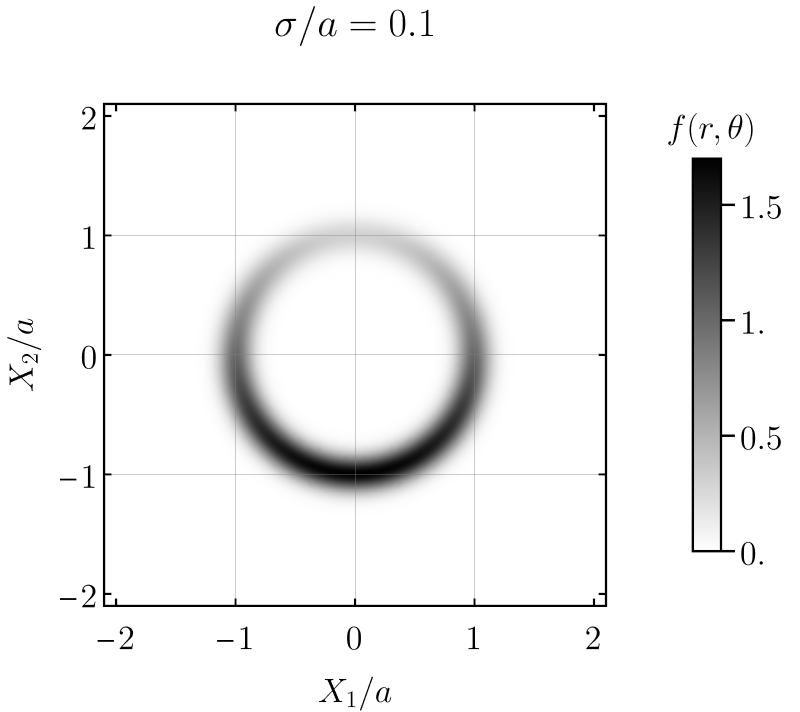}
    \caption{Example of a Gaussian ring with non-trivial angular behavior, given by \cref{eq: Gaussian example non trivial ang} with $\sigma/a=0.1$. The rescaled real and imaginary parts of the Fourier response are precisely the functions plotted in the first and second rows respectively of fig.~(\ref{fig: Gaussian different a comp}).}
    \label{fig: Gaussian density 2 ang behavior}
\end{figure}

Finally, for completeness we again note that once we have the individual angular mode integrals $\mathcal{J}_n$ one can easily sum their contributions up to get non-trivial angular behavior. For instance, consider the simple angular dependence given by
\begin{align}\label{eq: Gaussian example non trivial ang}
    f(r,\theta)=g_0(r)-0.7 g_1(r)\sin\theta
\end{align}
where $g_0(r), g_1(r)$ are given by \cref{eq: gaussian general gn} with $\sigma/a=0.1$ and whose density plot is shown in fig.~(\ref{fig: Gaussian density 2 ang behavior}).
The Fourier transform is then given by
\begin{align}
    \mathcal{F}[f](\rho,\phi)=2\pi\left[\mathcal{J}_0(\rho)+0.7 i \sin \phi \mathcal{J}_1(\rho)\right]
\end{align}
and so we see the real part is given precisely by the rescaled first row of fig.~(\ref{fig: Gaussian different a comp}) while the imaginary part is given by the rescaled second row of the same figure, with the rescaling now depending on the angle through a sine.

\subsection{Logarithmic Rings \label{sec: log ring}}

\subsubsection{Motivation}
We will now deal with a family of examples that were already highlighted as our representative demonstration in section~\ref{sec: demonstration}: rings with logarithmic divergences. In contrast to the annulus and Gaussian rings, the log-divergence rings' Fourier transform cannot be calculated analytically, at least to the best of our knowledge\footnote{It is possible there could exist a formulation of the relevant integrals using some exotic family of special functions but we don't investigate this possibility further -- at some point we must limit what special functions we do and do not allow ourselves to use, lest we slip into taxonomy.}. Before we begin, it is worthwhile to restate our physical motivation: one can show \cite{Logrings} \cite{PhotonRingReview} that near the photon ring of a black hole (assuming the accreting material is emissive, optically thin, and geometrically thick) the light intensity will appear to diverge logarithmically near a critical radius as in
\begin{align}\label{eq: f log near critical approx}
    f(r,\theta)\propto \log \left(\frac{c_{\pm}}{|r-a|}\right)
\end{align}
where the constant $c_{+}$ or $c_-$ depends on whether we approach the critical radius from the outside or the inside respectively. In reality this divergence will be eventually truncated by the non-zero absorption rate of the ambient matter, but we nevertheless expect it to provide a serviceable approximation for practical purposes. 

The question now is, how should we extend \cref{eq: f log near critical approx} to a full 2D function whose Fourier transform is well behaved? In particular, we must ensure the large $r$ tails of our function don't lead to divergences. We will consider two approaches. The first exhibits slowly decaying tails and is not motivated by realism but by pushing the focal expansion to its breaking point -- the presence of a divergence is already enough cause for concern in that regard, so we might as well see what else we can get away with. The second example will, in contrast, exhibit exponentially suppressed tails, rendering the analysis easier and somewhat more realistic. The first example will focus on divergences (approximately) even around the critical radius while in the second we will relax this and consider divergences with arbitrary degree of asymmetry.

\subsubsection{Example 1: Symmetric Logarithmic Divergence \label{sec: log rings against all odds}}

\begin{figure*}
    \centering
    \includegraphics[width=\linewidth]{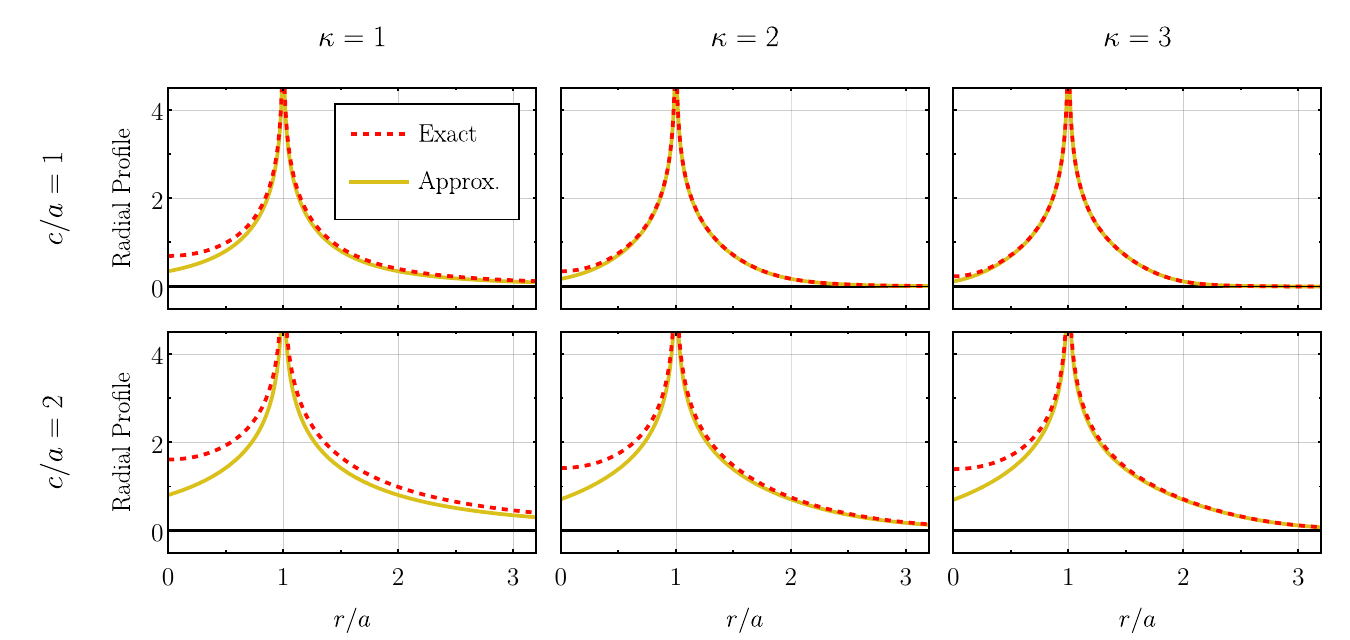}
    \caption{The radial profiles of several examples of logarithmic-divergence rings, with both the exact profile and the approximate extension to the full real line plotted. The exact and approximate expressions are given by \cref{eq: f log even} and  \cref{eq: gn log example 1 approx} respectively. The columns represent the three values $\kappa=1,2,3$ while the rows differ in the width parameter $c/a=1,2$. Note that for $\kappa=3$ and $c/a=1$ we already begin to see the limit \cref{eq: f log p inf limit} set in.}
    \label{fig: log example 1 radial profiles}
\end{figure*}

For this family of examples we will assume the coefficients $c_{\pm}$ in \cref{eq: f log near critical approx} are equal, so that the log divergence is symmetric, and choose to extend the function as\footnote{One could consider a similar family but with the log divergence being \textit{odd} around the critical radius and then consider superpositions of the odd and even cases to get asymmetric divergences, but that complicates the analytic calculations on the focal side. For instance, the 1D Fourier transform of $\text{sgn}(x)\log(1+1/x^2)$ needed for the focal expansion would require the introduction of the obscure family of Meijer G-functions, see \cite{MeijerGintro} \cite{MeijerGdetails}. It isn't all hopeless, by matching to the small and large asymptotics \cite{MeijerGasymptotics} of the particular G-function of interest we were able to approximate it to great accuracy by the combination 
$$\log\left[1+\frac{\omega^2}{\alpha}\right]\frac{1}{\omega}+\log\left[1+\frac{\alpha}{\omega^2}\right]\frac{\omega}{2}+\beta \omega e^{-\omega}$$
with $\alpha, \beta$ appropriately chosen constants and $\omega=2\pi\rho$. We don't include the details -- these extra difficulties obscure the already unclear behavior of the focal expansion we wish to showcase.}
\begin{align}\label{eq: f log even}
    \begin{split}
        f&(r,\theta)\equiv\\
        &\frac{\gamma(\theta)}{2\kappa}\left\{\log\left[1+\left(\frac{c}{r-a}\right)^{2\kappa}\right]+\log\left[1+\left(\frac{c}{r+a}\right)^{2\kappa}\right]\right\}
    \end{split}
\end{align}
where $\kappa\in \mathbb{N}-\{0\}$, $c>0$ are arbitrary constants and we include non-trivial angular dependence using
\begin{align}\label{eq: log example ang dependence}
    \gamma(\theta)=\sum_{n\in\mathbb{Z}}\gamma_n e^{in\theta}
\end{align}
as a multiplicative factor. Note the factor of $1/2\kappa$ outside the curly brackets ensures the proportionality constant in \cref{eq: f log near critical approx} is independent of $\kappa$. We chose $(r-a)^2$ instead of $|r-a|$ since the former can be extended to a holomorphic function, permitting the use of complex analysis in what follows. The second term in \cref{eq: f log even}, equal to the first reflected across the origin, lets us avoid introducing artificial asymptotics in numerical calculations. Not having said term would render the full image $f(r,\theta)$ discontinuous in its first derivative at the origin, as we're essentially cutting off the tail of the function at a point where its derivative is non-zero. As we know from the previous section, this can lead to asymptotics being overwhelmed by the artificial effects of the discontinuity. 

With these assumptions the $n$-th angular mode $g_n(r)$ is simply given by the sum inside the curly brackets of \cref{eq: f log even} rescaled by $\gamma_n$. Note the parameter $c$ describes a sense of width of the peak, see fig.~(\ref{fig: log example 1 radial profiles}) for examples. The second term in \cref{eq: f log even} is a small correction as far as the logarithmic peak is concerned and as such we won't worry about it when dealing with the focal expansion. We shift to doing so now as, again, there seems to be no analytic way of performing the integral of $r g_n(r)$ against $J_n$ -- one must resort to numerics for comparison.

As in the previous examples we begin by extending the angular mode profiles to the full real line. We avoid cutting off the tails and simply set
\begin{align}\label{eq: gn log example 1 approx}
    &g^{\text{ext.}}_n(r)\equiv\frac{\gamma_n}{2\kappa} \log\left[1+\left(\frac{c}{r-a}\right)^{2\kappa}\right]
\end{align}
as our approximate extensions. For a few examples comparing the exact radial profiles to this approximation see fig.~(\ref{fig: log example 1 radial profiles}).

To calculate the 1D Fourier transform of \cref{eq: gn log example 1 approx} we first integrate by parts to get\footnote{The dashed integral indicates the Cauchy principal value is to be taken around $z=0$.}
\begin{align}
\mathcal{F}\left[g_n\right](\rho)=e^{-2\pi i \rho a}\frac{i\gamma_n }{2\pi \rho }\,\dashint_{-\infty}^{\infty}\frac{e^{-2\pi i \rho c z}}{z (z^{2\kappa}+1)}\,\text{d}z.
\end{align}
Note the explicit phase factor indicating we should chose, unsurprisingly, the critical radius $a$ as the shift parameter. Performing the remaining integral via contour integration we obtain the function $\widehat{G}_n$ to be used in the focal expansion:
\begin{align}\label{eq: Gn log example 1}
    \widehat{G}_n(\rho)=\frac{\gamma_n}{2|\rho|}\left[1-\frac{1}{\kappa}\sum_{k=1}^{\kappa}e^{-2\pi c |\rho| \alpha_k}\cos\left(2\pi c|\rho|\beta_k\right)\right]
\end{align}
where we defined the auxiliary constants
\begin{align}
    \alpha_k&\equiv \sin\left(\frac{2k-1}{2\kappa}\pi\right),\quad \beta_k\equiv \cos\left(\frac{2k-1}{2\kappa}\pi\right).
\end{align}
As an example, for $\kappa=1,2$ we have
\begin{align}
        \widehat{G}_n\big|_{\kappa=1}(\rho)&=\frac{\gamma_n}{2|\rho|}\left[1-e^{-2\pi c |\rho| }\right],\label{eq: Gn log p=1}\\
        \widehat{G}_n\big|_{\kappa=2}(\rho)&=\frac{\gamma_n}{2|\rho|}\left[1- e^{-\sqrt{2}\pi c |\rho| }\cos\left(\sqrt{2}\pi c|\rho|\right)\right].\label{eq: Gn log p=2}
\end{align}
In particular, note that for any value of $\kappa$ the functions $\widehat{G}_n$ will have asymptotic tails decaying like $1/\rho$ which, after being multiplied by Bessel functions in the focal expansion, leads to asymptotic decays of $1/\rho^{3/2}$. The deviations from this asymptotic, universal regime take the form of a linear superposition of exponentially decaying and oscillatory terms, with different decay rates and frequencies.
We thus suspect the $1/\rho^{3/2}$ asymptotic decay is a universal signature of a log divergence peak and not of some other facet of our function \cref{eq: f log even} -- this statement will also be supported by the analysis in the next section. Note also that, strangely, this leading behavior does not depend at all on the width of the log divergence $c$ or the parameter $\kappa$ -- it is completely universal. The information about $c$ and $\kappa$ is located strictly near the origin and decays exponentially away from it. This is unfortunate, as we have seen the decay rate of $1/\rho^{3/2}$ can also be produced by, say, the annulus rings. So while investigations of the asymptotic decay rate can narrow down what type of ring we are dealing with, we manifestly see they cannot be the end all be all of any thorough analysis.

As a side note, the $\kappa\rightarrow \infty$ limit of the above calculations is well defined. One can easily see that in this limit and for $c<a$ the image function \cref{eq: f log even} becomes
\begin{align}\label{eq: f log p inf limit}
    \lim_{\kappa\rightarrow\infty} f(r,\theta)=\gamma(\theta)\Pi\left(\frac{r-a}{2c}\right)\log \left| \frac{c}{r-a}\right|
\end{align}
where $\Pi$ is a step function as in section~\ref{sec: annulus rings}. Fourier transforming the angular modes of \cref{eq: f log p inf limit} one will be led to 
\begin{align}\label{eq: log limit Ghat}
    \widehat{G}_n(\rho)=\gamma_n \frac{\text{Si}(2\pi\rho c)}{\pi \rho}
\end{align}
where $\text{Si}$ is the well studied special function
\begin{align}\label{eq: Si int def}
    \text{Si}(z)\equiv \int_0^z\frac{\sin t}{t}\,\text{d}t.
\end{align}
Alternatively, approaching the problem directly as the continuous limit of the sum in \cref{eq: Gn log example 1} one recovers the same answer but through the alternative integral representation \cite{abramowitz1965handbook}
\begin{align} \label{eq: Si alternative representation}
    \text{Si}(z)=\frac{\pi}{2}-\frac{1}{2}\int_0^{\pi}e^{-z \sin t}\cos(z\cos t)\,\text{d}t
\end{align}
and so we can view the above analysis as a derivation of \cref{eq: Si alternative representation} assuming \cref{eq: Si int def}.

There is a significant problem in using any of the functions given by \cref{eq: Gn log example 1} in the focal expansion: the absolute values. The first sign of trouble appears when we attempt to take derivatives of \cref{eq: Gn log example 1} to calculate its foci, the absolute values then lead to discontinuities.  It need not happen instantly, but once it does all successive derivatives will contain a delta function (if we are to dutifully follow the theory of distributions). For instance, \cref{eq: Gn log p=1} has a kink at the origin leading to a discontinuity in the first derivative there, while for \cref{eq: Gn log p=2} these problems are staggered by two derivatives, with the kink first appearing at the second derivative. The first kink appearance will get pushed to higher and higher derivative orders until, in the $\kappa\rightarrow \infty$ limit, we get the smooth function \cref{eq: log limit Ghat}. These kinks are directly a result of the large $r$ tails of \cref{eq: gn log example 1 approx}, which decay \textit{slightly} faster than $1/r^{2\kappa}$. Derivatives of the Fourier transform correspond to multiplying the original function by powers of the argument so we see the discontinuities come precisely from the Fourier integrals becoming distributional.

One might harbor hope that, maybe, we can simply ignore the distributional components of the foci -- they do not appear to generically cancel but perhaps they are simply an artifact of an overzealous application of distribution theory. The hope here is that the underlying Taylor-series matching we used to derive the focal expansion is still obeyed, despite the appearances. Alas, that too appears to fail. The entire machinery of focal operators relies, as we have derived it, on splitting the functions into even and odd \textit{power series} of the argument, the first containing only odd and the second only even powers. But here we have the expansion of \cref{eq: Gn log p=1}
\begin{align}\label{eq: Gn p=1 expansion}
    \begin{split}
        \left(\widehat{G}_n\big|_{\kappa=1}\right)&^+(\rho)=\widehat{G}_n\big|_{\kappa=1}(\rho)\\
        =&1-\frac{|\rho| }{2}+\frac{|\rho| ^2}{6}-\frac{|\rho| ^3}{24}+\frac{|\rho| ^4}{120}-\frac{|\rho| ^5}{720}+O\left(|\rho| ^6\right)
    \end{split}
\end{align}
which in no sense is an even power series, even after restricting to $\rho>0$, so taking even foci of it should not be well behaved. The series above also instantly discourages another attempt to salvage the focal expansion: removing the absolute values and treating \cref{eq: Gn p=1 expansion} like an actual power series, one might then try to actually split the even and odd powers and treat those as the even and odd parts of $\widehat{G}_n$ to be used in the focal expansion. The problem is these will be divergent at large $\rho$, for the same reason $\cosh \rho$, $\sinh \rho$ are divergent despite their difference being $e^{-\rho}$.

\begin{figure}
    \centering
    \includegraphics[width=\linewidth]{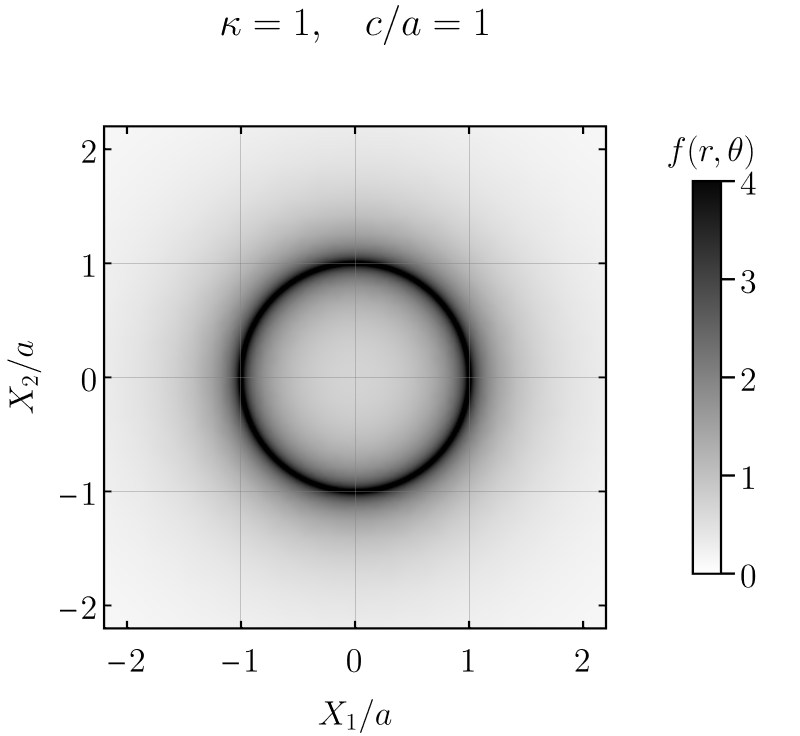}
    \caption{A density plot of an example logarithmic ring  with trivial angular dependence, with $f(r,\theta)$ given by \cref{eq: f log even} with $\kappa=1$, $c=a$,$\gamma_0=1$, and all other $\gamma_n$ vanishing. All function values above $f(r,\theta)=4$ are shown in black.}
    \label{fig: log example 1 density image}
\end{figure}

\begin{figure*}
    \centering
    \includegraphics[width=\linewidth]{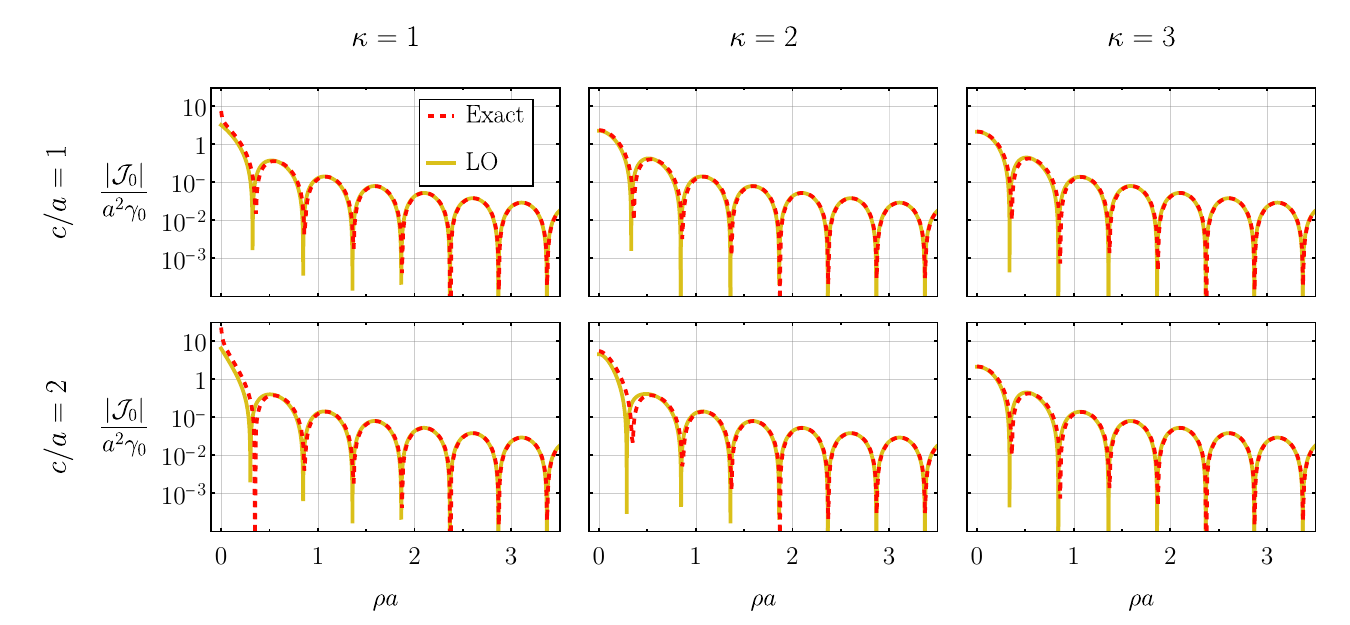}
    \caption{The Fourier transform of several symmetric log-divergence rings calculated numerically and via the focal expansion. The numerical results were calculated from the $n=0$ angular mode of \cref{eq: f log even} while the leading focal expansion was computed using \cref{eq: log example 1 leading focal}, both for several values of the log-peak parameters $\kappa, c$. Notice the vertical axes are logarithmic. We see broad agreement between the focal approximation and numerics save for a small range of arguments near the origin. This is an error not strictly of the focal expansion but of extending the log peak to the full real line -- see fig.~(\ref{fig: log example 1 half vs full}) and the surrounding discussion.}
    \label{fig: log example 1 grand comparison}
\end{figure*}
\begin{figure}
    \centering
    \includegraphics[width=\linewidth]{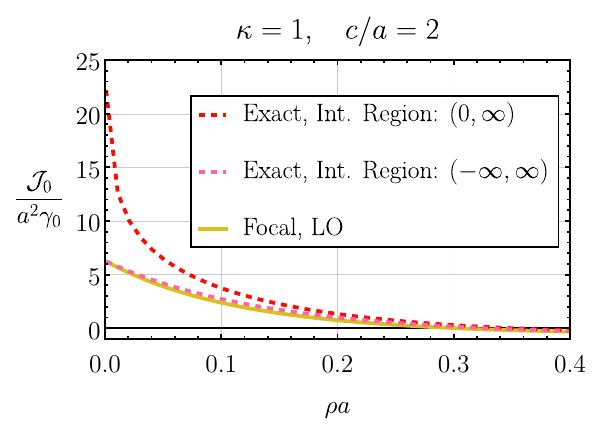}
    \caption{The $n=0$ angular mode of the Fourier response of the log peak (\cref{eq: f log even}) with $\kappa=1$ and $c=2$ (the first case from the left in the bottom row of fig.~(\ref{fig: log example 1 grand comparison})). Three functions were plotted: the focal expansion at leading order (yellow), the numerics for the integral \cref{eq: log example 1 half integral} with integration limits $0$ and $\infty$ (red, dashed), and finally the numerics for the integral \cref{eq: log example 1 full integral} over the full real line. The difference between the latter two integrals is precisely the step where we extend the peak function to the full real line, we see that this step accounts for the majority of the apparent error of the focal expansion. As such, for the remainder of this section we will investigate the comparison of the focal expansion to the full $\mathbb{R}$ integral.}
    \label{fig: log example 1 half vs full}
\end{figure}

\begin{figure*}
    \centering
    \includegraphics[width=\linewidth]{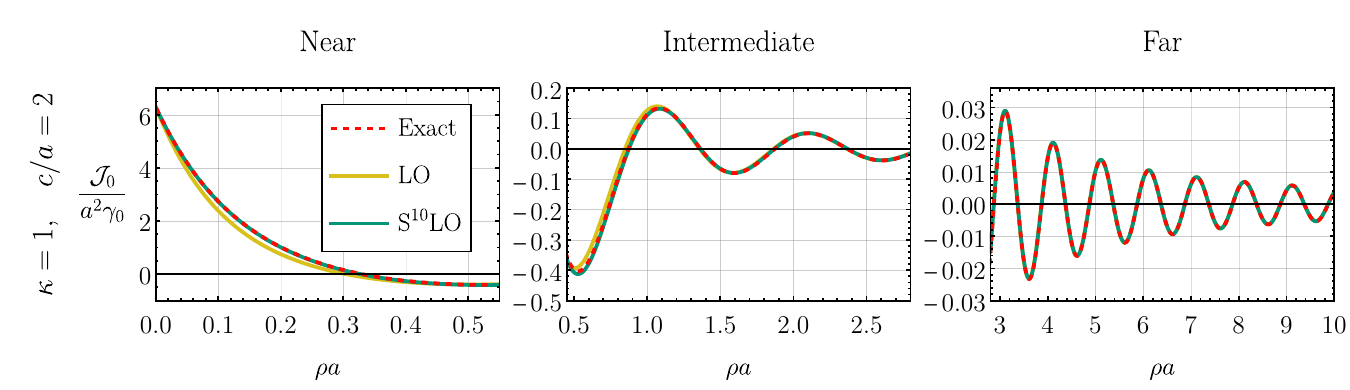}
    \caption{The comparison of the numeric result (dashed red) to the focal expansion (yellow, green) for the integral $\int_{-\infty}^{\infty}\text{d}r\, r J_0(2\pi\rho r) g^{\text{ext.}}_0(r)$ with a log peak given by $\kappa=1, c=2a$ (plotted in fig.~(\ref{fig: log example 1 half vs full})). Two orders of the focal expansion were plotted: leading (LO) and tenth subleading (S$^{10}$LO) orders. This particular higher focal order was picked as it is visually almost indistinguishable from the numerics. One should note, however, that the convergence is relatively slow due to the large width-to-radius ratio of 2. We see excellent agreement between the numerics and the focal expansion across all three $\rho$ regimes plotted -- even for just the leading focal order.}
    \label{fig: log example 1 p=1 c=2 comparison}
\end{figure*}
\begin{figure*}
    \centering
    \includegraphics[width=\linewidth]{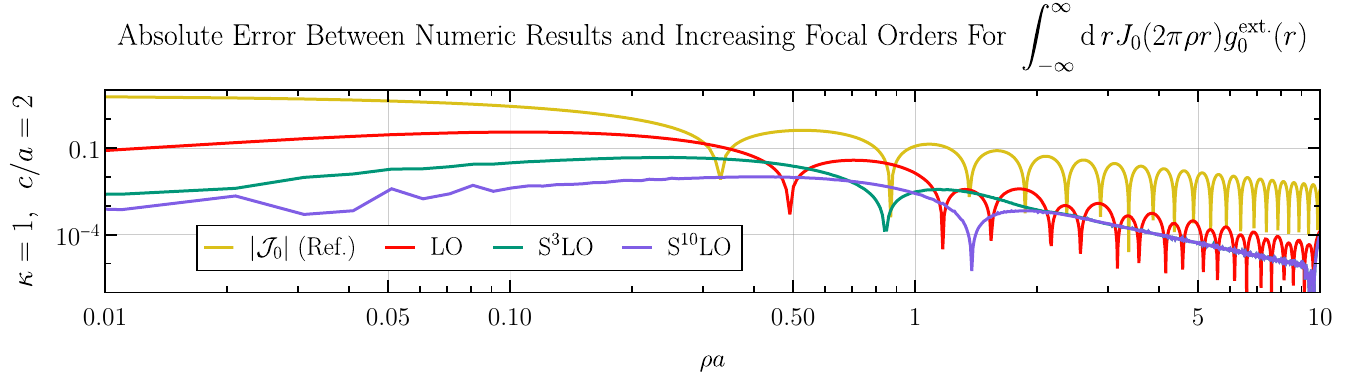}
    \caption{A log-log plot of the absolute errors between the integral $\int_{-\infty}^{\infty}\text{d}r\, r J_0(2\pi\rho r) g^{\text{ext.}}_0(r)$ and increasing orders of its focal approximation, where we have assumed $\gamma_0=1$. The full value of the integral was added for reference in yellow. We see improvement at higher focal orders, but it appears that at S$^3$LO we are already hitting the precision of our numerical evaluation of $\mathcal{J}_0$, especially at higher $\rho$ values. }
    \label{fig: log example 1 p=1 c=2 Errors}
\end{figure*}

\begin{figure*}
    \centering
    \includegraphics[width=\linewidth]{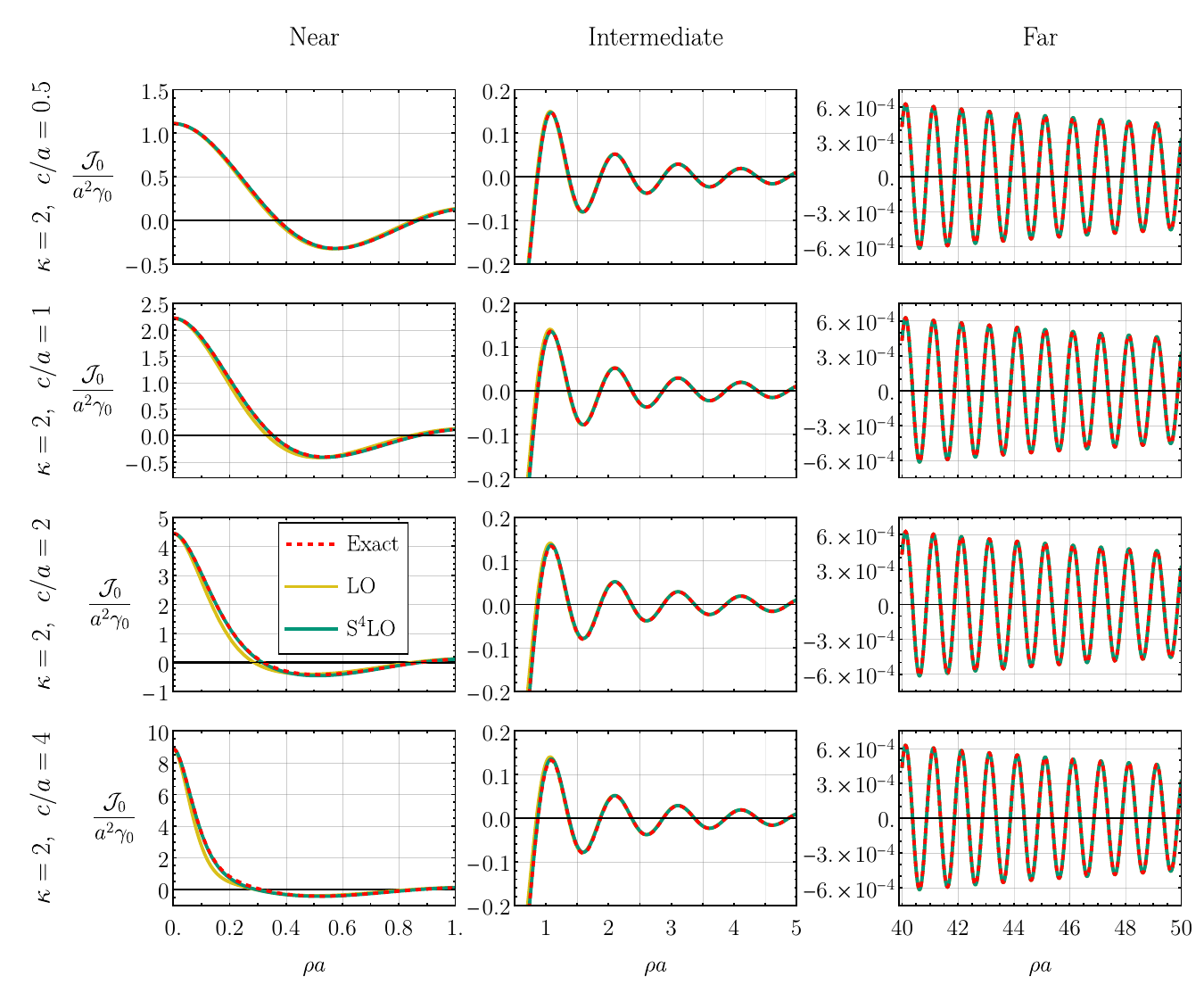}
    \caption{Comparison of the numerical evaluation (dashed red) of $\int_{-\infty}^{\infty}\text{d}r\, r J_0(2\pi\rho r) g^{\text{ext.}}_0(r)$ for a log peak with $\kappa=2$ and several values of $c$ to the focal expansion at LO (yellow) and S$^4$LO  (green). We see excellent agreement across all plots, with notable errors only visible around the origin for the LO at large width values. The fourth subleading order was chosen as that is when, visually, the discrepancy between the focal expansion and the numerics nearly disappears. Note that after around $\rho a\approx 1$ all four functions become nearly indistinguishable: this is the effect of exponentially decaying width information predicted by \cref{eq: Gn log example 1}.}
    \label{fig: log example 1 p=2 comparison}
\end{figure*}

\begin{figure*}
    \centering
    \includegraphics[width=\linewidth]{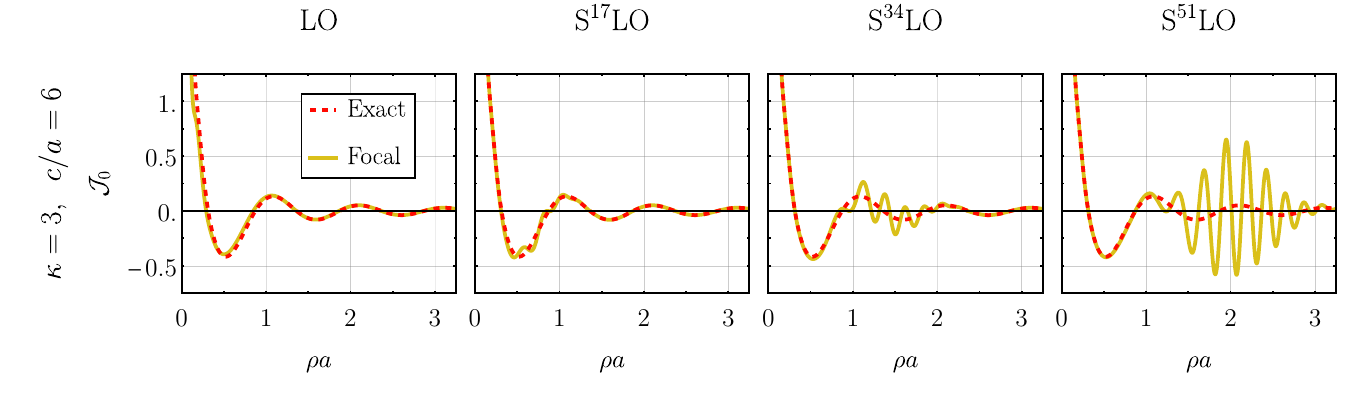}
    \caption{Comparison of the numerics of integral $\int_{-\infty}^{\infty}\text{d}r\, r J_0(2\pi\rho r) g^{\text{ext.}}_0(r)$ for a log peak with $\kappa=3$ and the relatively large width $c=6 a$ to the focal approximation at very large orders. The sequence of plots shows what we empirically found to be generic in situations where the focal expansion breaks down due to the width to radius ratio being large. A small but visible discrepancy starts out as a kink near the origin at LO, this can be seen in the top left of the first plot. As we increase the focal order this `worm' crawls down the function, becoming more oscillatory and growing in both length and amplitude. Eventually the amplitude of the worm eclipses the value of the function and the focal expansion breaks down along its length. Note, however, that \textit{after} the worm has passed a region the focal expansion seems to approximate it very well -- this is especially notable for the region near the origin. The same seems true of regions the worm is yet to crawl into. This suggests that, perhaps, one could use different focal orders depending on the value of the argument to obtain a good approximation if need be.}
    \label{fig: log example 1 breakdown}
\end{figure*}

We've taken care in laying out the arguments why the focal expansion should not work in the present case for two reasons. First, it should inform the reader of the possible problems one might encounter while applying the focal expansion out in the wild. The second, however, is far more important: despite all these belabored difficulties a naive application of the focal expansion works very well for the logarithmic rings. To be exact, considering that the function we plug into the focal expansion is $\widehat{G}_n+i/2\pi\, \widehat{G}'_n$ the even/odd parts of which are just  $\widehat{G}_n$ and $i/2\pi\, \widehat{G}'_n$ respectively, the LO focal order taken at face value is 
\begin{align}\label{eq: log example 1 leading focal}
    \begin{split}
            \mathcal{J}_n(\rho)\approx \widehat{G}_n(\rho)&J_{n}(2\pi a \rho)\\
        +\frac{1}{2\pi} \widehat{G}'_n(\rho)&\Bigg[J_{n+1}(2\pi a \rho)- \frac{n J_{n}(2\pi a \rho)}{2\pi a \rho}\Bigg].
        \end{split}
\end{align}

\noindent with $\widehat{G}_n$ given by \cref{eq: Gn log example 1}. The results for several values of $\kappa$ and $c$ are shown in fig.~(\ref{fig: log example 1 grand comparison}), we restricted ourselves to the $n=0$ case for brevity but one can check that the situation at higher angular orders is very similar. For concreteness, this would correspond to an angularly symmetric ring like the one shown in fig.~(\ref{fig: log example 1 density image}).

In general, we see broad agreement with the numerics across the full range of data with the only clearly visible discrepancies for small arguments, especially for $\kappa=1$. We must here differentiate between two types of errors: one comes from extending the function to the full real line, and the other from actual shortcomings of the focal expansion of the integral
\begin{align}\label{eq: log example 1 full integral}
    \mathcal{J}_n(\rho)\approx \int_{-\infty}^{\infty}\text{d}r\, r J_n(2\pi\rho r) g^{\text{ext.}}_n(r)
\end{align}
which we use to approximate the actual integral of interest that shows up in the Fourier transform
\begin{align}\label{eq: log example 1 half integral}
    \mathcal{J}_n(\rho)=\int_{0}^{\infty}\text{d}r\, r J_n(2\pi\rho r) g_n(r).
\end{align}
One can easily check most of the errors seen in fig.~(\ref{fig: log example 1 grand comparison}) are the fault of this extension procedure and not of the focal expansion -- see for instance fig.~(\ref{fig: log example 1 half vs full}).

Since our current motivation is investigating the validity of the focal expansion we will limit ourselves to plotting comparisons to the numerical results for \cref{eq: log example 1 full integral}, instead of the actual  \cref{eq: log example 1 half integral} used in the Fourier transform, for the remainder of this section. These results will still approximate $\mathcal{J}_n$ well, with significant errors coming mostly near the origin from the tail effects.

With that said, fig.~(\ref{fig: log example 1 p=1 c=2 comparison}) shows the numeric for the integral \cref{eq: log example 1 full integral} with $\kappa=1$, $c=2$ (which is the worst performing case in fig.~(\ref{fig: log example 1 grand comparison})) compared to its focal expansion. At higher orders the two are indistinguishable with the naked eye. In fig.~(\ref{fig: log example 1 p=1 c=2 Errors}) we also show the absolute errors for the first few focal orders, as in section~\ref{sec: annulus rings} we see higher orders improve the performance over the full range plotted\footnote{Although do note that at the higher orders we appear to be hitting the precision floor for our numerical calculation of \cref{eq: log example 1 full integral} -- due to the log divergence, slowly decaying tails, and oscillatory behavior these integrals are somewhat difficult to compute numerically (this is also the reason why we capped $\rho a$ at 10 in fig.~(\ref{fig: log example 1 p=1 c=2 Errors}), as opposed to the much larger ranges in the plots from section~\ref{sec: annulus rings}).}.

These results are significant since the case of $\kappa=1$ is the most poorly behaved out of the families in the current section; at $\rho=0$ the large $r$ tail of the integrand in \cref{eq: log example 1 full integral} decays just \textit{slightly} slower than $1/r$ making the integral just \textit{barely} convergent. This fills us with some optimism that the focal expansion can handle non-trivial subtleties -- integrable divergences, barely convergent tails and all -- on top of overcoming the apparent abstract, theoretical issues outlined above.

For further tests we move on to higher values of $\kappa$ for which the numerical calculations converge faster. In fig.~(\ref{fig: log example 1 p=2 comparison}) we show a comparison between numerical results and the focal expansion for $\kappa=2$ and several values of the width parameter $c$. We see excellent agreement across all of the widths and argument ranges, the only significant errors are visible near the origin for the leading focal order, but these appear gone by the time we get to fourth focal order. Note in particular we see that starting around $\rho a \approx 1$ all four functions seem to become approximately identical -- this is exactly the result of information about $c$ in \cref{eq: Gn log example 1} decaying exponentially away from the origin, as we predicted above. Further increasing the value of $\kappa$ generally leads to similarly good results, which was to be expected as doing so decreases the width of the peak in a subtle sense (since, in the $\kappa\rightarrow \infty$ limit, we end up with a compactly supported function).

We finish  this section with an example where the focal expansion clearly breaks. 
Based on the qualitative arguments in section~\ref{sec: Bessel function Kernels} we expect the focal expansion to start behaving poorly around $c/a\approx 2\pi\approx 6.3$. Figure~(\ref{fig: log example 1 breakdown}) shows the case of $\kappa=3$ and $c/a=6$ with increasingly large focal orders plotted. At leading order things look well, with only a barely noticeable appearance of a strange kink near the top left of the plot. That kink, however, is the focal expansions undoing. As we increase focal order it worms its way down the plot, taking the appearance of a wavelet, and growing both in length and amplitude. As far as we were able to check, its amplitude appears to grow without bound, with a sharp acceleration around the 50th focal order. It is not surprising the focal expansion has its limits, but the way this breakdown plays out is unusual. First of all note that, broadly speaking, the first few focal orders are still pretty serviceable approximations of the function in question. Only at higher orders does the behavior start becoming very poor, reminiscent of the behavior of an asymptotic expansion. Second, and more interestingly, note that as we increase focal order we do actually seem to increase our accuracy \textit{away} from the moving wavelet `worm'. It seems as if the focal expansion has no trouble giving us the correct large and small argument limits, but its attempts at fitting the region in between leads to errors that, when remedied with higher order terms, do not decrease uniformly but instead get pushed to higher argument values and amplified. The `worm' is regular enough that one can't help but wonder if it can be excised via some clever analytic procedure (other than, say, simply using a large focal order for $\rho a<1$ and then switching to a lower focal order for $\rho a>1$ to avoid the traveling worm).  More work is needed to understand this behavior and how universal it is when applying the focal expansion to generic functions, but we leave that for future work.

\subsubsection{Example 2: Asymmetric Divergence}

\begin{figure*}
    \centering
    \includegraphics[width=\linewidth]{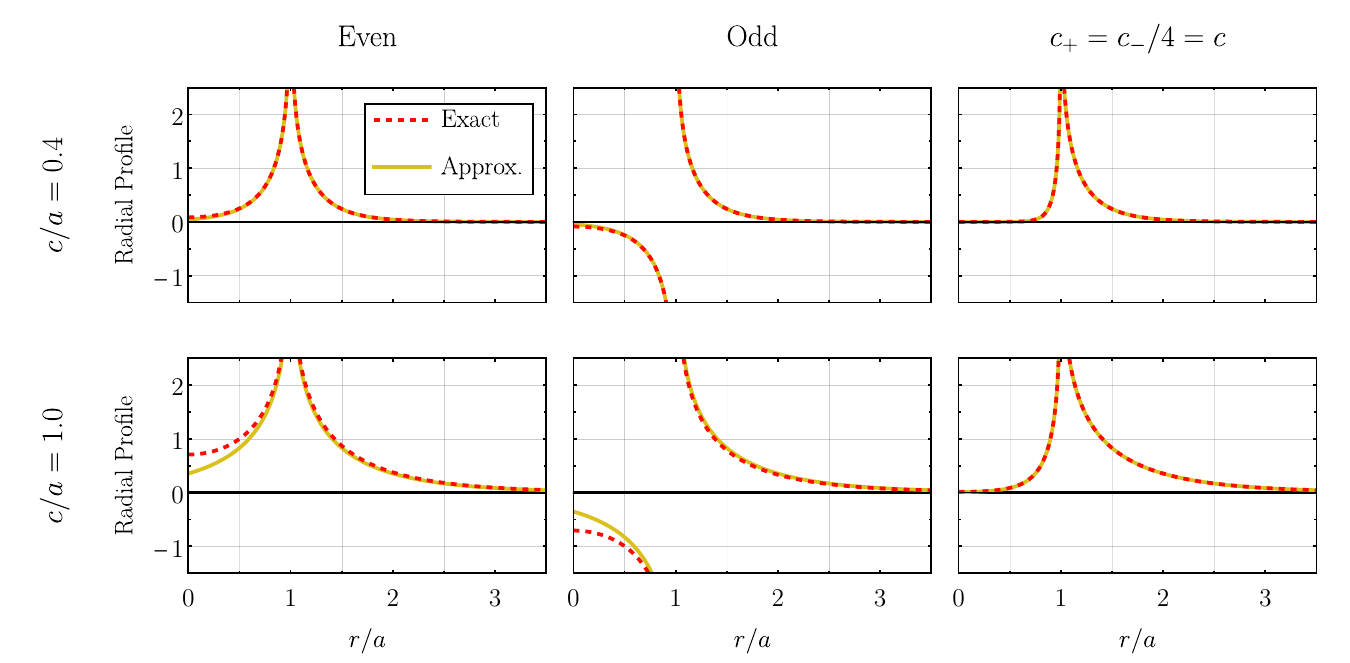}
    \caption{Several examples of even and odd log-divergence profiles modeled using the $K_0$ Bessel function, given by \cref{eq: f log example 2 even,eq: f log example 2 odd}, and also of the mixed profiles given by the superposition \cref{eq: f log example 2 mixed}. The exact profiles are plotted along with their approximations -- these are obtained by simply discarding the terms of \cref{eq: f log example 2 even,eq: f log example 2 odd} that are peaked at $r=-a$.}
    \label{fig: log example 2 radial profiles}
\end{figure*}

We now move to a different extension of the logarithmic divergence \cref{eq: f log near critical approx}. We denote by $K_0$ the modified Bessel function of the second kind of order zero. One can show that the following hold (assuming $z>0$ is a positive real number) \cite{BesselWatson}
\begin{align}
    K_0(z)&\big|_{z\ll 1}\approx\log\left(\frac{2 e^{-\gamma_E}}{z}\right),\\
    K_0(z)&\big|_{z\gg 1}\approx \sqrt{\frac{\pi}{2 z}} e^{-z},
\end{align}
with $\gamma_E\approx 0.577$ being the Euler gamma constant. This means we could extend our logarithmic divergence using $K_0$ to obtain a function with exponentially decaying tails, instead of the slower power-law tails of the previous section. What's more, the 1D Fourier transform of $K_0$ which we need for the focal expansion can be written in closed form using the identity \cite{BesselWatson}
\begin{align}\label{eq: K0 integral}
    \int_0^{\infty}e^{-i \omega x}K_0(x)\,\text{d}x=\frac{\pi/2-i \sinh^{-1}\omega}{\sqrt{\omega^2+1}}.
\end{align}
Another convenience of this extension is that since the lower limit of this integral is $0$, not $-\infty$, we can deal easily with log divergences of arbitrary asymmetry. This is in contrast to the previous section where our contour integration approach demanded the log divergence be even about the critical radius.

\begin{figure*}
    \centering
    \includegraphics[width=\linewidth]{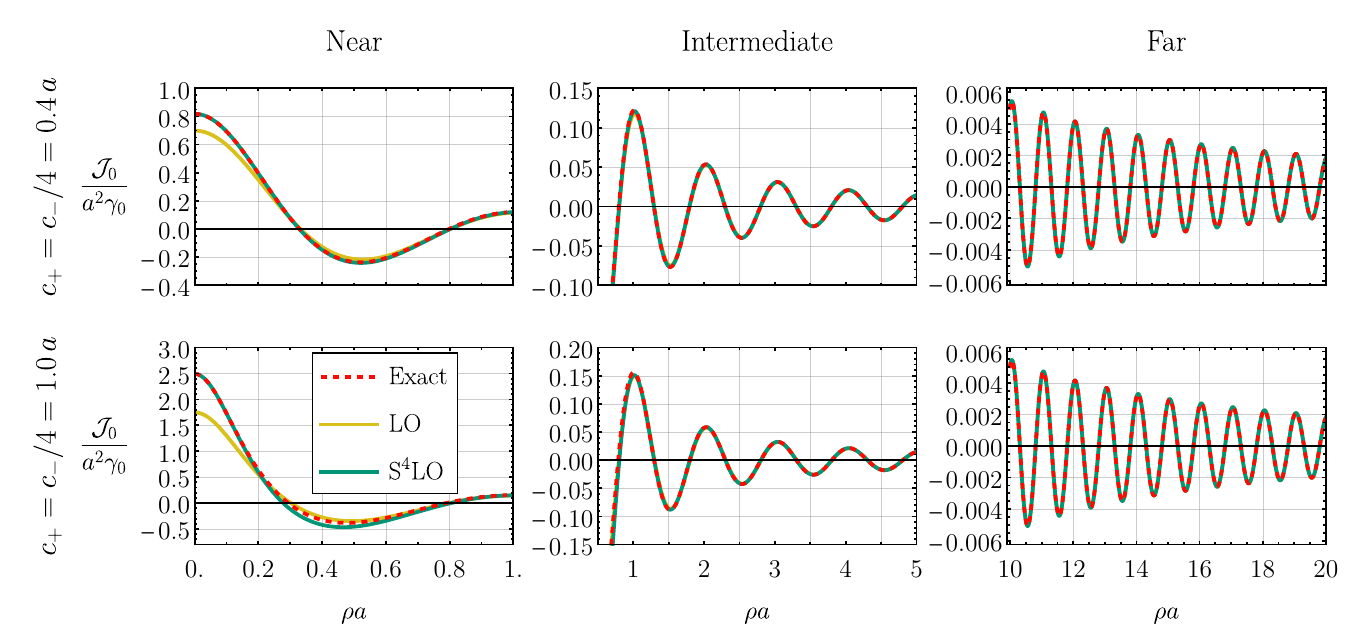}
    \caption{Comparison of the numeric (dashed red) and focal (yellow, green) results for $\mathcal{J}_0$ in the case of two mixed $K_0$ peaks with $c_+=c_-/4$ for two values of $c_+$. We see the focal LO \cref{eq: log example 2 leading focal order} performs very well except for regions very near the origin. Going up to S$^4$LO alleviates most of the discrepancies. Note also that, much like in the previous section, the two cases appear strikingly similar at larger arguments. This is due to the fact that only the asymptotics of the odd components of the mixed peak carry width information asymptotically, as shown by the limits \cref{eq: log example 2 leading even Gn asymptotic,eq: log example 2 leading odd Gn asymptotic}. If the odd component is much smaller compared to the even -- as is the case here -- the differences will be difficult to notice without going to very large $\rho$.}
    \label{fig: log example 2 J0 comparisons}
\end{figure*}

We begin by writing down the extensions for even and odd log divergent peaks:
\begin{align}
\begin{split}
        f^{e(c)}&(r,\theta)=\gamma(\theta)\Bigg[\\
        &K_0\left(2 e^{-\gamma_E}\left|\frac{r-a}{c}\right|\right)+K_0\left(2 e^{-\gamma_E}\left|\frac{r+a}{c}\right|\right)\Bigg]\label{eq: f log example 2 even}
\end{split}\\
    \begin{split}
        f^{o(c)}&(r,\theta)=\gamma(\theta)\Bigg[\\
        \text{sgn}&(r-a)K_0\left(2 e^{-\gamma_E}\left|\frac{r-a}{c}\right|\right)-K_0\left(2 e^{-\gamma_E}\left|\frac{r+a}{c}\right|\right)\Bigg]\label{eq: f log example 2 odd}
    \end{split}
\end{align}
where $\gamma(\theta)$ is given by \cref{eq: log example ang dependence} as before and where once again the extra parity flipped terms ensure no artificial asymptotics in numerical calculations\footnote{Note $r>0$ so the second sgn in \cref{eq: f log example 2 odd} evaluates to $-1$.}. One could just as easily consider functions for which the widths $c$ depend on the angular mode but we omit this generalization for clarity. A few examples of the radial profiles of the angular modes of \cref{eq: f log example 2 even,eq: f log example 2 odd} are shown in fig.~(\ref{fig: log example 2 radial profiles}). Near the critical radius $r=a$ these functions behave as
\begin{align}
    f^{e(c)}(r,\theta)\big|_{r\approx a}&\approx \gamma(\theta)\log\left|\frac{c}{r-a}\right|,\\
    f^{o(c)}(r,\theta)\big|_{r\approx a}&\approx \gamma(\theta)\text{sgn}(r-a)\log\left|\frac{c}{r-a}\right|.
\end{align}
Much like in the previous section, We chose our radial profile approximations for the focal expansion by simply dropping the symmetrized terms in \cref{eq: f log example 2 even,eq: f log example 2 odd} so that only the term giving the divergence at $r=a$ remains. These extend naturally to all of $\mathbb{R}$, taking the 1D Fourier transforms of these extensions with the help of \cref{eq: K0 integral} and choosing $a$ as the shift parameter one is then led to\footnote{The Euler constant $\gamma_E$ is not to be confused with the angular expansion coefficients $\gamma_n$.}
\begin{align}
    \widehat{G}^{e(c)}_n(\rho)=&\gamma_n\frac{\pi}{2} c e^{\gamma_E}\frac{1}{\sqrt{(\pi c e^{\gamma_E}\rho)^2+1}}\label{eq: Gn log example 2 even},\\
    \widehat{G}^{o(c)}_n(\rho)=&-i\gamma_n c e^{\gamma_E}\frac{\sinh^{-1}\left(\pi c e^{\gamma_E}\rho\right)}{\sqrt{(\pi c e^{\gamma_E}\rho)^2+1}}\label{eq: Gn log example 2 odd},
\end{align}
for the even and odd peaks respectively. Note that the asymptotics of \cref{eq: Gn log example 2 even,eq: Gn log example 2 odd} are
\begin{align}
    &\widehat{G}_n^{e(c)}\big|_{\rho\gg 1}\approx \gamma_n\frac{1}{2 \rho}\label{eq: log example 2 leading even Gn asymptotic},\\
    &\widehat{G}_n^{o(c)}\big|_{\rho\gg 1}\approx \gamma_n\frac{\gamma_E+\log(2\pi)+\log(c\rho)}{i \pi \rho},\label{eq: log example 2 leading odd Gn asymptotic}
\end{align}
so we see that, much like in the previous section, the leading asymptotic for the even peak contains no information about the width $c$ (although the disappearance is much slower than exponential in this case). On the other hand, the odd case decays slightly slower by an extra logarithmic factor which does not lose all of its $c$ dependence. Once again, when applied in the focal expansion \cref{eq: log example 2 leading even Gn asymptotic,eq: log example 2 leading odd Gn asymptotic} lead to $1/\rho^{3/2}$ decay due to an extra factor of $1/\sqrt{\rho}$ from the Bessels.

To make asymmetric peaks with $c_+\neq c_-$ we superimpose \cref{eq: f log example 2 even,eq: f log example 2 odd} as in
\begin{align}\label{eq: f log example 2 mixed}
    f^{(c_+,c_-)}=\frac{f^{e(c_+)}+f^{o(c_+)}}{2}+\frac{f^{e(c_-)}-f^{o(c_-)}}{2}.
\end{align}
A few examples of even, odd, and mixed peaks, along  with a comparison to their approximate $\mathbb{R}$ extensions, are shown in fig.~(\ref{fig: log example 2 radial profiles}). By linearity we see the functions to be used in the focal expansion are likewise superpositions of \cref{eq: Gn log example 2 even,eq: Gn log example 2 odd}:
\begin{align}\label{eq: Gn log example 2 mixed}
    \widehat{G}_n^{(c_+,c_-)}=\frac{\widehat{G}_n^{e(c_+)}+\widehat{G}_n^{o(c_+)}}{2}+\frac{\widehat{G}^{e(c_-)}-\widehat{G}^{o(c_-)}}{2}.
\end{align}
Note that, as expected, these functions don't have the problems with absolute values that we had to deal with in the previous section, meaning we can apply the focal expansion as usual with no tweaks. We state the leading focal order here for completeness:
\begin{figure*}
    \centering
    \includegraphics[width=\linewidth]{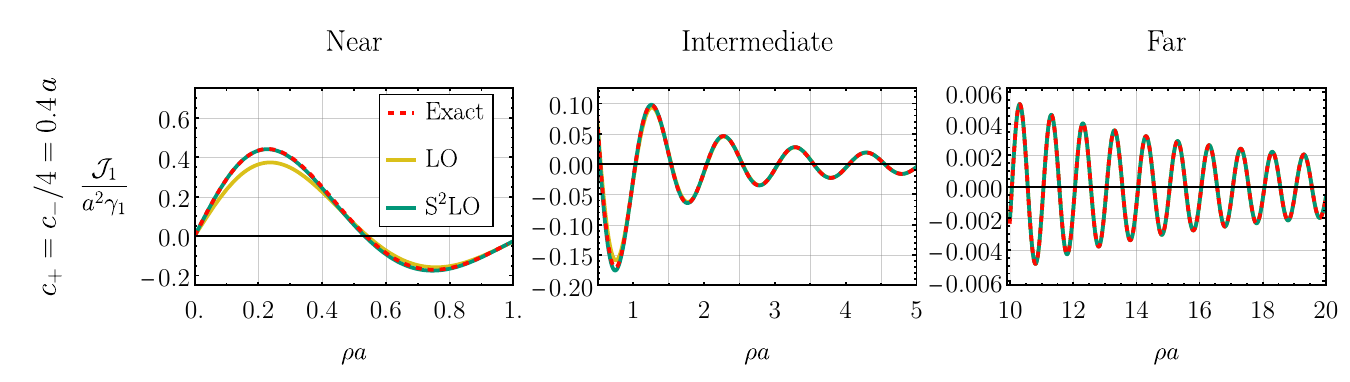}
    \caption{Comparison of the numeric results for $\mathcal{J}_1$ in the case of a mixed $K_0$ peak with $c_+=c_-/4=0.4$. We see the focal LO \cref{eq: log example 2 leading focal order} performs very well except for regions very near the origin. Going up to S$^2$LO alleviates most of the discrepancies.}
    \label{fig: log example 2 J1 comparisons}
\end{figure*}
\begin{align}\label{eq: log example 2 omega pm def}
    \omega_{\pm}&\equiv \pi c_{\pm} e^{\gamma_E},
\end{align}
\begin{align}\label{eq: log example 2 leading focal order}
    \begin{split}
        \mathcal{J}_n(\rho)\approx& \frac{a\gamma_n}{4}\left[\frac{\omega_+}{\sqrt{\omega_+^2\rho^2+1}}+\frac{\omega_-}{\sqrt{\omega_-^2\rho^2+1}}\right]J_n(2\pi a\rho)\\
        -\frac{a\gamma_n}{2\pi}&\left[\frac{\omega_+ \sinh^{-1}(\omega_+\rho)}{\sqrt{\omega_+^2\rho^2+1}}-\frac{\omega_- \sinh^{-1}(\omega_-\rho)}{\sqrt{\omega_-^2\rho^2+1}}\right]\\
        \times&\left[J_{n+1}(2\pi a\rho)-\frac{n J_n(2\pi a \rho)}{2\pi a \rho}\right].
    \end{split}
\end{align}

\begin{figure}
    \centering
    \includegraphics[width=\linewidth]{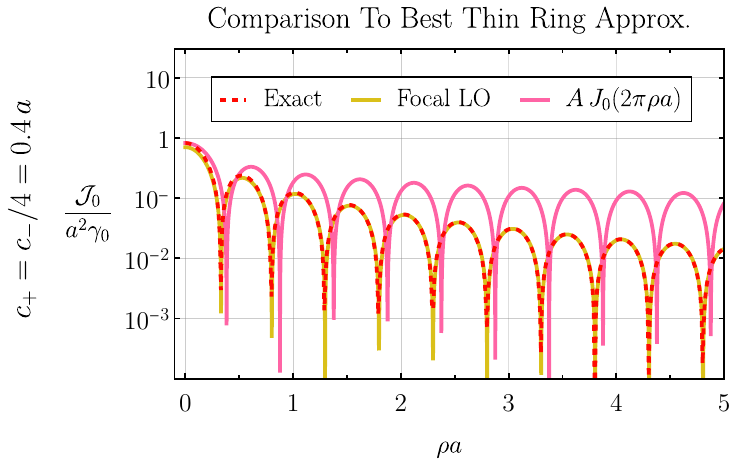}
    \caption{Comparison of numerical results (dashed red) and the LO focal expansion (yellow) for a mixed $K_0$ ring to the best (determined by matching at the origin) $J_0$ approximation (magenta) corresponding to a circular delta function ring. The numerics and the focal expansion are the same here as in the first row of fig.~(\ref{fig: log example 2 J0 comparisons}), note the vertical axis is logarithmic. We see that the $J_0$ approximation pretty quickly becomes insufficient as we move away from the origin in terms of both the falloff rate and the phase -- the latter of which is slightly but visibly off.}
    \label{fig: log Example 2 thin ring comparison}
\end{figure}

The performance of the focal expansion for the mixed peaks of fig.~(\ref{fig: log example 2 radial profiles}) and $n=0$ is shown in fig.~(\ref{fig: log example 2 J0 comparisons}). As we have by now come to expect, the leading focal expansion provides a reasonable approximation and we get an excellent result by adding a few more focal orders. The improvement is mostly near the origin -- above $\rho\approx 1/a$ the LO seems more than sufficient. The situation is very much similar for higher angular modes, see fig.~(\ref{fig: log example 2 J1 comparisons}) where we have plotted $\mathcal{J}_1$ for the smaller $c$ value from fig.~(\ref{fig: log example 2 J0 comparisons}).
We note that even for the smaller widths we have plotted, like the $c_+=.4$ above, approximating the divergence as just a circular delta function $\delta(r-a)$ is insufficient. Delta function circles of arbitrary angular dependence always decay asymptotically like $1/\sqrt{\rho}$, much slower than our $1/\rho^{3/2}$, so they will not provide a useful approximation beyond the immediate neighborhood of the origin -- see fig.(\ref{fig: log Example 2 thin ring comparison}).
\begin{figure}
    \centering
    \includegraphics[width=\linewidth]{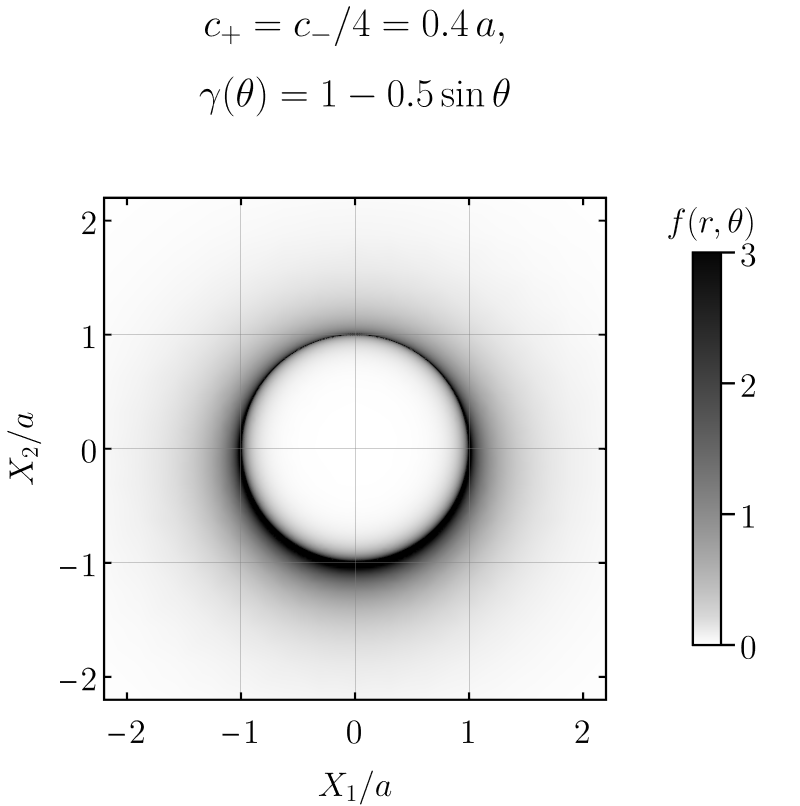}
    \caption{An example of an asymmetric logarithmic ring with non-trivial angular dependence. All function values above $f(r,\theta)=3$ are shown in black.}
    \label{fig: log example 2 density}
\end{figure}

Finally, we close the section by noting that with \cref{eq: log example 2 leading focal order} (and higher orders) in hand one can easily add the $\mathcal{J}_n$'s to create non-trivial angular dependence. For instance, taking once again $c_+=4c_-=0.4 a$ and
\begin{align}
    \gamma(\theta)=1-\frac{1}{2}\sin\theta
\end{align}
produces the ring shown in fig.~(\ref{fig: log example 2 density}) whose Fourier response is given by 
\begin{align} \label{eq: log example 2 F ang dep}
    \mathcal{F}[f](\rho,\phi)=2\pi\left[\mathcal{J}_0(\rho)+\frac{i}{2}\sin\phi \,\mathcal{J}_1(\rho)\right].
\end{align}
The real part of \cref{eq: log example 2 F ang dep} is then precisely the first row of fig.~(\ref{fig: log example 2 J0 comparisons}) rescaled by $2\pi$ while the imaginary part is precisely fig.~(\ref{fig: log example 2 J1 comparisons}) rescaled by the $\phi$-dependent factor $\pi\sin\phi$.

\subsection{Centrally Peaked Example}

\begin{figure}
    \centering
    \includegraphics[width=\linewidth]{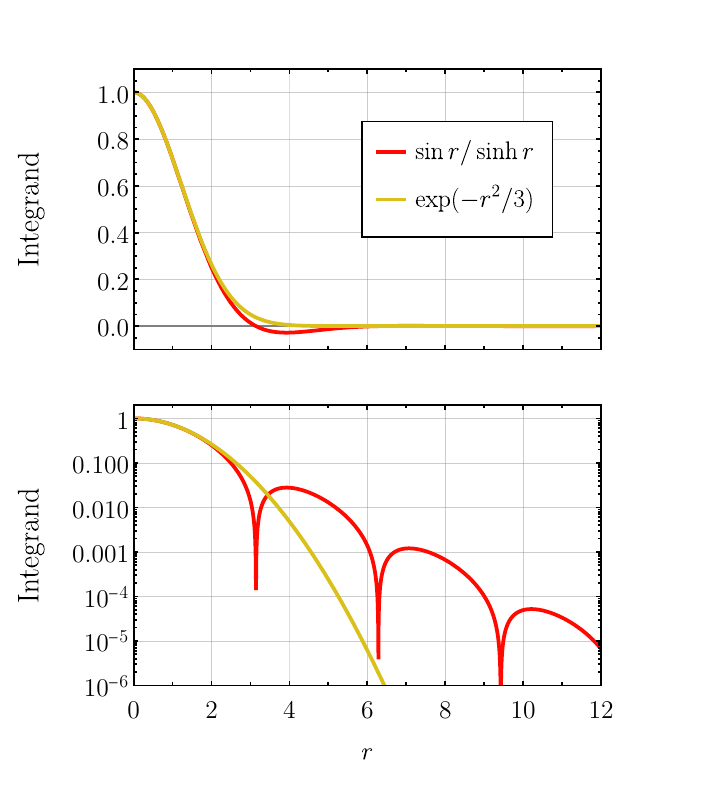}
    \caption{The comparison of the function $\sin r/\sinh r$ to its best Gaussian fit near the origin. Note the Gaussian captures most of the weight but not the asymptotic behavior, as seen on the lower logarithmic plot.}
    \label{fig: peak example profile comp}
\end{figure}

In this example we address the accuracy of the procedure for removing weight from the origin, outlined in section~\ref{section: peak at origin}. The integral we will be investigating is\footnote{Many thanks to Serhii Kryhin for suggesting this example.}
\begin{align}\label{eq: peak at origin integral example}
    \mathcal{I}_n(\rho)\equiv \int_{0}^{\infty}\frac{\sin r}{\sinh r}J_n(2\pi\rho r)\,\text{d}r.
\end{align}
A quick calculation shows that near the origin 
\begin{align}
    \frac{\sin r}{\sinh r}\approx 1-\frac{r^2}{3}
\end{align}
and so we choose to approximate the peak with the Gaussian $e^{-r^2/3}$. The remainder 
\begin{align}\label{eq: peak at origin example remainder}
    R(r)\equiv \frac{\sin r}{\sinh r}-e^{-r^2/3}
\end{align}
is peaked away from the origin and so we may use the focal expansion on it -- see fig.~(\ref{fig: peak example profile comp}) where the integrand (excluding the Bessel) is plotted along with the Gaussian approximation.
By \cref{eq: gaussian fit} we see that the Gaussian contributes to the integral as
\begin{align}\label{eq: peaked example gaussian contribution}
    \begin{split}
        \mathcal{G}_n(\rho)&\equiv \int_0^{\infty}e^{-r^2/3}J_n(2\pi \rho r)\,\text{d}r\\
        &=\frac{\sqrt{3\pi}}{2}e^{-3\pi^2\rho^2/2}I_{n/2}\left(\frac{3\pi^2\rho^2}{2}\right).
    \end{split}
\end{align}
Based on the top plot of fig.~(\ref{fig: peak example profile comp}) we expect that the Gaussian alone will be a serviceable approximation for our integral. But we are currently less interested in practicalities than in testing the focal expansion. This example is intended as a stress test of the origin-subtraction prescription: can the remaining (small) correction be captured analytically, using the focal expansion, without compromising the uniform behavior in $\rho$?

We must start by extending the remainder $R(r)$ to the full $\mathbb{R}$.  This is a subtle problem. As it stands, $R(r)$ is an even function so if we simply let $r<0$ in the expression above then we will get \textit{two} peaks, one at $r=1$ and the second at $r=-1$. We need to separate these two peaks somehow, and do it in a way that will make the 1D Fourier transform analytically tractable. One approach is to write

\begin{figure}[H]
    \centering
    \includegraphics[width=\linewidth]{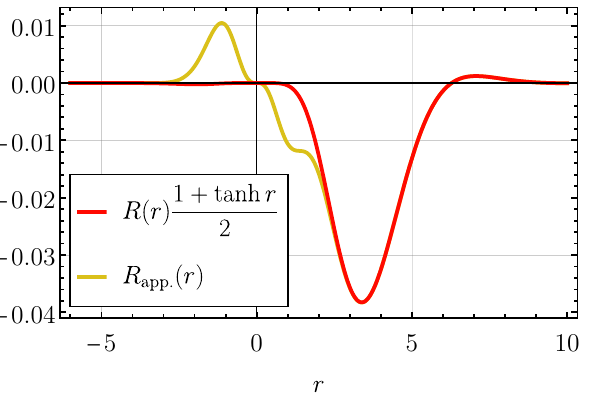}
    \caption{Comparison of the positive-$r$ peak of the remainder $R(x)$, excised using $(1+\tanh r)/2$, to the approximation \cref{eq: peak at origin example remainder approx} we made in order to be able to perform the relevant integrals analytically. Note that we approximate the peak itself well, but have added two small artificial peaks near the origin.}
    \label{fig: peak example remainder comp}
\end{figure}

\begin{figure*}
    \centering
    \includegraphics[width=\linewidth]{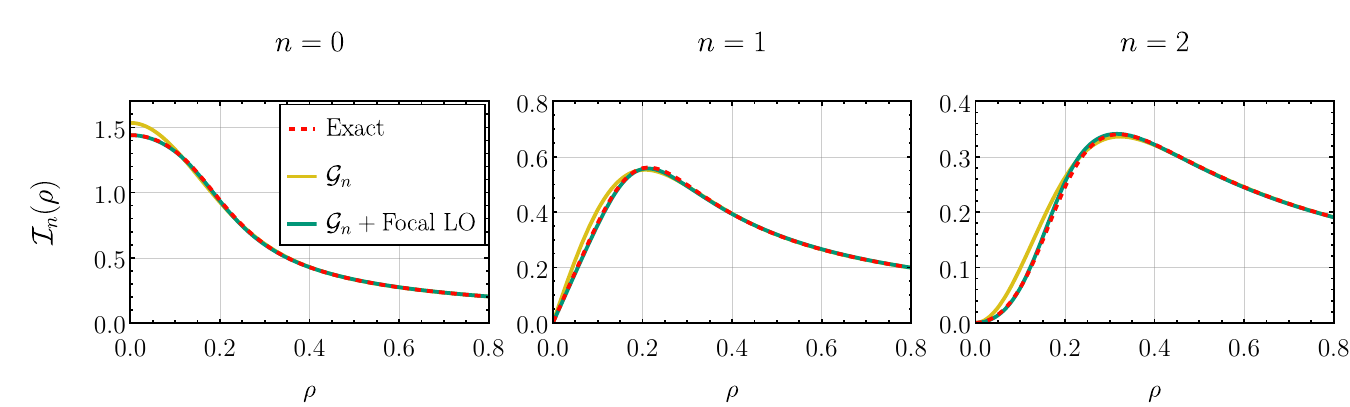}
    \caption{Comparison of the numerical results (dashed red) for the integral $\int_{0}^{\infty}\frac{\sin r}{\sinh r}J_n(2\pi\rho r)\,\text{d}r$ at three values of $n$ to the best Gaussian approximation (given by \cref{eq: peaked example gaussian contribution}) with and without the extra focal LO contribution (green and yellow respectively). We see that the small errors exhibited by the pure Gaussian approach are almost entirely remedied by the LO focal approximation -- see fig.~(\ref{fig: peak example remainder Ftr comp}) below for a clearer picture of the remaining errors.}
    \label{fig: peak example full Ftr comp}
\end{figure*}

\begin{align}
    R(r)=R(r)\,\frac{1+\tanh r}{2}+R(r)\,\frac{1-\tanh r}{2}
\end{align}
and keep just the first term, which (approximately) picks out the $r=1$ peak. The benefit to doing this is that $\tanh r \sin r/\sinh r=\sin r/\cosh r$ and so we can do that part of the 1D Fourier transform integral of $R(r)$ analytically. The Gaussian component causes trouble as it appears $\mathcal{F}[e^{-r^2/3}\tanh r]$ has no closed form solutions. Here we introduce another approximation: one can find that to reasonable accuracy
\begin{align}
    \tanh(r) \approx \text{Erf}\left(\frac{\sqrt{\pi}}{2}r\right).
\end{align}
The reason for performing this approximation is that $\mathcal{F}[e^{-r^2/3}\text{Erf}(\text{const.}\times r)]$ \textit{can} be calculated analytically. All together we take the approximation
\begin{align}\label{eq: peak at origin example remainder approx}
    R_{\text{app.}}(r)\equiv\frac{\sin r}{\sinh r}\frac{1+\tanh(r)}{2}-e^{-r^2/3}\frac{1+\text{Erf}\left(r\sqrt{\pi}/2\right)}{2}
\end{align}
the comparison of which to the exact remainder is shown in Fig.~(\ref{fig: peak example remainder comp}). We see our approximation has clumsily added two small peaks (odd around the origin). These are a small but significant when compared to the main peak, meaning we likely won't get any significant performance improvements after leading focal order, but that is the price we have paid to make our analysis analytically tractable.

We now calculate the 1D Fourier transform of \cref{eq: peak at origin example remainder approx}, the relevant identities can be shown through a mixture of standard integration tricks and are
\begin{align}
    \begin{split}\label{eq: sin/sinh Ftr}
        \mathcal{F}&\left[\sin r/\sinh r\right](\rho)=\\
        &\frac{\pi}{2}\left[\tanh\left(\frac{\pi}{2}\left(2\pi\rho+1\right)\right)-\tanh\left(\frac{\pi}{2}\left(2\pi\rho-1\right)\right)\right],
    \end{split}
\end{align}
\begin{align}
    \begin{split}\label{eq: sin/cosh Ftr}
        \mathcal{F}&\left[\sin r/\cosh r\right](\rho)=\\
        &\frac{i\pi}{2}\left[\frac{1}{\cosh\left(\frac{\pi}{2}\left(2\pi\rho+1\right)\right)}-\frac{1}{\cosh\left(\frac{\pi}{2}\left(2\pi\rho-1\right)\right)}\right],
    \end{split}
\end{align}
\begin{align}
    \begin{split}\label{eq: Gaussian Erf Ftr}
        \mathcal{F}&\left[e^{-r^2/3}\text{Erf}\left(r\sqrt{\pi}/2\right)\right](\rho)=\\
        &-i\sqrt{3\pi}e^{-(\sqrt{3}\pi\rho)^2}\text{Erfi}\left(\frac{\sqrt{3}\pi\rho}{\sqrt{1+4/3\pi}}\right),
    \end{split}
\end{align}
where $\text{Erfi}(z)\equiv \text{Erf}(iz)/i$ is the complex error function (despite the standard but confusing name, it is real for real arguments). Putting everything together we have
\begin{align} \label{eq: peak at origin remainder hat}
    \widehat{R}_{\text{app.}}(\rho)=\frac{\text{(\ref{eq: sin/sinh Ftr})}}{2}+\frac{\text{(\ref{eq: sin/cosh Ftr})}}{2}-\frac{\text{(\ref{eq: Gaussian Erf Ftr})}}{2}-\frac{\sqrt{3\pi}}{2}e^{-3\pi^2\rho^2}
\end{align}
where we refereed to expressions \cref{eq: sin/cosh Ftr,eq: sin/sinh Ftr,eq: Gaussian Erf Ftr} solely by their number for brevity. What's troubling about this expression is that, for the first time, there is no obvious upfront phase factor to tell us what to choose for the shift parameter. 
Remember, however, that we have shown the shift parameter need not be determined with great precision. Looking at fig.~(\ref{fig: peak example remainder comp}) it appears the apex of the main peak is positioned above $r=3$ but a little under $r=3.5$. As an example, suppose $a=\pi$. This means that the function to be used in the focal expansion is
\begin{figure*}
    \centering
    \includegraphics[width=\linewidth]{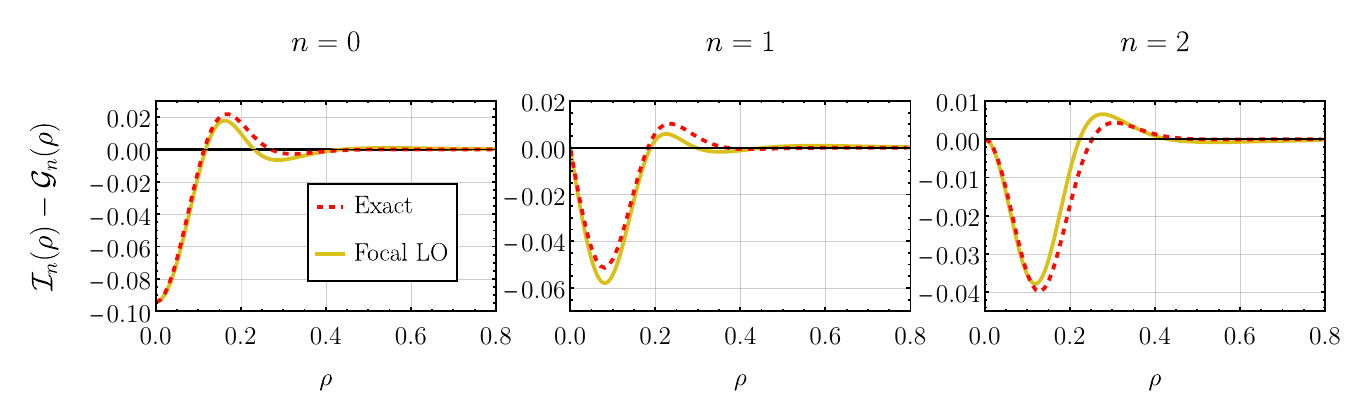}
    \caption{Comparison of the residual of the numerically calculated integral $\int_{0}^{\infty}\frac{\sin r}{\sinh r}J_n(2\pi\rho r)\,\text{d}r$ (dashed red) obtained after subtracting the Gaussian contribution (given by \cref{eq: peaked example gaussian contribution}) to the focal expansion LO (yellow). We see the LO is a serviceable approximation of the function at hand. We have not plotted higher focal orders as they seem to suffer from the same subtle `worm' issue of fig.~(\ref{fig: log example 1 breakdown}). This could be due to the fact that, as we can see in fig.~(\ref{fig: peak example remainder comp}), the width to position ratio of our remainder appears somewhat large, especially with the artificial peaks our approximation adds. Nevertheless, the focal expansion still provides a good approximation despite these issues.}
    \label{fig: peak example remainder Ftr comp}
\end{figure*}
\begin{align}
    \widehat{G}(\rho)\equiv e^{2\pi^2 i \rho}\widehat{R}_{\text{app.}}(\rho)
\end{align}
with the remainder factor given by the cumbersome \cref{eq: peak at origin remainder hat}. One then calculates $\widehat{G}^{\pm}$ (this is simple since every function appearing in \cref{eq: peak at origin remainder hat} has unambiguous parity) and plugs them into the leading focal order \cref{eq: summary LO} with $a=\pi$. The end result is cumbersome so we do not state it here and move straight to comparing it to numerics. Figure~(\ref{fig: peak example full Ftr comp}) compares the numerics to the $\mathcal{G}_n$ and $\mathcal{G}_n$ plus leading focal order approximations. We see that the focal LO seems to take care of what little discrepancy there was between $\mathcal{I}_n$ and $\mathcal{G}_n$. To get a clearer picture, we have also directly compared $\mathcal{I}_n-\mathcal{G}_n$ calculated numerically to the focal LO in fig.~(\ref{fig: peak example remainder Ftr comp}). Here we do see some moderate shortcomings of the leading focal order. One can check that adding subleading focal orders doesn't appear to significantly help -- part of the problem certainly stems from the artificially introduced small peaks seen in fig.~(\ref{fig: peak example remainder comp}) but it could also be that the width of the peak is relatively large leading to poor focal convergence\footnote{What makes us suspect this second comment might be true is that the subleading focal orders help in \textit{some} places (notably the secondary peaks around $\rho \approx 0.2$) but for larger argument we get behavior somewhat reminiscent of the `worm' of fig.~(\ref{fig: log example 1 breakdown}).}, it is difficult to say without doing more calculations (the two could also be connected: the small peaks do appear to increase the sense of width of the function). We see, however, that overall $\mathcal{G}_n$ plus leading focal order are more than enough to explain the behavior of $\mathcal{I}_n$.

We close this section by once again noting the purpose of these calculations was less issues of practicality -- the integral \cref{eq: peak at origin integral example} can be calculated numerically with ease -- but of principle and exposition. We believe the methods we present are sufficiently general (and Fourier transforms sufficiently widespread) that they might find non-trivial and insightful application in research problems. 

\subsection{Delta Function Rings \label{sec: delta function rings}}

\subsubsection{General Calculations}

For our final set of examples we analyze the case of rings of zero thickness -- those whose radial profile is a Dirac delta. We  specialize to curves whose radial distance from the origin can be written as a positive function of the angle $R(\theta)>0$ (so, in particular, we do not allow the shape to `back up' in angle). The function to be Fourier transformed is then
\begin{align}
    f(r,\theta)=\gamma(\theta)\delta(r-R(\theta))
\end{align}
where we allowed for the intensity to vary with angle as $\gamma(\theta)$. As before, we assume the intensity function has a Fourier series
\begin{align} \label{eq: intensity profile decomp}
    \gamma(\theta)=\sum_{n\in\mathbb{Z}}\gamma_n e^{i n \theta}
\end{align}
so that the angular modes of the image are given by
\begin{align} \label{eq: gn for delta function ring}
    g_n(r)=\frac{1}{2\pi}\sum_{m\in\mathbb{Z}}\gamma_m\int_0^{2\pi}e^{i(m-n)\theta}\delta(r-R(\theta))\,\text{d}\theta.
\end{align}
Because we assumed $R(\theta)>0$, \cref{eq: gn for delta function ring} already gives us a valid extension of $g_n$ to the full real line as the function simply vanishes for $r\leq 0$. We may then move straight to calculating the focal expansion.

One might be tempted to immediately use the delta function to evaluate the $\theta$ integral but that is not the only way to proceed. The quantity relevant for the focal expansion is not $g_n$ itself but its 1D Fourier transform,
\begin{align}\label{eq: gnhat delta ring}
   \begin{split}
        \widehat{g}_n(\rho)=&\frac{1}{2\pi}\sum_{m\in\mathbb{Z}}\gamma_m\\
        \times &\int_0^{2\pi}\text{d}\theta \int_{\mathbb{R}}\text{d}r\, \delta(r-R(\theta))\,e^{i(m-n)\theta-i2\pi r \rho}.
   \end{split}
\end{align}
We see then there are two distinct ways forward: we can use the Dirac delta to evaluate either the $\theta$ or the $r$ integral first. Doing the latter results in
\begin{align}\label{eq: gnhat angular delta ring}
       \begin{split}
        \widehat{g}_n(\rho)=&\frac{1}{2\pi}\sum_{m\in\mathbb{Z}}\gamma_m
        \int_0^{2\pi}\text{d}\theta \,e^{i(m-n)\theta-i2\pi \rho R(\theta) },
   \end{split}
\end{align}
while the former results in 
\begin{align}\label{eq: gnhat radial delta ring}
   \begin{split}
        \widehat{g}_n(\rho)=&\frac{1}{2\pi}\sum_{m\in\mathbb{Z}}\gamma_m \int_{\mathbb{R}}\text{d}r\,e^{-i2\pi r \rho} \sum_{j}\frac{e^{i(m-n)\theta_j(r)}}{|R'(\theta_j(r))|},
   \end{split}
\end{align}
where $\theta_j(r)$ are the angles at which $R(\theta_j)=r$. Note that for fixed $j$ the solution $\theta_j(r)$ can only exist in some range of $r$ and so must be omitted from the $j$-sum in \cref{eq: gnhat radial delta ring} when integrating outside of said range.

In terms of simplicity, \cref{eq: gnhat angular delta ring} appears preferable to \cref{eq: gnhat radial delta ring} but both approaches are valid -- which one to choose depends on which integral is easier to perform analytically for a given curve. However, generically neither integral will have a closed form solution\footnote{If $\mathbb{R}(\theta)$ has a $\theta$-Fourier series then \cref{eq: gnhat angular delta ring} can be written in terms of \textit{generalized Bessel functions} -- see \cite{genBessel,genBesselold} -- but these are more complex and less well understood than standard Bessel functions. We leave this direction for possible future work.} and one is forced to use some approximation scheme. It is important to pause on this point. Once again, the price we pay for the uniform validity of the focal expansion is that we are still required to evaluate certain 1D integral analytically. We exchanged a difficult 2D integral for a series of 1D integrals, but if we can't uniformly approximate those 1D integrals then the final expansion will not be uniformly valid either.

Nevertheless, for completeness we perform an asymptotic expansion of \cref{eq: gnhat radial delta ring,eq: gnhat angular delta ring}. These are oscillatory integrals and one can employ many standard (but not generically uniformly-valid) approaches for approximating them \cite{BenderOrszag}; we use stationary phase. The result, naturally identical for both forms of $\widehat{g}_n$, is given by: 
\begin{align}\label{eq: gnhat stat phase}
    \begin{split}
        \widehat{g}_n(\rho)&\Big|_{\text{stat. phase}}\approx \sum_{m\in\mathbb{Z}}\gamma_m\\\times&\sum_k \frac{e^{i(m-n)\theta_{k}-i2\pi R(\theta_{k})\rho}}{2\pi \sqrt{|\rho R''(\theta_{k})|}}e^{-i\pi/4\, \text{sgn}(\rho R''(\theta_k))}
    \end{split}
\end{align}
where $\text{sgn}$ is the sign function and the $k$ sum is over all solutions to 
\begin{align}\label{eq: Rprime cond}
    R'(\theta_{k})=(m-n)/2\pi\rho
\end{align}
for fixed $m$. One then uses either the exact $\widehat{g}_n$ if available, or the stationary phase approximation if not, in the focal expansion. Note that we may \textit{not} simply take \cref{eq: Rprime cond} to vanish in the large $\rho$ limit -- for any value of $\rho$ there are values of $n, m$ for which \cref{eq: Rprime cond} is non-negligible (although for large enough values the solution set of $\theta_k$ will be empty if our curve is smooth). 

The approximate expression \cref{eq: gnhat stat phase} looks cumbersome already, and would be even more so after plugging it into the leading order focal expansion, but if we want to get the actual Fourier response of our curve we must remember we still need to sum over all angular modes $n$. In the previous examples we were able to reassure ourselves that only a finite number of these modes needed to be kept to obtain a good overall approximation but that is no longer the case here. The non-smoothness of the Dirac delta results in non-trivial resummations when adding up all of the angular modes. The easiest way to see this is to note that, based on \cref{eq: gnhat stat phase}, naively every Dirac delta ring should have asymptotic response decaying like $1/\rho^{3/2}$ (with the extra $1/\sqrt{\rho}$ coming from the focal expansion Bessels) but that is \textit{not} correct, as shown by \cite{Gralla} and as we will see in the next sections' example.

These non-trivialities, uniquely accentuated due to the Dirac deltas, get at the heart of what is and is not natural in the focal expansion. Ours is an approach anchored in spherical coordinates and so it is beset by difficulties when dealing with objects that depart from spherical symmetry. We can overcome that when finitely many angular modes suffice but that tends to lead to `fuzzy' non-circular shapes as we saw back in section~\ref{sec: making shapes} -- the infinite precision of delta function rings lays bare this particular shortcoming. There is another related issue: things clear in cartesian coordinates are very much not so in the focal expansion. For instance, the function $R(\theta)$ will in general transform in a highly non-trivial manner under cartesian translations of the underlying function $f(r,\theta)$. We know from Fourier analysis that a translation by a 2D vector $y$ must amount to multiplication of the Fourier transform by $e^{i2\pi k \cdot y}$ but it is not at all obvious how this fact is encoded in the  focal expansion, if at all. This situation should be once again compared to the analysis of delta function curves in \cite{Gralla} which does not suffer from such issues and where cartesian behavior like translations and rescaling along the axes are manifest. We suspect these expected cartesian behaviors are still present in the focal expansion but may arise due to non-trivial resummations of all the terms -- we leave investigations of this for future work\footnote{The one cartesian sanity check we can easily perform involves the shifted Dirac delta $\delta^2(\Vec{r}-\Vec{a})=\frac{1}{a}\delta(r-a)\delta(\theta-\varphi)$. One can show that choosing shift parameter to be $a$ yields the $\rho$ independent $\widehat{G}_n(\rho)=e^{-i\varphi n}/2\pi$ which collapses the focal expansion to $\mathcal{J}_n=\frac{e^{-i\varphi n}}{2\pi}J_n(2\pi a \rho)$ and, upon an application of \cref{eq: Anger-Jacobi}, results in the expected $e^{-i 2\pi \Vec{r}\cdot \Vec{a}}$.}. There is, however, one particularly important example that happens to appeal to the focal expansions' strength, we turn to it now.

\begin{figure*}
    \centering
    \includegraphics[width=\linewidth]{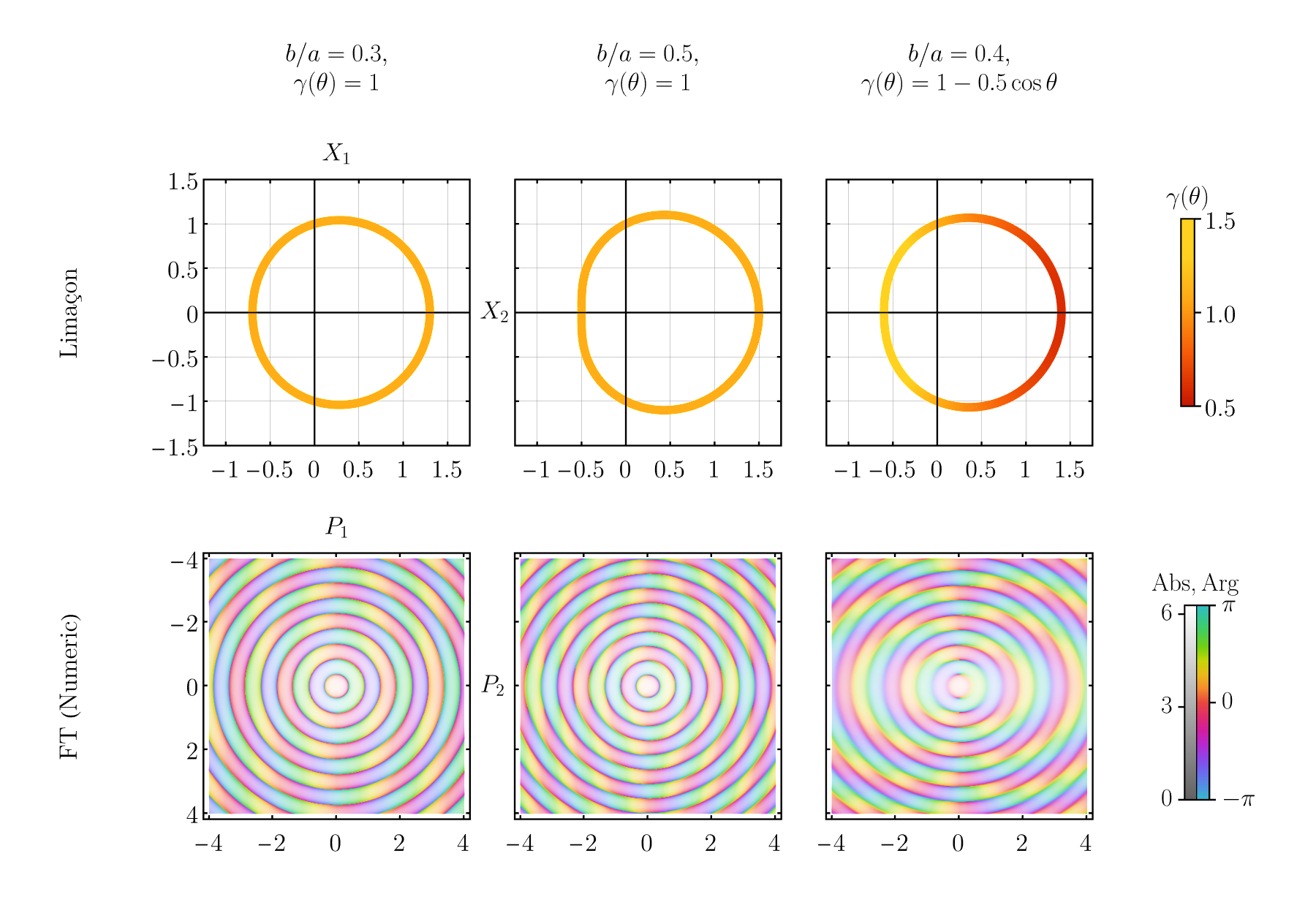}
    \caption{Several examples of limaçons with different values of $b/a$, including one with non-trivial angular variation along the perimeter. The second row shows the 2D Fourier transforms (calculated numerically) of those same limaçons plotted using domain coloring: the color encodes the complex phase of the Fourier transform at a particular point, while the overall brightness (how white a point is) is determined by the absolute value of the Fourier transform at that same point. For instance, near the center all three plots are relatively white indicating a large absolute value, while in the troughs between the concentric `waves' the colors are the most intense (with the least amount of white mixed in that would dull them out), indicating very small absolute value.}
    \label{fig: Limacon examples}
\end{figure*}

\subsubsection{The Limaçon\label{sec: limacon}}

\begin{figure*}
    \centering
    \includegraphics[width=\linewidth]{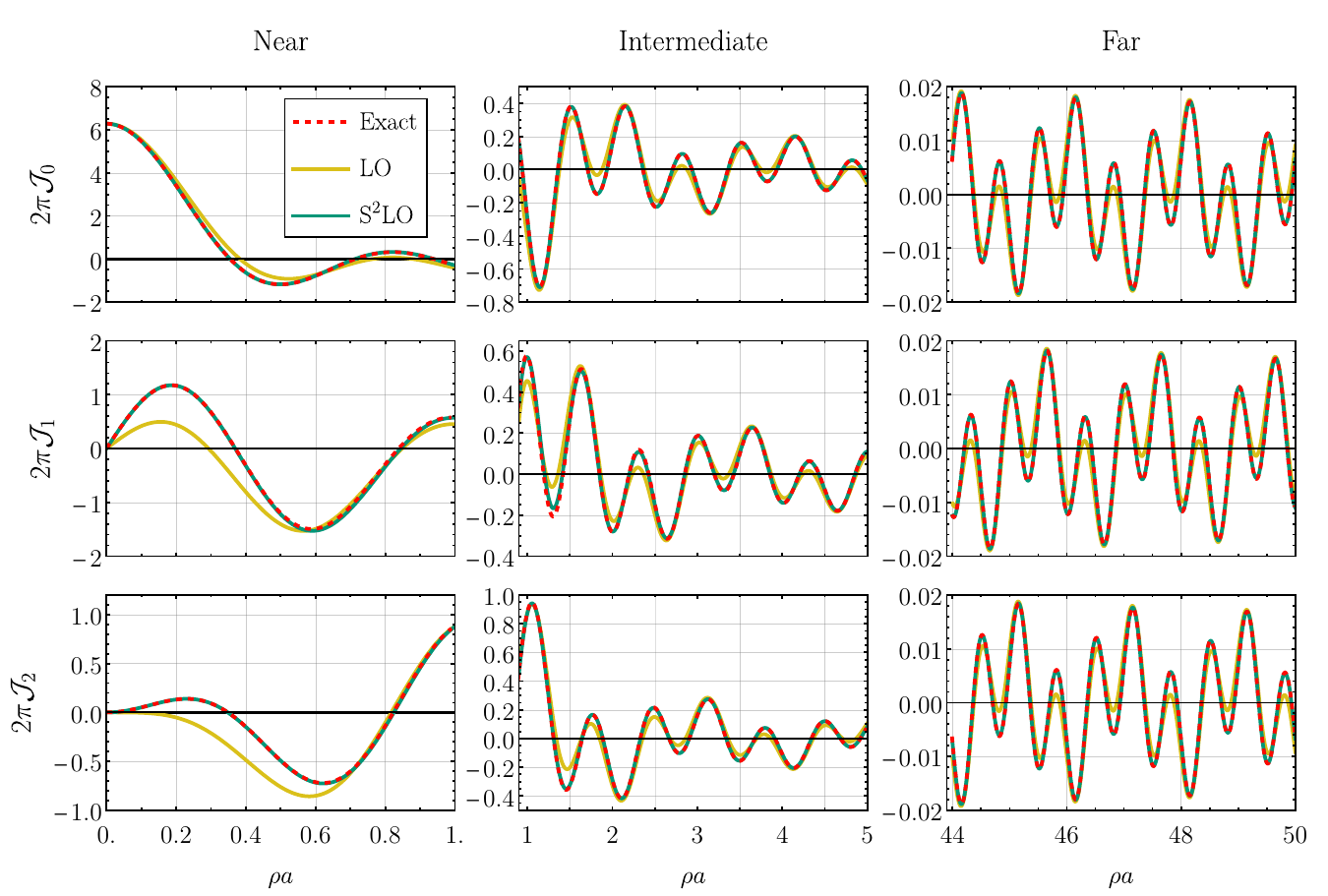}
    \caption{Comparisons of the numerical evaluations of the integrals $\mathcal{J}_0$, $\mathcal{J}_1$, and $\mathcal{J}_2$ to the LO and S$^2$LO focal expansion for the middle limaçon of fig.~(\ref{fig: Limacon examples}) with $b/a=0.5$. The LO provides a serviceable and the S$^2$LO an excellent approximation to all three integrals across the full range of arguments.}
    \label{fig: limacon J0 J1 J2}
\end{figure*}

\begin{figure*}
    \centering
    \includegraphics[width=\linewidth]{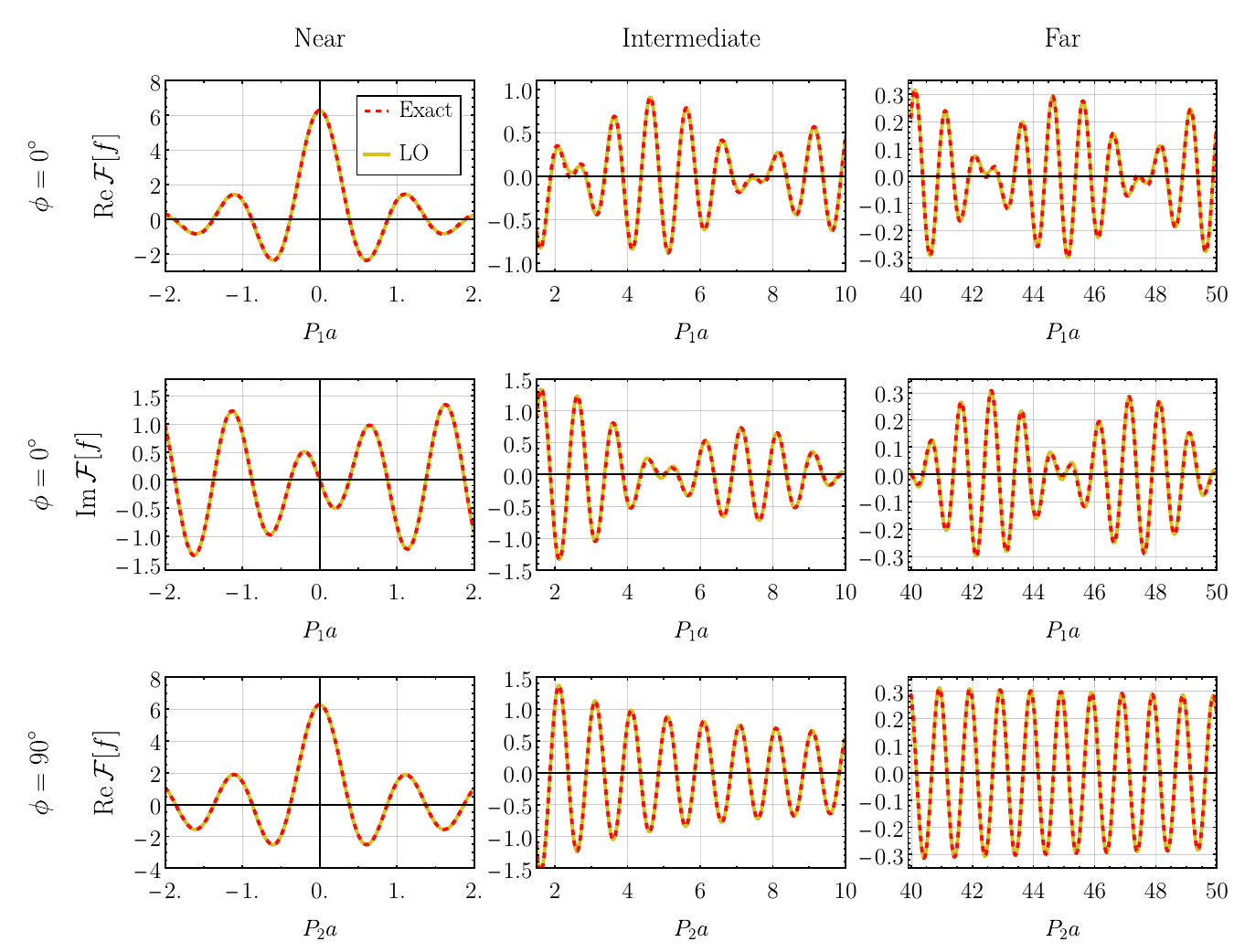}
    \caption{Comparison of the numerically calculated Fourier transform of a limaçon (dashed red) with $b/a=0.1$ to the resummed leading order focal expansion (yellow, given by \cref{eq: limacon final}) along the vertical and horizontal axes ($\phi=90^{\circ}$ and $\phi=0^{\circ}$ respectively). The real and imaginary parts are plotted separately, the imaginary part along $\phi=90^{\circ}$ has been omitted as it vanishes (both numerically, up to precision, and analytically) along that direction. We see excellent agreement among all of the plots.}
    \label{fig: limacon resummed comparison b=0.1}
\end{figure*}

\begin{figure*}
    \centering
    \includegraphics[width=\linewidth]{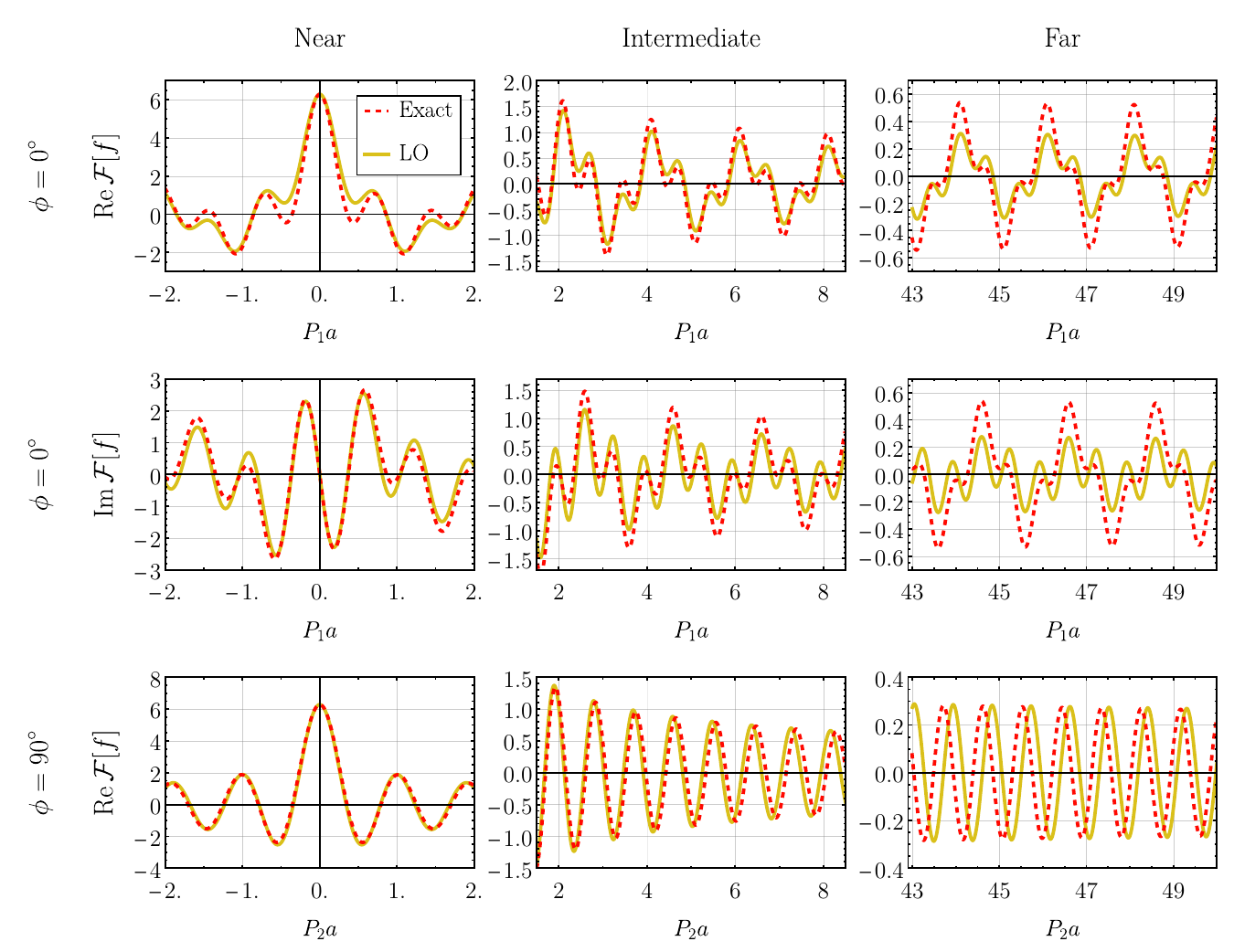}
    \caption{Comparison of the numerically calculated Fourier transform of a limaçon with $b/a=0.5$ (dashed red, corresponding to the middle column of fig.~(\ref{fig: Limacon examples})) to the resummed leading order focal expansion (yellow) along the vertical and horizontal axes ($\phi=90^{\circ}$ and $\phi=0^{\circ}$ respectively). The real and imaginary parts are plotted separately, the imaginary part along $\phi=90^{\circ}$ has been omitted as it vanishes (both numerically, up to precision, and analytically) along that direction. The agreement for large values of $\rho$ along $\phi=0^{\circ}$ is poor, as one would expect from the relatively large $b/a$, but there is no clear evidence of a phase of frequency discrepancy -- most errors appear to be in amplitude modulation. The situation is very different along the $\phi=90^{\circ}$ direction where the amplitude appears to match excellently but there is a slight frequency mismatch between the numeric and focal expressions. We suspect this is due to non-trivial resummations among all angular modes, as there was no sign of this discrepancy in the plots of fig.~(\ref{fig: limacon J0 J1 J2}). This is likely an issue unique to delta function rings as they cannot be accurately approximated by any finite number of angular modes even before taking the Fourier transform.}
    \label{fig: limacon resummed comparison b=0.5}
\end{figure*}

\begin{figure*}
    \centering
    \includegraphics[width=\linewidth]{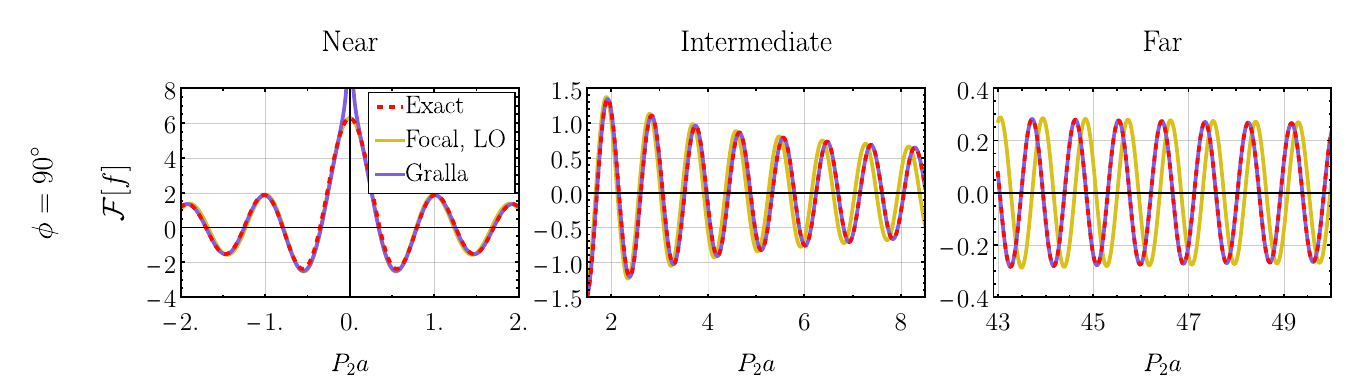}
    \caption{Comparison of the numerically calculated Fourier transform of a limaçon with $b/a=0.5$ (dashed red, corresponding to the middle column of fig.~(\ref{fig: Limacon examples})) along the vertical ($\phi=90^{\circ}$) direction to the resummed leading order focal expansion (yellow, given by \cref{eq: limacon final}) and the Gralla approximation (blue, given by \cref{eq: Gralla}), the last calculated using results from \cite{Gralla}. We see the Gralla approach breaks down near the origin but provides an excellent approximation for larger values of $\rho$. In particular, it does not suffer the frequency discrepancy of the resummed focal expansion.}
    \label{fig: Gralla}
\end{figure*}

The limaçon is the curve given by
\begin{align} \label{eq: limacon}
    R(\theta)=a+b \cos(\theta+\delta)
\end{align}
where $a,b,\delta$ are real constants. We assume $a>0$ and $a/b>1$, the latter condition coming from requiring the limaçon to be a smooth, non-self-intersecting curve \cite{Dimple}. This family of curves corresponds arguably to the simplest function $R(\theta)$ one can write down (beaten only by the circle, which is a special case of \cref{eq: limacon}). It is remarkable then that such a simple function provides an excellent approximation for the apparent shape of the photon ring for spinning black holes, as shown in \cite{Limacon_approx}. For a few examples of limaçons see fig.~(\ref{fig: Limacon examples}), where we have also shown domain coloring plots of their Fourier transforms for reference.

The simplicity of \cref{eq: limacon} in polar coordinates plays into the advantages of the focal expansion. The integral \cref{eq: gnhat angular delta ring} that we must evaluate is now
\begin{align}
\begin{split}
            \widehat{g}_n(\rho)=\frac{e^{-i2\pi \rho a}}{2\pi}\sum_{m\in\mathbb{Z}}&\gamma_m e^{i(n-m)\delta}\\
        \times&\int_0^{2\pi}\text{d}\theta \,e^{i(m-n)\theta-i2\pi \rho b \cos\theta }.
\end{split}
\end{align}
where we continue to assume our limaçon has a non-trivial intensity profile given by \cref{eq: intensity profile decomp}. The integral above is the $(n-m)$th Fourier coefficient of $e^{-i2\pi \rho b \cos\theta }$, but from the Anger-Jacobi expansion \cref{eq: Anger-Jacobi} we immediately see this is simply $(-i)^{m-n}J_{m-n}(2\pi \rho b)$. This means that 
\begin{align}
        \widehat{g}_n(\rho)=&e^{-i2\pi \rho a}\sum_{m\in\mathbb{Z}}\gamma_m e^{i(n-m)(\delta+\pi/2)}
       J_{m-n}(2\pi \rho b).
\end{align}
We see that $a$ should be chosen as the shift parameter (justifying the notation) so that
\begin{align}
    \widehat{G}_n(\rho)=\sum_{m\in\mathbb{Z}}\gamma_m e^{i(n-m)(\delta+\pi/2)}
       J_{m-n}(2\pi \rho b),
\end{align}
meaning that the leading focal order is
\begin{align} \label{eq: limacon mathcalJn}
    \mathcal{J}_n(\rho)\approx &\sum_{m\in \mathbb{Z}}\gamma_m e^{i(n-m)\delta}i^m\left(h_{n,m}+(-1)^mh_{n,m}^* \right)(\rho),
\end{align}
\begin{align}\label{eq: limacon hnm}
    \begin{split}
        &h_{n,m}(\rho)\equiv\\
        &\frac{(-i)^n a}{2}\left(J_n+\frac{i}{2}J_{n-1}-\frac{i}{2}J_{n+1}\right)(2\pi \rho a)\, J_{n-m}(2\pi \rho b),
    \end{split}
\end{align}
where we defined an axillary function $h_{n,m}$ for brevity and where $*$ denotes the complex conjugate. In other words, the angular modes of the Fourier transform of a limaçon will be approximately given by a linear superposition of products of two Bessel $J$ functions. In particular, when the angular dependence is trivial $\gamma(\theta)=1$ and the rotation angle $\delta=0$ vanishes we have
\begin{align}
    \widehat{G}_n(\rho)=(-i)^n J_{n}(2\pi \rho b)
\end{align}
and the leading focal order simplifies to
\begin{align}
    \begin{split}
        \mathcal{J}_{2n}&\big|_{\gamma=1, \delta=0}(\rho)\approx(-1)^{\frac{n}{2}} a\, J_n(2\pi\rho b)\, J_n(2\pi \rho a),\\
        \mathcal{J}_{2n+1}&\big|_{\gamma=1, \delta=0}(\rho)\approx \\   
            &(-1)^{\frac{n-1}{2}}\frac{a}{2}\,J_n(2\pi\rho b)\,\left(J_{n-1}-J_{n+1}\right)(2\pi\rho a).
    \end{split}
\end{align}
Figure~(\ref{fig: limacon J0 J1 J2}) shows the numerics for $\mathcal{J}_0$, $\mathcal{J}_1$, and $\mathcal{J}_2$ of the limaçon seen in the middle column of fig.~(\ref{fig: Limacon examples}) compared to the Focal LO and S$^2$LO. We see that, once again, the leading order provides a serviceable and the subsubleading order an excellent approximation across all of the plots.

Looking at the general case with arbitrary intensity variation, \cref{eq: limacon mathcalJn,eq: limacon hnm}, it appears there is no obvious way to disentangle the angular intensity information (represented by factors of $\gamma_m$) from the information describing the shape of the curve (represented by the constants $a,b,\delta$). This is emblematic of a larger issue; In fact, it would be surprising if the two \textit{could} be separated. For a generic ring-like function $f$ there is no canonical way of separating angular intensity variation and ring shape: if the ring appears to our eyes to have some non-circular shape, it can nevertheless be described as a circular ring with appropriately chosen radial profiles for each of the angular modes $g_n(r)$. This is relevant even for a delta function curve as the radial profiles of its angular modes \cref{eq: gn for delta function ring} will not in general be delta functions themselves. It is true that for many situations it would be abundantly clear for the human observer to separate out angular intensity dependence from shape, like in our limaçon, but the point is that this language is not mathematically natural  and so we should not expect our calculations to yield to it. 

There is, surprisingly, a little more we can do with the limaçon focal expansion. Due to the relative simplicity of the leading order $\mathcal{J}_n$ it is actually possible to analytically perform the sum over $n$ in \cref{eq: summary ang decomp of F[f]}. One must repeatedly use Graf's addition theorem \cite{BesselWatson}:
\begin{align}
    \sum_{k\in\mathbb{Z}}J_{N+k}(x)J_k(y) e^{ik(\varphi+\pi)}=\frac{(x+y e^{-i \varphi})^N}{\omega_{x,y}(\varphi)^N}J_N(\omega_{x,y}(\varphi))
\end{align}
where we define
\begin{align}
    \omega_{x,y}(\varphi)\equiv \sqrt{x^2+y^2+2 xy\cos \varphi}.
\end{align}
Performing the sum over angular modes in \cref{eq: summary ang decomp of F[f]} and simplifying one obtains
\begin{align}\label{eq: limacon final}
\begin{split}
    \mathcal{F}\left[f\right](\rho,\phi)\approx \pi\sum_{m\in \mathbb{Z}}&\gamma_mi^m e^{-im\delta}\\
    \times&\big[\Omega_m(\phi+\delta)+(-1)^m\Omega_m^*(\pi-\phi-\delta)\big]\\
    &\phantom{filler}
\end{split}
\end{align}
where we have defined the auxiliary functions

\begin{align}\label{eq: limacon Omega}
    \begin{split}
        &\Omega_m(\varphi)\equiv \left(\frac{-\omega_{a,b}(\varphi)}{b+a e^{-i \varphi}}\right)^m\times\\&\Bigg[a J_{m}+\frac{i \omega_{a,b}(\varphi)}{2}\Big(J_{m-1}-J_{m+1}\Big)\Bigg]\Big(2\pi \rho \, \omega_{a,b}(\varphi)\Big).
    \end{split}
\end{align}
Note in particular that we have recovered the expected $1/\sqrt{\rho}$ asymptotic decay behavior, even though the individual angular modes all decay as $1/\rho$. This is all thanks to the non-trivial resummation performed by Graf's theorem.

Comparing \cref{eq: limacon final} to numerics one will notice a surprising problem. At first, for small values of $b$, it seems that the focal approximation performs well -- see fig.~(\ref{fig: limacon resummed comparison b=0.1}) where we have plotted the real and imaginary components of the Fourier transform along the directions $\phi=0^{\circ}$ and $\phi=90^{\circ}$ case $b=0.1 a$ from the first column  of fig.~(\ref{fig: Limacon examples}) (the plot of the imaginary component for $\phi=90^{\circ}$ is missing as it vanished identically along this direction). This is not a surprising result as for small $b$ the limaçon can be shown to approximate a shifted circle.

Pushing the resummed leading focal expansion \cref{eq: limacon final} to the larger value of $b=0.5 a$ results in large but somewhat expected amplitude errors along the $\phi=0^{\circ}$ direction -- see the first two rows of fig.~(\ref{fig: limacon resummed comparison b=0.5}). Along the orthogonal $\phi=90^{\circ}$ direction, however, we get an asymptotically catastrophic frequency mismatch: our predicted frequency appears just barely too large, making our peaks lag behind those of the numeric results by an amount increasing with distance from the origin.
This behavior is particularly surprising as we saw no evidence of any frequency errors in any of the plots for the individual angular modes in fig.~(\ref{fig: limacon J0 J1 J2}). We suspect that non-trivial resummations occur among the subleading focal orders to result in this slight frequency shift -- as mentioned above, we expect delta function rings to be unique in that they severely accentuate these types of resummation issues. It would be very interesting to see if one can apply a similar derivation using Graf's theorem to resum higher focal orders over all angular modes and whether doing so removes this frequency discrepancy. We leave this for future work and let this particular angular resummation figure as a cautionary tale for the time being: the focal expansions of sharply peaked functions still work well for individual angular modes, but adding all these modes together is not a trivial matter.

Before we conclude, it is important to note the above mentioned frequency discrepancy does not appear when one uses existing asymptotic techniques. For instance, according to Gralla \cite{Gralla} for $b/a=0.5$ along the $\phi=90^{\circ}$ line we should approximate
\begin{align}\label{eq: Gralla}
    \mathcal{F}\left[f\right](\rho,90^{\circ})\approx \frac{C_G}{\sqrt{\rho a}}\cos\left(\pi d_G \rho a -\pi/4\right)
\end{align}
where one can find the constants to be
\begin{align}
    C_G&=\sqrt[4]{6+\frac{7 \sqrt{3}}{2}}\\
    d_G&=\sqrt{\frac{9}{4}+\frac{3 \sqrt{3}}{2}}.
\end{align}
To find these one needs to solve $0=\frac{d}{d \theta}R(\theta)\sin\theta$, which is analytically tractable in the present simple case but can prove quite difficult in general without numerical methods. The comparison between \cref{eq: Gralla}, the resummed focal expansion, and the numerical result is shown in fig.~(\ref{fig: Gralla}). We see that Gralla's calculation breaks down near the origin, as is typical of asymptotic methods, but provides an excellent approximation for larger arguments. This raises the question of whether this alternative approach can be combined with the focal expansion in some generic way, mitigating the downsides of both approaches in the process. We leave such considerations for future work.

Taken together, the examples in this section support the main practical claims of the focal expansion: for smooth ring-like profiles with up to moderate width-to-radius ratio, the LO approximation is often already accurate over a wide range of $\rho$, and successive subleading orders improve the result without an apparent finite-$\rho$ breakdown. The Gaussian and logarithmic examples also highlight two important implementation details: (i) the extension of $g_n(r)$ to the full real line should not introduce artificial non-smoothness at $r=0$, since such discontinuities can dominate large-$\rho$ behavior, and (ii) the shift parameter need only be chosen within a reasonable neighborhood of the dominant ring radius for the phase to remain correct. Finally, the delta-function examples emphasize that zero-thickness features can require nontrivial resummations over angular modes, and may benefit from hybrid treatments that combine the focal expansion with complementary asymptotic methods such as stationary phase.

\section{Concluding Remarks}
\label{sec:conclusion}

In this work, we derived a new approximation -- the ``focal expansion'' -- for the Fourier transform of radially concentrated functions, focusing on the 2D case of ``ring-like'' images. We demonstrated that it provides accurate analytic approximations to the corresponding 2D Fourier transforms (interferometric visibilities) for several families of images. 
The focal expansion is written in terms of Bessel-type oscillatory factors set primarily by the ring radius, modulated by the one-dimensional Fourier transform of the radial profile and the angular image modes. 
Hence, the focal expansion offers a clean separation between angular and radial information. In many of the examples considered here, truncation of the focal expansion at leading order accurately reproduces the correct global behavior over the full range of spatial frequency, even for non-circular rings and those with fractional thickness of order unity. 

As presented, this method is formulated primarily as a forward-modeling tool: it approximates the Fourier transform of a known, ring-like image model. A natural next step is to develop an inverse (reconstruction) counterpart that maps oscillatory Fourier-domain data to a ring-like image model while retaining the separation of angular and radial structure. Such an inverse formulation could be particularly valuable for interferometric applications where the measured data are naturally sparse samples in Fourier space, especially for space-VLBI missions that sample baselines dominated by a black hole photon ring, such as the Black Hole Explorer \cite{BHEX_Concept,BHEX_PhotonRing}.

On the analytic side, several questions remain open. Establishing convergence criteria and practical error estimates would clarify the domain of validity of the approximation and strengthen its reliability. In particular, Section~\ref{sec: log rings against all odds} shows unexpectedly good numerical performance in a regime where the assumptions used in the derivation appear to be strained, while the worm-like nonuniform breakdown illustrated in fig.~(\ref{fig: log example 1 breakdown}) is atypical compared to standard truncation errors and may point to a distinct mechanism of failure. Appendix~\ref{app: counterexamples} further indicates that straightforward generalizations to non-Bessel kernels can perform poorly, motivating modified kernels or resummations that preserve the favorable uniform behavior observed in Bessel-type cases.

Several extensions appear especially promising. First, specializing the framework to half-integer-order Bessel kernels (i.e., spherical Bessel functions) could simplify implementations in odd spatial dimensions, as discussed in Appendix~\ref{app: higher dim}. Second, the apparent leading-order frequency offset observed for the lima\c{c}on example in Sec.~\ref{sec: limacon} suggests that related reorganizations may apply to more general oscillatory integrals on $S^1$; understanding this effect may open a route to uniformly accurate approximations for generalized Bessel-type integrals.

In short, we expect the focal expansion to provide a practical analytic tool for modeling and interpreting Fourier-domain signatures of ring-like sources, especially for studies of black holes with radio interferometry, and we hope it stimulates further development of uniform approximation methods for oscillatory integrals in broader settings.

\begin{acknowledgments}

We wish to thank Samuel Buckley-Bonanno for helpful discussion and feedback and Serhii Kryhin for suggesting the integral in the example of section~\ref{section: peak at origin}. We acknowledge financial support from the National Science Foundation (AST-2307887). This publication is funded in part by the Gordon and Betty Moore Foundation, Grant GBMF12987. This work was supported by the Black Hole Initiative, which is funded by grants from the John Templeton Foundation (Grant \#62286) and the Gordon and Betty Moore Foundation (Grant GBMF-8273) - although the opinions expressed in this work are those of the authors and do not necessarily reflect the views of these Foundations.
\end{acknowledgments}

\appendix
\section{Angular Decomposition in Higher Dimensions \label{app: higher dim}}

Using our methods for higher dimensional Fourier transforms necessitates the use of the generalization of spherical harmonics to an arbitrary number $N$ of dimensions. For an elementary introduction see \cite{Spherical_Harmonics_in_higher_dim}, we only summarize the necessary details.

Let $\Vec{X}\in \mathbb{R}^N$ and $\hat{r}\equiv \Vec{X}/r$. A spherical harmonic $Y_{n,l}(\hat{r})$ of degree $n\in\mathbb{N}$ in $N$ dimensions is a restriction to the unit sphere $S^{N-1}$ of a polynomial $H_{n,l}(\Vec{X})$ in the variables $X_1,\cdots,X_N$ such that
\begin{align}
    H_{n,l}(\lambda \Vec{X})=\lambda^n H_{n,l}(\Vec{X})
\end{align}
for any $\lambda\in\mathbb{R}$ and
\begin{align}
    \Delta_N H_{n,l}=0
\end{align}
where $\Delta_N$ is the $N$-dimensional Laplacian. For a fixed $n$ there are only 
\begin{align}
    \frac{2n+N-2}{n}\binom{n+N-3}{n-1}
\end{align}
spherical harmonics with that degree and so the allowed values for the index $l$ come from a finite set. For $n\neq m$ and any two $l,l'$ the spherical harmonics $Y_{n,l}$, $Y_{m,l'}$ are naturally orthogonal when integrated over the sphere $S^{N-1}$ with the standard measure $d\Omega_{N-1}$. For fixed $n$ one can then orthonormalize over the finite set of $l$ values to ensure any two spherical harmonics are orthonormal. 

Having set all that up, the set of all spherical harmonics (for all values of $n$ and $l$) forms an orthonormal basis for functions on the unit sphere $S^{N-1}$. This means that any function $f(\Vec{X})$ can be expanded as
\begin{align}\label{eq: higher dim f decomposition}
    f(\Vec{X})=\sum_{n=0}^{\infty}\sum_l g_{n,l}(r)Y_{n,l}(\hat{r}).
\end{align}
Using this expansion in the Fourier transform of $f$ written in generalized spherical coordinates yields
\begin{align}\label{eq: higher dim Ftr raw}
    \begin{split}
        \mathcal{F}\left[f\right](\Vec{P})=&\sum_{n,l}\int_0^{\infty}\text{d}r\,r^{N-1}g_{n,l}(r)\\
        \times&\int_{S^{N-1}} \text{d}\Omega_{N-1}(\hat{r})\,Y_{n,l}(\hat{r})e^{-2\pi i \Vec{X}\cdot \Vec{P}}.
    \end{split}
\end{align}
Now, note that the Jacobi-Anger expansion \cref{eq: Anger-Jacobi} is symmetric under $\theta\rightarrow -\theta$ and so it can be written as 
\begin{align}
    \begin{split}
        e^{iz\cos\theta}=&\sum_{m\in \mathbb{Z}}i^mJ_m(z)\cos(m\theta)\\
        =&\sum_{m\in \mathbb{N}}(2-\delta_{m,0})i^mJ_m(z)T_{m}\left(\cos\theta\right)
    \end{split}
\end{align}
where the $T_m$ are Chebyshev polynomials of the first kind. Using this in \cref{eq: higher dim Ftr raw} yields
\begin{align}\label{eq: higher dim Ftr raw Chebyshev}
    \begin{split}
        \mathcal{F}\left[f\right](\Vec{P})=&\sum_{n,m,l}(-i)^m\int_0^{\infty}\text{d}r\,r^{N-1}g_{n,l}(r)J_m(2\pi r\rho)\\
        \times&(2-\delta_{m,0})\int_{S^{N-1}} \text{d}\Omega_{N-1}(\hat{r})\,Y_{n,l}(\hat{r})T_{
        m}\left(\hat{r}\cdot \hat{\rho}\right)
    \end{split}
\end{align}
where $\hat{\rho}\equiv \Vec{P}/\rho$ (recall that $\rho\equiv ||\Vec{P}||$). One must now generalize the family of Legendre polynomials (figuring prominently in three dimensional spherical analysis) to an arbitrary number of dimensions. We call this family $P_{n}^{(N)}$. These are more commonly known, after the rescaling
\begin{align}
    C_n^{\left(\frac{N-2}{2}\right)}(t)=\binom{n+N-3}{n}P_{n}^{(N)}(t),
\end{align}
as Gegenbauer or ultraspherical polynomials \cite{NIST,abramowitz1965handbook} ($t$ here is a generic variable). With these polynomials one can now transform the spherical integral in \cref{eq: higher dim Ftr raw Chebyshev} using the Hecke-Funk Theorem, which states that for any unit vector $\hat{k}$ and an arbitrary function $h:\mathbb{R}\rightarrow \mathbb{R}$
\begin{align}\label{eq: Hecke Funk}
   \begin{split}
        \int_{S^{N-1}}&\text{d}\Omega_{N-1}(\hat{r})\,h(\hat{r}\cdot \hat{\rho})Y_{n,l}(\hat{r})\\
        =&\Omega_{N-2}Y_{n,l}(\hat{\rho})\int_{-1}^1\text{d}t\, h(t) P_{n}^{(N)}(t)\, (1-t^2)^{\frac{N-3}{2}}
   \end{split}
\end{align}
where 
\begin{align}
    \Omega_{N-2}=\frac{2\pi^{\frac{N-1}{2}}}{\Gamma\left(\frac{N-1}{2}\right)}
\end{align}
is the area of $S^{N-2}$. We see that to apply \cref{eq: Hecke Funk} in \cref{eq: higher dim Ftr raw Chebyshev} we simply replace the function $f$ by the Chebyshev $T_m$. We must therefore evaluate the integrals
\begin{align}
    \eta_{m,n}^{(N)}\equiv \int_{-1}^1\text{d}t\,T_{m}(t)P_{n}^{(N)}(t)\,(1-t^2)^{\frac{N-3}{2}}.
\end{align}
The generalized Legendre polynomials are actually orthonormal over the measure $(1-t^2)^{\frac{N-3}{2}}\,\text{d}t$ and interval $t\in[-1,1]$ so if we could expand a generic Chebyshev $T_m$ in terms of the ultraspherical polynomials we would be essentially done. The key is to note the following properties of ultraspherical polynomials \cite{NIST}:
\begin{align}
    C^{(\mu)}_n(t)=\sum_{l=0}^{\lfloor n/2 \rfloor} \frac{\lambda+n-2l}{\lambda}\frac{(\mu)^{\uparrow}_{n-l} (\mu-\lambda)^{\uparrow}_l}{(\lambda+1)^{\uparrow}_{n-l}l!}C^{(\lambda)}_{n-2l}(t),
\end{align}
\begin{align}
    \lim_{\mu\rightarrow 0}\frac{n+\mu}{\mu}C^{(\mu)}_n(t)=2T_n(t)-\delta_{n,0},
\end{align}
\begin{align}
    \int_{-1}^{1}\text{d}t\, \left[C^{(\lambda)}_n(t)\right]^2 (1-t^2)^{\lambda-1/2}=\frac{2^{1-2\lambda}\pi \Gamma(n+2\lambda)}{(n+\lambda)\Gamma(\lambda)^2 n!}.
\end{align}
Using these one can show that
\begin{align}\label{eq: higher dim eta p final}
    \begin{split}
        m=&0: \qquad \eta_{0,n}^{(N)}=\delta_{0,n} 2^{3-N}\pi\frac{\Gamma(N-2)}{\Gamma\left(\frac{N-2}{2}\right)\Gamma\left(\frac{N}{2}\right)}\\
        m>&0: \qquad \eta_{m,n}^{(N)}=4\pi m\, \delta_{m,n}^{\text{mod }2} \frac{(N-3)!}{2^N \Gamma\left(\frac{N}{2}-1\right)^2 }\\
        &\qquad \qquad \qquad \times\frac{\left(\frac{m+n}{2}-1\right)! \left(1-\frac{N}{2}\right)_{\frac{m-n}{2}}^{\uparrow}}{\left(\frac{m-n}{2}\right)! \left(\frac{N}{2}-1\right)^{\uparrow}_{\frac{m+n}{2}+1}}
    \end{split}
\end{align}
where we used $\delta^{\text{mod} 2}_{m,n}$ to indicate the expression vanishes unless $m,n$ are of the same parity. Note that due to the factor of $\left(\frac{m-n}{2}\right)!$ in the denominator, $\eta^{(N)}_{m,n}$ vanishes unless $n\leq |m|$. 

Bringing \cref{eq: higher dim Ftr raw Chebyshev,eq: Hecke Funk,eq: higher dim eta p final} together we see that the Fourier transform is
\begin{align}\label{eq: higher dim Ftr final}
    \begin{split}
        \mathcal{F}\left[f\right](\Vec{P})=&\Omega_{N-2}\sum_{m\in\mathbb{N}}\sum_{n=0}^{m}\sum_l  (-i)^m \eta_{m,n}^{(N)} Y_{n,l}(\hat{\rho})\\
        &\times(2-\delta_{m,0}) \int_0^{\infty}\text{d}r\,r^{N-1}g_{n,l}(r)J_m(2\pi r\rho)
    \end{split}
\end{align}
where we accounted for the fact that $\eta_{m,n}^{(N)}$ vanishes for $n>|m|$ by adding summation limits. The remaining Bessel integral in \cref{eq: higher dim Ftr final} can then be approximated using precisely the same methods as derived in section~\ref{sec: The Method} and summarized in section~\ref{subsec: summary}. The higher dimensionality simply manifests as a different power of $r$ in the integrands (which our method can handle without difficulty) and the fact that the sum over angular modes is more complicated, accounting for the extra spherical degrees of freedom in higher dimensions.

Note also that for $N=2$ one must understand \cref{eq: higher dim eta p final} in a limiting sense, doing so one will obtain
\begin{align}
    \Omega_0\,\eta^{(2)}_{m,n}=\pi \delta_{m,n}(1+\delta_{0,n}).
\end{align}
Noting then that for $N=2$ we have $Y_{n,l}=e^{i n l \phi}$ with $\{l\}={+1,-1}$ for $n>0$ and $\{l\}=1$ for $n=0$, we recover the expression \cref{f hat decomposed into sums of integrals} derived in the main text.

Note, however, that the form of \cref{eq: higher dim Ftr final} in $N>2$ dimensions is somewhat unwieldy. In practical applications one will have to truncate the $m$ sum and while in $N=2$ the terms with fixed $m$ correspond neatly with the $m$th order angular modes, the same is not true in higher dimensions: note for instance that the $n=0$ angular mode $Y_{0}$ contributions come in at \textit{every} $m$ order. Unless the coefficient functions $g_{n,l}$ decay rapidly with growing $n$ this means one might encounter non-trivial resummations in this approach even at leading angular mode order. The rapid decay condition should be true of smooth functions encountered in practice but one must be mindful of this possible pitfall when applying the focal expansion for higher dimensional Fourier transforms.

There is an alternative to \cref{eq: higher dim Ftr final} but it comes at a cost. We could skip the Jacobi-Anger step and apply the Hecke-Funk theorem directly to \cref{eq: higher dim Ftr raw}, this will yield
\begin{align}\label{eq: higher dim Ftr raw alt}
    \begin{split}
        \mathcal{F}\left[f\right](\Vec{P})=&\Omega_{N-2}\sum_{n,l}Y_{n,l}(\hat{\rho})\int_0^{\infty}\text{d}r\,r^{N-1}g_{n,l}(r)\\
        \times&\int_{-1}^1\text{d}t\, e^{-2\pi i \rho r t}P_n^{(N)}(t)\, (1-t^2)^{\frac{N-3}{2}}.
    \end{split}
\end{align}
Surprisingly, the integral over $t$ can be evaluated using the identity \cite{NIST}
\begin{align}
    \int_{-1}^1\text{d}t\, e^{i\omega t} C_{n}^{(\lambda)}(t)\, (1-t^2)^{\lambda-\frac{1}{2}}=\frac{2\pi i^n \Gamma(n+2\lambda)}{n! 2^{\lambda} \Gamma(\lambda)}\frac{J_{n+\lambda}(\omega)}{\omega^{\lambda}}
\end{align}
which plays the role of Jacobi-Anger in introducing Bessel functions into our analysis. In the end we get
\begin{align}\label{eq: higher dim Ftr final alt}
    \begin{split}
        \mathcal{F}\left[f\right](\Vec{P})=&\frac{2^{3-N}\pi\Omega_{N-2}}{(\pi\rho)^{\frac{N}{2}-1}}\frac{(N-3)!}{\Gamma\left(\frac{N}{2}-1\right)}\sum_{n,l}(-i)^n Y_{n,l}(\hat{\rho})\\
        &\times\int_0^{\infty}\text{d}r\,r^{\frac{N}{2}}g_{n,l}(r) J_{n-1+N/2}(2\pi \rho r).
    \end{split}
\end{align}
In even dimensions, \cref{eq: higher dim Ftr final alt} is often preferable to \cref{eq: higher dim Ftr final} because the angular modes are cleanly separated and the remaining integrals are all integer-order Bessel integrals precisely of the kind we have studied in this paper with the focal expansion. In odd dimensions, however, we get integrals over half-integer order Bessel functions which we have not thoroughly studied. That is not to say we necessarily expect the focal expansion to fail for half-integer order Bessel kernels, but the analysis of section~\ref{sec: Bessel function Kernels} would need to be repeated for this new case. In addition, some comparisons to numerics and simple analytically workable examples (in the spirit of section~\ref{sec: examples}) would need to be done to gain confidence in the convergence properties and behavior of our approach for this class of kernels. We leave this avenue for future work.

\section{Solving the General Focal Recurrence} \label{app: solving the general focal recurrence}

In this appendix we investigate ways to solve the recurrence relations \cref{eq: rec A-,eq: rec A+} for infinite subsets of the $A_{l,n}^{\pm}$ coefficients, instead of the iterative approach summarized at the end of section \ref{subsec: The Focal Expansion}. We first define the symbol
\begin{align}
    x_{\pm}=\begin{cases}
        x,\quad &\text{for } +,\\
        x+1,\quad &\text{for } -,
    \end{cases}
\end{align}
for any quantity $x$ (this subscript is not to be confused with the superscript, e.g. $A^{\pm}$, corresponding to the even/odd parts of the focal expansion \cref{eq: Focal Expansion}) and use this to rewrite the recurrence relations as 
\begin{align} \label{eq: rec main}
       X^{\pm}_{k,m}=&\sum^{m-k_{\pm}}_{n=0}\sum_{l=0}^{k}  A_{l,n}^{\pm} M^{\pm}_{l,n,k,m}
\end{align}
where we have also defined 
\begin{align}
\begin{split} \label{eq: M def}
    M^{\pm}_{l,n,k,m}=&\alpha_{l+n+m-k}(2\pi a)^l 2^{2(n+l)+0_{\pm}}\\
    &\times (k)^{\downarrow}_l (l+n+m-k)^{\downarrow}_{2n+l+0_{\pm}},
\end{split}\\
\text{\phantom{filler}}\nonumber\\
     X^{\pm}_{k,m}=&(-1)^k i^{0_{\pm}} \alpha_m (2m)^{\downarrow}_{2k+0_{\pm}}.
\end{align}
Solving the full recurrence for all of the coefficients at once remains daunting but it is possible to find $A_{l,n}^{\pm}$ column-by-column or row-by-row. We follow the column-by-column approach for a reason that will become clear shortly.

Assume we have found the first $c-1$ columns so that $A_{l,n}$ with $n\leq c-1$ is known. Now take $m=k_{\pm}+c$. We define
\begin{align} \label{eq: Z def}
\begin{split}
        &Y^{\pm}_k(c)\equiv X^{\pm}_{k,k_{\pm}+c}-\sum_{n=0}^{c-1}\sum_{l=0}^k A^{\pm}_{l,n}M^{\pm}_{l,n,k,k_{\pm}+c},\\
    &N^{\pm}_{k,l}(c)\equiv M^{\pm}_{l,c,k,k_{\pm}+c},\quad B^{\pm}_l(c)\equiv A^{\pm}_{l,c}.
\end{split}
\end{align}
With this \cref{eq: rec main} becomes
\begin{align} \label{eq: vec finally 2}
        \begin{split}
            Y^{\pm}_{k}(c)=&\sum_{l=0}^k N^{\pm}_{k,l}(c) B^{\pm}_l(c),\\
        \Vec{Y}^{\pm}(c)=&\mathbf{N}^{\pm}(c) \Vec{B}^{\pm}(c),
        \end{split}
\end{align}
where we naturally extended $N^{\pm}_{k,l}\left(c\right)$ into an infinite lower-triangular matrix $\mathbf{N}(c)$ and $Y^{\pm}_k\left(c\right), B^{\pm}_l\left(c\right)$ into infinite vectors $\Vec{Y}^{\pm}\left(c\right), \Vec{B}^{\pm}\left(c\right)$. The vector $\Vec{Y}^{\pm}(c)$ is known by assumption so to find the $A^{\pm}_{l,c}=B^{\pm}_l(c)$ we must calculate the matrix inverse $\left[\mathbf{N}^{\pm}(c)\right]^{-1}$: we have thus reduced our problem to finding inverses of arbitrary infinite lower-triangular matrices. 

\begin{figure}[ht] 
\centering
  \begin{minipage}[c]{0.42\linewidth}
    \[
      \quad\mathbf{T} =
      \begin{bmatrix}
        \textcolor{red}{T_{11}} & 0 & 0 & 0\\
        \textcolor{blue}{T_{21}} & \textcolor{red}{T_{22}} & 0 & 0\\
        \textcolor{blue}{T_{31}} & \textcolor{blue}{T_{32}} & \textcolor{red}{T_{33}} & 0\\
        \textcolor{blue}{T_{41}} & \textcolor{blue}{T_{42}} & \textcolor{blue}{T_{43}} & \textcolor{red}{T_{44}}
      \end{bmatrix} \mathlarger{\mathlarger{\Rightarrow}}
    \]
  \end{minipage}\hfill
\begin{minipage}[c]{0.55\linewidth}
\begin{tikzpicture}[baseline=(current bounding box.center)]
\node[mynode](R4) at (0,0){$\scriptstyle{R_4}$};
\node[mynode](R3) at (0,-1cm){$\scriptstyle{R_3}$};
\node[mynode](R2) at (0,-2cm){$\scriptstyle{R_2}$};
\node[mynode](R1) at (0,-3cm){$\scriptstyle{R_1}$};
\node[mynode](C4) at (2cm,0){$\scriptstyle{C_4}$};
\node[mynode](C3) at (2cm,-1cm){$\scriptstyle{C_3}$};
\node[mynode](C2) at (2cm,-2cm){$\scriptstyle{C_2}$};
\node[mynode](C1) at (2cm,-3cm){$\scriptstyle{C_1}$};
\draw[red, thick] (R4)-- node[above]{$\mathsmaller{1/T_{44}}$}(C4);
\draw[blue, thick, dashed] (R4)-- node[above, xshift=.1cm]{$\mathsmaller{T_{43}}$}(C3);
\draw[blue, thick, dashed] (R4)-- (C2);
\draw[blue, thick, dashed] (R4)-- (C1);
\draw[red, thick] (R3)-- (C3);
\draw[blue, thick, dashed] (R3)-- (C2);
\draw[blue, thick, dashed] (R3)-- (C1);
\draw[red, thick] (R2)-- (C2);
\draw[blue, thick, dashed] (R2)-- node[below, xshift=-.1cm,]{$\mathsmaller{T_{21}}$} (C1);
\draw[red, thick] (R1)-- node[below]{$\mathsmaller{1/T_{11}}$} (C1);
\end{tikzpicture}
\end{minipage}
\caption{A $4\times 4$ lower-triangular matrix $\mathbf{T}$ and its graph $G(T)$ with several edge weights listed.}
\label{fig: matrix graph}
\end{figure}
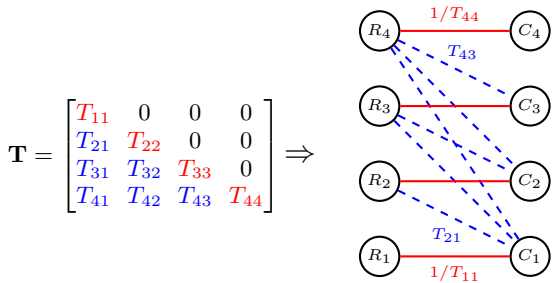

We now describe closed form expressions for our matrix inverses, following closely section two of \cite{TriInverse}. Consider first a finite dimensional, $p\times p$ lower-triangular matrix $\mathbf{T}$ with entries $T_{ij}$ and non-zero diagonal entries (a sufficient and necessary condition for the existence of $\mathbf{T}^{-1}$). We can model $\mathbf{T}$ as a graph $G(T)$ with vertices
$$\{R_1,R_2,\cdots,R_p\}\cup \{C_1,C_2\cdots,C_p\}$$
representing the rows and columns respectively and with edges only between pairs $\{R_i, C_j\}$ with $j\leq i$, corresponding to the entries $T_{ij}$ which do not vanish in general -- see fig.~ (\ref{fig: matrix graph}). We single out the edges corresponding to the diagonal entries $D\equiv \{\{R_i,C_i\}|\;1\leq i\leq p\}$ (marked in red on fig.~(\ref{fig: matrix graph})), and define the \textit{weight} of an edge $E_{ij}=\{R_i, C_j\}$ to be $1/T_{ii}$ if $E_{ij}\in D$ and $T_{ij}$ otherwise.

Now, we define a path $P$ on a graph to be a sequence of vertices with every consecutive pair connected by an edge and where each vertex appears only once. The weight of a path $w(P)$ is then the product of the weights of all the individual edges that make up $P$. Finally, define $\mathcal{P}_{ij}$ to be the set of all paths connecting $C_i$ and $R_j$ such that every other edge, as well as the first and last ones, lie in the diagonal subset $D$ (we allow single edge between $C_i$ and $R_i$). With this elaborate setup an elementary proof by induction in matrix size (see theorem 1 in \cite{TriInverse}) shows that the inverse of $\mathbf{T}$ is given by
\begin{align} \label{eq: inverse graph}
     T^{-1}_{ij}=\sum_{P\in \mathcal{P}_{ij}}(-1)^{(l(P)-1)/2}w(P)
\end{align}
where $l(P)$ is the number of edges in the path $P$ and $\mathcal{P}_{ij}$ is non-empty only when $j\leq i$. To get an explicit expression in terms of matrix elements $T_{ij}$ note that we may intuitively think of $G(T)$ as a ladder -- the rungs are the diagonal-element edges $D$ (solid red lines in fig.~(\ref{fig: matrix graph})). The paths in $\mathcal{P}_{ij}$ start at $C_i$ on the right side of the $i$-th rung and are forced to then move to the $R_i$ on left by our definition. However, all non-$D$ edges of $R_i$ (dashed blue lines in fig.~(\ref{fig: matrix graph})) connect to the right sides of strictly lower rungs in the ladder--  meaning we must move downwards to the right and the process repeats. We can therefore categorize the paths in $\mathcal{P}_{ij}$ by the number and size of strictly downwards steps taken while moving from rung $i$ to rung $j$. In the end this reasoning lets us write \cref{eq: inverse graph} explicitly as\footnote{We have not found \cref{eq: inverse explicit} written in this form; it follows straightforwardly from \cite{TriInverse}. The relative obscurity of this explicit formula might be owed to the fact that evaluating inverses of finite lower-triangular matrices is easily done iteratively via the method of forward substitution.}
\begin{align} \label{eq: inverse explicit}
    T^{-1}_{ij}=\sum_{u=0}^{i-j}\quad\sum_{j=k_0<k_1<\cdots <k_u=i}(-1)^u \frac{\prod_{v=0}^{u-1}T_{k_{v+1}k_v}}{\prod_{v=0}^{u}T_{k_v k_v}},
\end{align}
with the first sum vanishing whenever $i<j$ and the second vanishing whenever $u=0$ and $i-j\neq 0$. This expression must extend to infinite lower-triangular matrices as well. First note that the space of lower-triangular matrices of fixed size (infinite or not) is closed under inverses and products. In particular products are always well defined: every entry in a product matrix is given by a sum with only finitely many non-zero terms so there are no issues of convergence. One can then see that for lower-triangular $\mathbf{T}$ the square submatrices of $\mathbf{T},\mathbf{T}^{-1}$ stretching from $(i,j)=(0,0)$ to some $(i,j)=(n,n)$ must be inverses of each other, this means that \cref{eq: inverse explicit} must apply to all such submatrices of $\mathbf{T}$ and so also to the whole infinite, parent matrix as well. In the intuitive picture of fig.~(\ref{fig: matrix graph}) all this simply corresponds to the fact that we can only take steps down the ladder, but now the ladder extends infinitely high.

We can finally begin solving \cref{eq: vec finally 2}. Inserting the definitions \cref{eq: Z def,eq: M def} into \cref{eq: inverse explicit} and simplifying we get
\begin{align}
\begin{split} \label{eq: N inv explicit}
        &\left[\mathbf{N}^{\pm}(c)\right]^{-1}_{ij}=\frac{1}{\alpha_{i_{\pm}+2c}(8\pi a)^{i}2^{(2c)_{\pm}}\, j!\, (i_{\pm}+2c)!}\\
   \quad &\times \sum_{u=0}^{i-j}\quad\sum_{j=k_0<k_1<\cdots <k_u=i}\;\;\;\;\prod_{v=0}^{u-1}\frac{(-1)^u}{(k_{v+1}-k_v)!}.
\end{split}
\end{align}
Note that the sums of \cref{eq: inverse explicit} have reduced to an $\alpha$ independent quantity that we can evaluate once and for all -- essentially due to the fact that in \cref{eq: M def} for $m=k+c_{\pm}$ the alpha factors lose all $k$ dependence. This is also the reason for taking the column-by-column approach -- one can show that for the row-by-row version of \cref{eq: N inv explicit} non-trivial dependence on $\alpha_i$ remains within the sums.

The remaining sum in \cref{eq: N inv explicit} can be evaluated as an identity involving Stirling numbers of the second kind \cite{NIST} but we present a short proof using generating functions for completeness. First define the variables $d_i\equiv k_i-k_{i-1}$ and note
\begin{align}
    \begin{split} \label{eq: evaluating the sum 1}
        &\sum_{u=0}^{i-j}\quad\sum_{j=k_0<k_1<\cdots <k_u=i}\;\;\;\;\prod_{v=0}^{u-1}\frac{(-1)^u}{(k_{v+1}-k_v)!}\\
        &= \sum_{u=0}^{i-j}(-1)^u \sum_{\substack{d_1+\cdots+d_u=(i-j) \\ d_m\geq 1}}\;\;\frac{1}{d_1!\, \cdots d_u!}.
    \end{split}
\end{align}
The second sum of the last line in \cref{eq: evaluating the sum 1} gives precisely the coefficient multiplying $x^{(i-j)}$ in the Taylor expansion of $\left(e^x-1\right)^u$, where the $-1$ serves to enforce our $d_m\geq 1$. Note also that for $u> (i-j)$ the relevant coefficient is always zero, meaning we can extend our sum over $u$ to all natural numbers.  We therefore seek the $x^{(i-j)}$ coefficient in the Taylor expansion of
\begin{align}
    \sum_{u\geq 0} (-1)^u (e^x-1)^u = \frac{1}{1-(1-e^x)}=e^{-x}
\end{align}
which is simply $(-1)^{(i-j)}/(i-j) !$. With this we can write \cref{eq: N inv explicit} as
\begin{align}
    \begin{split} \label{eq: N inv explicit final}
        &\left[\mathbf{N}^{\pm}(c)\right]^{-1}_{ij}=\frac{(-1)^{i-j}}{2^{(2c)_{\pm}}\,\alpha_{i_{\pm}+2c}(i_{\pm}+2c)!\,(8\pi a)^{i}\, j!\,  (i-j)!}.
\end{split}
\end{align}

We are finally ready to find the focal coefficients. From \cref{eq: Z def} we see that we must initialize the procedure at $c=0$ with 
\begin{align}
    Y^{\pm}_k(0)=&(-1)^k i^{0_{\pm}}(2(k_{\pm}))!\; \alpha_{k_{\pm}}.
\end{align}
Doing so and using \cref{eq: N inv explicit final} immediately lets us calculate $B_l^{\pm}(0)=A^{\pm}_{l,0}$, which in turn allows us to calculate $Y_l^{\pm}(1)$ and then $B_l^{\pm}(1)=A^{\pm}_{l,1}$, and so on, proceeding column by column. The results for the first two columns are:
\begin{align} \label{eq: A0 exact}
    A_{l,0}^{\pm}=&\frac{(-1)^l\; i^{\,0_{\pm}}}{(8\pi a)^l\; 2^{0_{\pm}}\, (l_{\pm})!}\sum_{j=0}^l \frac{(2(j_{\pm}))!}{j!\, (l-j)!}\frac{\alpha_{j_{\pm}}}{\alpha_{l_{\pm}}},\\
    \begin{split} \label{eq: A1 exact}
        A_{l,1}^{\pm}=&-\frac{A_{l,0}^{\pm}}{4 (l_{\pm}+2)}\frac{\alpha_{l_{\pm}+1}}{\alpha_{l_{\pm}+2}}\\
    +&\frac{(-1)^l\, i^{\,0_{\pm}}}{2^{2_{\pm}}(2_{\pm})!(8 \pi a)^l (l_{\pm}+2)!}\sum_{j=0}^l\frac{(2(j_{\pm})+2)!}{j! (l-j)!}\frac{\alpha_{j_{\pm}+1}}{\alpha_{l_{\pm}+2}}.
    \end{split}
\end{align}
It's elementary but increasingly tedious to find analytic expressions for higher order columns. It might be possible to extend this reasoning inductively to obtain a closed form expression for all of the columns at once, but we do not pursue this further here. We note, however, that \cref{eq: A0 exact,eq: A1 exact} agree with the values listed in \cref{eq: A iterative sol example} obtained via the more rudimentary entry-by-entry method. We also get expressions in agreement with \cref{eq: A Bessel} for the explicit case of $K=J_0$ after evaluating the sums with Sister Celine's method \cite{AB}.

Note also that \cref{eq: A0 exact,eq: A1 exact} require that none of the $\alpha_k$'s vanish; this embodies the invertibility assumption we put on $\mathbf{N}^{\pm}(c)$, the $l$-th diagonal element of which is proportional to $\alpha_{2c+l_{\pm}}$. We also directly see why the present matching procedure cannot immediately be applied to higher order Bessel function kernels, as in those cases some $\alpha_k$ vanish.

\section{Alternative Form for Bessel Kernels \label{app: alt form of bessel kernels}}

In this appendix we rewrite the Bessel function sums of \cref{eq: focal expansion even bessel,eq: focal expansion odd bessel} to make both small and large $\rho$ behavior of the terms manifest. To understand the problem and how to remedy it consider the sum multiplying the $\left(\partial\right)^+_1 \widehat{f}^+$ term in the focal expansion of $\mathcal{J}_{4,a}[f]$:
\begin{align}\label{eq: exp term example}
    J_1(z)-\frac{24J_2(z)}{z}+\frac{120 J_3(z)}{z^2}
\end{align}
where we set $2\pi a \rho\equiv z$. As it stands, all three terms come in at the same order for small $z$, but we have a lot of freedom to manipulate \cref{eq: exp term example} via the recurrence \cref{eq: Bessel recurrence 1/x}. We can write it as
\begin{align} \label{eq: rec appendix}
    J_n(z)=\frac{2(n+1)}{z}J_{n+1}(z)-J_{n+2}(z).
\end{align}
Note that the first term on the right hand side comes in at $O(z^n)$ -- just like the original $J_n$ -- but the second is a strictly higher order correction at $O(z^{n+2})$. In other words, we can push the leading order onto terms composed of higher order Bessel functions divided by powers of the argument, while corrections will come in via even higher-order Bessels divided by \textit{lower} powers of the argument. The powers of $1/z$ have the added benefit of being well-behaved for large arguments, so doing this will not alter the clarity of which term in \cref{eq: exp term example} is leading for large $z$.

This suggests a strategy: we will push \textit{all} leading $O(z^1)$ behavior of \cref{eq: exp term example} to term proportional to $J_3(z)/z^2$ via repeated use of \cref{eq: rec appendix}. We will need to use the recurrence once for the $J_2(z)/z$ term
\begin{align} \label{eq: J1/z example}
    \frac{J_2(z)}{z}=\frac{6}{z^2}J_3(z)-\frac{J_4(z)}{z}
\end{align}
but twice on the $J_1(z)$ term (once on the original term and then again on every resulting term):
\begin{align}
        J_1(z)=&\frac{4}{z}J_{2}(z)-J_{3}(z) \label{eq: J0 example 1}\\
              =&\frac{24}{z^2}J_3(z)-\frac{12}{z}J_4(z)+J_5(z).\label{eq: J0 example 2}
\end{align}
Note that we have to apply the recurrence again to each term in \cref{eq: J0 example 1} if we want to avoid any two terms in the expansion of \cref{eq: exp term example} having the same order; $J_3(z)$ in \cref{eq: J0 example 1} and $J_4(z)/z$ in \cref{eq: J1/z example} both are $O(z^3)$ for instance. Putting \cref{eq: J1/z example,eq: J0 example 2} together we get
\begin{align}\label{eq: example final}
    J_1(z)-\frac{24J_2(z)}{z}+\frac{120 J_3(z)}{z^2}=J_5(z)+\frac{12 J_4(z)}{z}
\end{align}
where the $J_3/z^2$ term has vanished completely. We thus see that the naive guess of the leading order of \cref{eq: exp term example} being $O(z^1)$ was wrong: we get non-trivial cancellations that mean the leading term is actually $O(z^3)$. We see \cref{eq: example final} makes \textit{both} the small $z$ and large $z$ behaviors completely manifest, with no possible hidden cancellations remaining.

We now generalize this reasoning. The first step is writing down the form of an arbitrary $J_m(z)$ after applying \cref{eq: rec appendix} an arbitrary number $N$ of times. One can show by induction in $N$ that the required expression is
\begin{align}\label{eq: Bessel recurrence N times}
    \begin{split}
        J_m(z)=\sum_{j=0}^N\binom{N}{j}(m+j+1)^{\uparrow}_{N-j}\frac{(-1)^j  J_{N+m+j}(z)}{(z/2)^{N-j}}.
    \end{split}
\end{align}
We must now determine the appropriate number of iterations $N$ of the recurrence \cref{eq: rec appendix} to use for each of the $l$ sums in \cref{eq: focal expansion even bessel,eq: focal expansion odd bessel}. Just like we did in \cref{eq: exp term example}, we want to push all leading order small argument dependence to the term with the largest power of $1/z$. 

Consider for example the sum multiplying $\left(\partial\right)^+_k\widehat{f}^+$ in the even Bessel focal expansion \cref{eq: focal expansion even bessel}
\begin{align}\label{eq: bessel sumA1}
    \sum_{l=0}^n \frac{(2k+2l-1)!!}{k!}\frac{(-2)^l}{1+\Omega_{l,n}}\binom{n+l}{2l}\frac{J_{k+l}(z)}{(z/2)^l}
\end{align}
where we again set $z=2\pi a \rho$. As it stands, all terms come in at the same $O(z^k)$ order, we want this leading order to be represented only by the $l=n$ term $J_{k+n}(z)/z^n$. This means we should use \cref{eq: Bessel recurrence N times} to write
\begin{align}\label{eq: bessel sumA2}
    \begin{split}
        &\frac{J_{k+l}(z)}{z^l}=\\
        &\quad \sum_{j=0}^{n-l}\binom{n-l}{j}(k+l+j+1)^{\uparrow}_{n-l-j}\frac{(-1)^j  J_{n+k+j}(z)}{(z/2)^{n-j}}.
    \end{split}
\end{align}
Note that \cref{eq: bessel sumA2} completely removes all $l$ dependence from the Bessel function orders and powers of $z$. This means that upon plugging \cref{eq: bessel sumA2} into \cref{eq: bessel sumA1} the sum over $l$ becomes a hypergeometric sum with $n,j,k$ as free parameters -- to get the desired expansion we just need to evaluate it. 

A completely analogous argument holds for all of the other $l$ sums of Bessel functions in \cref{eq: focal expansion even bessel,eq: focal expansion odd bessel} -- we pick the appropriate application of \cref{eq: Bessel recurrence N times}, plug it in, and evaluate the resulting hypergeometric sum over the index $l$. We skip the details but list said hypergeometric sums as they are non-standard; each one can be proved with Zeilberger's algorithm \cite{AB}. The two identities corresponding to the $l$ sums in \cref{eq: focal expansion even bessel} are
\begin{align}
\begin{split}
        \sum_{l=0}^{n-j} &(-4)^l\frac{(k+1/2)^{\uparrow}_l (n+l-1)!}{(2l)! (n-l-j)! (k+l+j)!}\\
    =&\frac{(-1)^{n-j}\;j!}{(k+n)! \;n}\binom{2n}{2j}(k)^{\downarrow}_{n-j},
\end{split}\\
&\text{\phantom{filler}}\nonumber\\
\begin{split}
        \sum_{l=0}^{n-j}&(-4)^l\frac{(k+1/2)^{\uparrow}_l (n+l-1)!}{(2l)! (n-l-j)! (k+l+j+1)!}\\
        =&\frac{(-1)^{n-j}\; j!}{(k+n+1)!\; n}\binom{2n+1}{2j+1}(k+1)^{\downarrow}_{n-j}\, p_{n,k}(j),
\end{split}\\
&\text{\phantom{filler}}\nonumber\\
p_{n,k}&(j)\equiv \frac{2j(1+k+n)-(2n^2-k-1)}{(1+k)(1+2n)},\label{eq: p def}
\end{align}
for the $(\partial)_k^+\widehat{f}^+$ and $(\partial)_k^-\widehat{f}^-$ terms respectively, while the two corresponding to the $l$ sums in \cref{eq: focal expansion odd bessel} are
\begin{align}
    \begin{split}
        \sum_{l=0}^{n-j}&(-4)^l\frac{(k+1/2)^{\uparrow}_l (n+l)!}{(2l)! (n-l-j)! (k+l+j+1)!}\\
        =&\frac{(-1)^{n-j}\; j!}{(k+n+1)!}\binom{2n+1}{2j+1}(k)^{\downarrow}_{n-j},
    \end{split}\\
    &\text{\phantom{filler}}\nonumber\\
    \begin{split}
        \sum_{l=0}^{n+1-j}&(-4)^l\frac{ (k+1/2)^{\uparrow}_l(n+l-1)!(n(n+1)+l/2)}{(2l)! (k+l+j)!(n+1-l-j)!}\\
        =&\frac{(-1)^{n+1-j}\;j!}{(k+n+1)!}\binom{2n+2}{2j}(k+1)^{\downarrow}_{n-j+1}\, q_{n,k}(j),
    \end{split}\\
    &\text{\phantom{filler}}\nonumber\\
    q_{n,k}&(j)\equiv\frac{2j(3/2+k+n)-(2n^2+3n+1)}{(1+k)(2n+2)},\label{eq: q def}
\end{align}
for the $(\partial)_k^+\widehat{f}^+$ and $(\partial)_k^-\widehat{f}^-$ terms respectively. Using these identities in the resummation process described above lets us write the alternative forms for the focal expansions \cref{eq: focal expansion odd bessel,eq: focal expansion even bessel}
\begin{align}
    \begin{split}\label{eq: focal expansion even bessel alt}
        \mathcal{J}_{2n,a}&\left[f\right](\rho)=\sum_{k=0}^{\infty}\frac{(2k-1)!!}{k!\;(-4 \pi a)^k}\Bigg{\{ }\\
    &\left(\left(\partial\right)^+_k \widehat{f}^+\right)(\rho)\Bigg[ \sum_{j=0}^{n}\frac{J_{k+n+j}(2\pi a\rho)}{(\pi a \rho)^{n-j}}\binom{2n}{2j}(k)^{\downarrow}_{n-j}\Bigg]\\
    -i&\left(\left(\partial\right)^-_k \widehat{f}^-\right)(\rho)\Bigg[\sum_{j=0}^{n}\frac{J_{k+n+j+1}(2\pi a\rho)}{(\pi a \rho)^{n-j}}\\
    &\text{\phantom{aaaaaaaaaaaaaa}}\times\binom{2n+1}{2j+1}(k+1)^{\downarrow}_{n-j}\;p_{n,k}(j)\Bigg]\Bigg{\}},\\
    &\text{\phantom{filler}}
    \end{split}
\end{align}
\begin{align}
    \begin{split}\label{eq: focal expansion odd bessel alt}
        \mathcal{J}_{2n+1,a}&\left[f\right](\rho)=\sum_{k=0}^{\infty}\frac{(2k-1)!!}{k!\;(-4 \pi a)^k}\Bigg{\{ }\\
    &\left(\left(\partial\right)^+_k \widehat{f}^+\right)(\rho)\Bigg[ \sum_{j=0}^{n}\frac{J_{k+n+j+1}(2\pi a\rho)}{(\pi a \rho)^{n-j}}\\
    &\text{\phantom{aaaaaaaaaaaaaa}}\times\binom{2n+1}{2j+1}(k)^{\downarrow}_{n-j}\Bigg]\\
    -i&\left(\left(\partial\right)^-_k \widehat{f}^-\right)(\rho)\Bigg[\sum_{j=0}^{n+1}\frac{J_{k+n+j+1}(2\pi a\rho)}{(\pi a \rho)^{n+1-j}}\\
    &\text{\phantom{aaaaaaaaaa}}\times\binom{2n+2}{2j}(k+1)^{\downarrow}_{n-j+1}\;q_{n,k}(j)\Bigg]\Bigg{\}},\\
    &\text{\phantom{filler}}
    \end{split}
\end{align}
with $p_{n,k}$ and $q_{n,k}$ defined in \cref{eq: p def,eq: q def}. Remarkably, one can simplify this even further. Changing summation variables $j= n-m$ in both $j$ sums of \cref{eq: focal expansion even bessel alt} and the first $j$ sum of \cref{eq: focal expansion odd bessel alt}, while making the replacement $j= n+1-m$ in the last $j$ sum of \cref{eq: focal expansion odd bessel alt} we get
\begin{align}\label{eq: focal expansion alt master}
    \begin{split}
        \mathcal{J}_{N,a}&\left[f\right](\rho)=\sum_{k=0}^{\infty}\frac{(2k-1)!!}{k! (-4\pi a)^k}\Bigg\{\\
        &\left(\left(\partial\right)^+_k \widehat{f}^+\right)(\rho)\Bigg[\sum_{m=0}^{\min(\lfloor N/2\rfloor,k)}\frac{J_{N+k-m}(2\pi a \rho)}{(\pi a \rho)^m}\\
        &\text{\phantom{aaaaaaaaaaaaaaa}}\times\binom{N}{N-2m}(k)^{\downarrow}_m\Bigg]\\
        -i&\left(\left(\partial\right)^-_k \widehat{f}^-\right)(\rho)\Bigg[\sum_{m=0}^{\min(\lceil N/2 \rceil,k+1)}\frac{J_{N+k-m+1}(2\pi a\ \rho)}{(\pi a \rho)^m}\\
        &\text{\phantom{aaaaaaaaaaa}}\times\binom{N+1}{N-2m+1}(k+1)^{\downarrow}_mQ_{N,k}(m)\Bigg]\Bigg\}
    \end{split}
\end{align}
where
\begin{align}\label{eq: Q def}
    Q_{N,k}(m)\equiv 1-m \frac{N+2(k+1)}{(k+1)(N+1)}
\end{align}
and $N$ is an arbitrary non-negative integer. Note in particular that now we need not state separately the results for even and odd Bessel kernels.

The expression \cref{eq: focal expansion alt master} has precisely the transparent limiting behavior we had hoped for: the terms corresponding to the largest $m$ give us the leading behavior for small $\rho$, while the term corresponding to $m=0$ give us the leading behavior for large $\rho$. This form of our focal expansion also has the added benefit of naturally including the edge $\mathcal{J}_{0,a},\mathcal{J}_{1,a}$ cases without the need of special conditions like those in \cref{eq: Omega,eq: Omega prime}.

\section{Breakdown of the General Focal Expansion \label{app: counterexamples}}

In this appendix we briefly highlight two non-Bessel kernels for which the general focal expansion \cref{eq: Focal Expansion} does not yield meaningful results.

The first kernel we chose is the Cauchy distribution
\begin{align}
    K_C(z)\equiv\frac{1}{1+z^2}
\end{align}
well known for having undefined mean, variance, and all other higher moments. Here we have not found all of the focal coefficients $A^{\pm}_{k,n}$ but instead used the algorithmic approach described at the end of section~\ref{subsec: The Focal Expansion} to find all coefficients with $0\leq k\leq 10$ and $0\leq n\leq 10-k$ (the $n$-dependent upper bound on $l$ results naturally in this method). Here we list the first several of these coefficients for completeness:
\begin{figure}
    \centering
    \includegraphics[width=\linewidth]{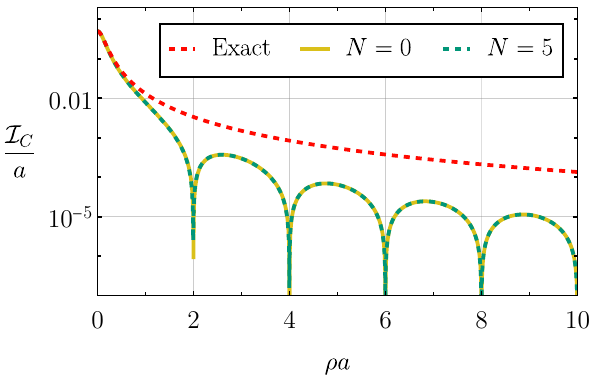}
    \caption{The comparison of the exact Cauchy kernel integral \cref{eq: appendix cauchy integral exact} to the $N=0, 5$ orders of its focal expansion (for the definition of $N$th focal order in this case see the text). Not only does the focal expansion provide a poor approximation beyond $\rho a \approx 1$, it also appears that subsequent focal orders are far too small to change the situation in any way. We see virtually no change in the value of the function when going from $N=0$ to $N=5$.}
    \label{fig: Cauchy appendix}
\end{figure}
\begin{figure}
    \centering
    \includegraphics[width=\linewidth]{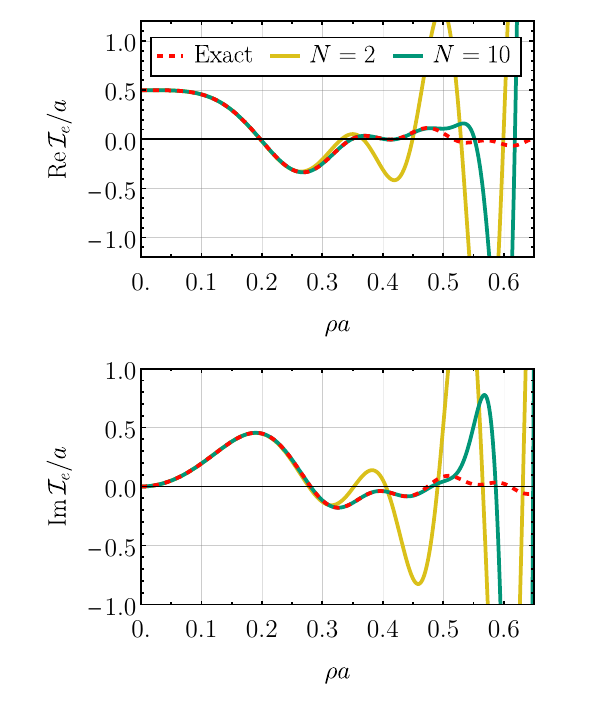}
    \caption{Comparison of the real and imaginary parts of the exact oscillatory integral \cref{eq: Fresnel appendix} to its $N=2,10$ order focal approximations (for the definition of $N$th order in this case see the text). We see that, unlike in fig.~(\ref{fig: Cauchy appendix}), successive focal orders do appear to significantly improve the performance of the focal expansion. This improvement, however, cannot hide the fact that for larger arguments the focal expansion becomes divergent -- we appear to have lost the key asymptotic accuracy of the focal expansion we saw in the case of Bessel kernels. As such, the focal expansion in this case is of no practical use -- a Taylor series will perform the same task with significantly less complexity involved.}
    \label{fig: exp appendix}
\end{figure}
\begin{align}
    \begin{split} \label{eq: A examples Cauchy}
        A^+_{0,0}&=1,\quad  A^+_{0,1}=A^+_{0,2}=0, \\
        A^+_{1,0}&=-\frac{1}{8\, a \pi}, \quad A^+_{1,1}=\frac{3}{64 \pi a},\quad  A^+_{1,2}=\frac{1}{1536 \pi a}, \\
        A^+_{2,0}&=\frac{21}{256 \pi^2 a^2},\quad A^+_{2,1}= -\frac{137}{6144 \pi^2 a^2}, \\
        &\text{\phantom{filler}}\\
        A^-_{0,0}&=i,\quad  A^-_{0,1}=A^-_{0,2}=0, \\
        A^-_{1,0}&=-\frac{11 i}{16\, a \pi}, \quad A^-_{1,1}=\frac{25 i}{768 \pi a},\quad  A^-_{1,2}=\frac{13 i}{46080 \pi a}, \\
        A^-_{2,0}&=\frac{337 i}{768 \pi^2 a^2},\quad A^-_{2,1}= -\frac{947 i}{30720 \pi^2 a^2}.
    \end{split}
\end{align}
The major difference between this case and the Bessel function kernels is that here the sum over $n$ in the focal expansion \cref{eq: Focal Expansion} does not appear to truncate and we are therefore dealing with a doubly infinite sum. Whenever choosing a finite order one must then choose some truncation scheme for both sums, we opt for the one suggested by the algorithmic approach: at order $N$ we sum all terms with $0\leq k\leq N$ and $0\leq n \leq N-k$.

For the function $f(x)$ being integrated we pick a simple top hat of width $w$ and height 1 so the integral we are approximating is
\begin{align}\label{eq: appendix cauchy integral exact}
    \begin{split}
        \mathcal{I}_C(\rho)&\equiv \int_{a-w/2}^{a+w/2}\frac{dx}{1+(2\pi\rho x)^2}\\
        &=\frac{\tan^{-1}\left(2\pi \rho \left(a+\frac{w}{2}\right)\right)-\tan^{-1}\left(2\pi \rho \left(a-\frac{w}{2}\right) \right)}{2\pi \rho}
    \end{split}
\end{align}
where we have also evaluated it exactly. To get the focal expansion we use the expression for $\widehat{f}$ from section~\ref{sec: annulus general calc}. In fig.~(\ref{fig: Cauchy appendix}) below we have plotted a comparison between the exact result \cref{eq: appendix cauchy integral exact} and the focal expansion at $N=0, 5$ orders for the case of $w/a=0.5$. 
We see that we obtain the right small argument behavior but the focal expansion then soon loses relevance. What's more, the subleading corrections are so small as to have virtually no effect on the accuracy. We make no further comment on this strange behavior other than to raise it as a cautionary counter-example.

The second non-Bessel kernel we consider is the quadratically chirped phase
\begin{align}
    K_e(z)=e^{i z^2}.
\end{align}
Calculating the focal coefficients algorithmically one will again find that the sum over $n$ in \cref{eq: Focal Expansion} does not appear to truncate. The several first coefficients are:
\begin{align}
    \begin{split} \label{eq: A examples exp}
        A^+_{0,0}&=1,\quad  A^+_{0,1}=A^+_{0,2}=0, \\
        A^+_{1,0}&=\frac{i-2}{8a\pi}, \quad A^+_{1,1}=\frac{i}{8 \pi a},\quad  A^+_{1,2}=0, \\
        A^+_{2,0}&=\frac{11-4i}{128 \pi^2 a^2},\quad A^+_{2,1}=-\frac{1+6i}{64 \pi^2 a^2}, \\
        &\text{\phantom{filler}}\\
        A^-_{0,0}&=i,\quad  A^-_{0,1}=A^-_{0,2}=0, \\
        A^-_{1,0}&=-\frac{1+6i}{8\pi a}, \quad A^-_{1,1}=-\frac{1}{8\pi a},\quad  A^-_{1,2}=0, \\
        A^-_{2,0}&=\frac{12+59 i}{128 \pi^2 a^2},\quad A^-_{2,1}= \frac{10-i}{64 \pi^2 a^2}.
    \end{split}
\end{align}
We choose the same $f(x)$ as in the previous example in the name of consistency. Here too one can evaluate the integral we're after analytically using special functions:
\begin{align}\label{eq: Fresnel appendix}
    \begin{split}
        \mathcal{I}_e(\rho)\equiv& \int_{a-w/2}^{a+w/2}e^{i (2 \pi \rho x)^2}\text{d}\,x\\
        =&\frac{C\left(\sqrt{2 \pi } (w-2 a) \rho \right)+C\left(\sqrt{2 \pi
   } (w+2a) \rho \right)}{2 \sqrt{2 \pi } \rho }\\
   &+i \frac{S\left(\sqrt{2 \pi }
   (w-2a) \rho \right)+S\left(\sqrt{2 \pi } (w+2a) \rho
   \right)}{2 \sqrt{2 \pi } \rho }
    \end{split}
\end{align}
where $C(z), S(z)$ are the Fresnel integrals
\begin{align}
    S(z)=\int_0^z\sin(x^2)\text{d}x,\qquad C(z)=\int_0^z\cos(x^2)\text{d}x.
\end{align}
In fig.~(\ref{fig: exp appendix}) we have compared the exact result \cref{eq: Fresnel appendix} to the focal expansion with $w/a=0.5$ (where the $N$th order here is understood as truncating both infinite sums in the focal expansion in the exact same manner as in the previous example).
We see the situation here is less dire than for the Cauchy distribution as there's significant improvement in accuracy as we increase the order. The problem here is that the focal expansion clearly begins diverging for larger values, making it no better than a simple Taylor series. 

It remains to be seen if the general focal expansion is useful beyond the case of Bessel function kernels. We have seen in section~\ref{sec: examples} it works very well for integer order Bessel kernels and so intuitively we suspect that it will also work well for functions `close' (in some sense) to integer order Bessels, but the two counter-examples of this appendix temper our hopes for generic kernels. At the very least, we see no reason to expect the `pincer maneuver' reasoning present in the case of Bessel kernels is replicated for other generic kernels. Perhaps one can tweak the focusing operators or the exact form of our expansion based on what family of kernels one is dealing with, but we leave all such analysis to future work.

\bibliography{apssamp.bib}

\end{document}